\documentclass[fleqn,10pt]{wlscirep}
\usepackage[utf8]{inputenc}
\usepackage[T1]{fontenc}
\usepackage{eucal}
\usepackage{multirow}
\usepackage{tabularx}
\usepackage{caption}
\usepackage{rotating}
\usepackage{color}
\usepackage{float}
\usepackage{lmodern}
\usepackage{bm}
\usepackage{amssymb}
\usepackage{comment}

\newcommand{\Jx}{J(\bm{x})}
\newcommand{\qx}{q(\bm{x})}

\usepackage{wasysym}
\usepackage{amsthm}

\usepackage{soul}
\usepackage{subfigure}

\usepackage{algorithm}
\usepackage{algpseudocode}

\newcommand{\Bx}{\bm{x}}
\newcommand{\Ba}{\bm{a}}
\newcommand{\BJ}{\bm{J}}
\newcommand{\Btheta}{\bm{\theta}}

\title{Probing reaction channels via reinforcement learning}

\author[1]{Senwei Liang}
\author[2,3]{Aditya N. Singh}
\author[1]{Yuanran Zhu}
\author[2,3,4,5]{David T. Limmer}
\author[1,$^\dag$]{Chao Yang}
\affil[1]{Applied Mathematics and Computational Research Division, Lawrence Berkeley National Laboratory, Berkeley, CA 94720, USA}
\affil[2]{Department of Chemistry, University of California Berkeley, Berkeley, CA 94720, US}
\affil[3]{Chemical Sciences Division, Lawrence Berkeley National Laboratory, Berkeley, CA 94720, USA}
\affil[4]{Materials Science Division, Lawrence Berkeley National Laboratory, Berkeley, CA 94720, USA}
\affil[5]{Kavli Energy Nanoscience Institute at Berkeley, Berkeley, CA 94720, USA}
\affil[$^\dag$]{Correspondence should be addressed to: cyang@lbl.gov}

\begin{abstract}
We propose a reinforcement learning based method to identify important configurations that connect reactant and product states along chemical reaction paths. By shooting multiple trajectories from these configurations, we can generate an ensemble of configurations that concentrate on the transition path ensemble. This configuration ensemble can be effectively employed in a neural network-based partial differential equation solver to obtain an approximation solution of a restricted Backward Kolmogorov equation, even when the dimension of the problem is very high. The resulting solution, known as the committor function, encodes mechanistic information for the reaction and can in turn be used to evaluate reaction rates.

\end{abstract}

\begin{document}
\flushbottom
\maketitle

\thispagestyle{empty}

\section{Introduction}
\label{sec:intro}

The study of rare reactive events is a fundamental topic within the field of chemical physics.\cite{peters2017reaction} These reactions are dynamical processes that can be characterized by the transition of a molecular system from one collection of meta-stable atomic configurations, i.e., a reactant, to another, i.e., a product. This metastability arises from the features of the potential energy surface, where metastable states are configurations near the local minimizer  separated by high energy saddle points. When the saddle points lie along the typical transition paths between reactants and products, they are referred to as transition states. The direct numerical studies through molecular dynamics (MD) simulations are typically prohibitive computationally because these reactions are rare relative to the timescales of the typical thermal fluctuations of the system\cite{chandler1978statistical}. 

A primary quantity of interest that has been employed to infer a mechanistic understanding of reactive events is the committor, a function that maps the phase space of the system to the probability of  reacting.\cite{bolhuis2002transition,geissler1999kinetic} This function encodes the ideal reaction coordinate, and the computation of this function can be framed as a multidimensional optimization problem.\cite{weinan2002string,vanden2006towards,vanden2010transition} The high-dimensionality of this problem poses difficulties in its inference, however a multitude of methods have been developed over the past decade that have made the computation of this function tractable. Some notable examples include the string method\cite{weinan2002string,vanden2006towards}, diffusion maps\cite{coifman2008diffusion,thiede2019galerkin,evans2022computing,trstanova2020local} and neural networks\cite{ma2005automatic,rotskoff2022active,li2019computing,khoo2019solving,hasyim2021supervised,strahan2022forecasting,strahan2023inexact,jung2023machine,singh2023variational,zhu2023learning}.

This optimization of the committor can be framed as two separate but intertwined problems -- (1) finding configurations with high reactive densities and (2) fitting a nonlinear ansatz to solve the optimization problem on those configurations to compute the committor. To solve the first problem, one needs a measure of the reactive density that depends on the committor itself, and to solve the second problem, one needs to access those configurations. One way to solve the first problem is to obtain an ensemble of reactive trajectories via Transition Path Sampling~\cite{dellago2002transition,bolhuis2021transition,bolhuis2002transition,dellago2006transition}, a Monte-Carlo based method that enables generation of new reactive trajectories from old ones. While this method provides access to the ideal set of configurations to compute the committor on, it often suffers from low acceptance rates resulting in long decorrelation times between subsequent trajectories, and further refinements based on approximations of the committor are required to enable efficient sampling. Beyond Transition Path Sampling, a wide range of methods have been developed in the past two decades to access rare but important configurations that encode information of reactive events. Such examples include but are not limited to metadynamics\cite{barducci2008well,barducci2011metadynamics}, weighted ensemble method\cite{huber1996weighted,zwier2015westpa} and variational enhanced sampling\cite{valsson2014variational}.  These methods rely on applying a carefully chosen bias potential to the original Hamiltonian along some low-rank ansatz of the reaction coordinate, often referred to as an order parameter to generate more atomic configurations near the transition state. While these methods can obtain accurate estimates of thermodynamic and kinetic observables, they strongly hinge on the overlap between the order parameter and the reaction coordinate. As such, the ensembles of configurations are often limited by the choice of the order parameter.

In this paper, we use a reinforcement learning (RL) based method to obtain an ensemble of configurations with high reactive densities. This is done through a two-step method in which we first find saddle points along the reactive probability density, which we refer to as \textit{connective configurations}. These configurations can be thought of as the maxima of reactive probability along each reactive channel of the reactive process. This optimization procedure allows for the discovery of saddle points along each reactive channel when multiple reactive pathways exist. The optimization proceeds through an RL algorithm where an agent moves from one state to another by taking an action in order to achieve a certain goal. Each action is associated with a reward, and which action to take depends on a policy function designed to guide the agent toward the goal. In the context of searching for connective configurations, a state is an atomic configuration, an action simply moves the agent from one configuration to another according to a policy that is updated (or learned) over time.  The optimal policy, which is obtained after performing several RL episodes with each episode consisting of a sequence of actions, would allow us (the agent) to move from any arbitrary configuration toward a connective configuration. In our RL algorithm, the reward function, which we will describe in detail in section~\ref{sec:method}, is chosen as a proxy to the true objective function to be maximized. While such RL based methods have been previously used to identify transition states~\cite{zhang2021deep} and perform transition path sampling~\cite{das2021reinforcement,das2022direct,lelievre2022generative} or cloning \cite{das2019variational,rose2021reinforcement}, the proposed method has a different objective and yields different quantities of interest. Once we have obtained these configurations, we perform shooting operations from these points to obtain a set of configurations along a relatively small subspace in the configuration space, which we refer to as \textit{reaction channels}. The notion of reaction channels is similar to the concept of transition tubes introduced in Ref.~\cite{vanden2010transition}. Both are expected to concentrate on the reactive path ensemble.

Finally, once we obtain these configurations within the reaction channels, we train a feed-forward neural network (NN) to solve the backward-Kolmogorov equation (BKE). While a range of methods have been proposed to approximate a committor using an NN, the method demonstrated in this work is unique in that it solves the exact BKE rather than the variational form\cite{khoo2019solving,rotskoff2022active,hasyim2021supervised,li2019computing} or the Feynman-Kac form\cite{strahan2022forecasting,strahan2023inexact}. This form of optimization is more accurate, however similar to other methods, it is strongly sensitive to the configurations that it is trained on. We find that training on samples obtained from RL based method proposed in this work outperforms optimization on samples generated from a grid-based or high-temperature based sampling method by at least one order of magnitude. Beyond the accuracy of the committor, the fidelity of this method is also demonstrated through accurate estimates of reactive rates, that are often non-trivial to converge. 

The rest of the paper is organized as follows. In Section~\ref{sec:pre}, we introduce terminologies and concepts that are relevant to the approach we take to study chemical reactions. We outline the basic scheme we use with some detail in Section~\ref{sec:method}.
Section~\ref{sec:rlenv} is devoted to the details of the RL algorithm we use to identify connective configurations. In particular, we discuss how to define the reward function and design an effective policy.  We discuss how to use NN to obtain the values of the committor function at selected configurations generated within reaction channels in Section~\ref{sec:reactchannelrate}.  We demonstrate the effectiveness of our approach using a few examples in Section~\ref{sec:num}. We conclude the paper with some additional perspectives in Section~\ref{sec:conclu}. Some of the computational details and discussions are provided in the appendix.

\section{Preliminaries}
\label{sec:pre}
In this section, we introduce terminologies and concepts needed to present our algorithms in subsequent sections. Specifically, we define a transition path and reactive trajectory associated with an overdamped Langevin dynamics, the committor function associated with transition paths, and describe how the reaction rate constant can be calculated from the committor function. We then show how the committor function can be computed by using a feed-forward neural network.  
\subsection{Transition path and reactive trajectory}
\label{sec:preTPT}
Let $\Bx(t) = (x_1(t),x_2(t),...,x_d(t)) \in \mathbb{R}^d$ be the coordinates of a system that satisfy an overdamped Langevin dynamics defined by
\begin{align}
\gamma_i\dot{x_i}(t)=-\frac{\partial V(\Bx(t))}{\partial x_i}+\xi_i(t), \ \  i=1,\cdots, d,
\label{eqn:overdamplangvin}
\end{align}
where $V:\Omega\to \mathbb{R}$ is a potential energy function, $\gamma_i$ is the friction coefficient for $x_i$, $\xi_i(t)$ is white noise with mean $\langle \xi_i(t) \rangle$ and variance $\langle \xi_i(t) \xi_j(t') \rangle = 2\beta^{-1} \gamma_i \delta(t-t')\delta_{ij}$ and $\beta$ is the inverse of the product of temperature and Boltzmann's constant. 

Let $A$ and $B$ be two disjoint metastable regions of interest in the configuration space $\Omega\subset \mathbb{R}^d$. 
They correspond to regions surrounding two distinct local minima of the potential energy $V(\Bx)$.  We are interested in trajectories that start in $A$ and terminate in $B$. Along such a trajectory,  $\Bx(t)$ must escape out of one metastable region and cross over a transition region $\Omega\setminus(A\cup B)$ before reaching another metastable region. Such a trajectory is often referred to as a \textit{transition path}.
For chemical systems, a transition region is associated with a chemical reaction. A transition path is also referred to as a \textit{reactive trajectory}. 

The probability density of $\Bx$ follows the Boltzmann-Gibbs distribution $p(\Bx)=\exp(-\beta V(\Bx))/Z$, where $Z=\int_\Omega \exp(-\beta V(\Bx)){\rm d} \Bx$ is a normalization constant or the partition function. The probability of observing $\Bx$ in a transition region relative to the probability of observing $\Bx$ in a metastable region is very low. As a result, a transition path or a reactive trajectory is a rare event and generally requires a very long period of simulation time to observe.  

\textbf{Committor function.} 
Reactive trajectories are not unique. In fact, a chemical reaction is often characterized by an ensemble of reactive trajectories. Along each reactive trajectory, we are particularly interested in configurations that are equally likely to evolve (or commit) to either one of the metastable regions $A$ and $B$. Such likelihood can be characterized by what is known as a \textit{committor function}.
To give a precise definition of a committor function $\qx$, let $\tau_D(\Bx)$ be the first hitting time of region $D$ when the dynamics is initiated from $\Bx$, i.e. $\tau_D(\Bx)$ represents the time it takes for the chemical system to first enter region $D$ when the dynamics starts from $\Bx$. A committor function, denoted by $q(\Bx)$, is defined as the probability that a trajectory $\Bx(t)$, starting from a point $\Bx(0)=\Bx$ within a given set $\Omega$, reaches $B$ before it reaches $A$, namely, $q(\Bx)={\rm Prob}\left(\tau_B<\tau_A \mid \Bx(0)=\Bx\right)$. It is well known that $q(\Bx)$ satisfies the backward Kolmogorov equation (BKE)
\begin{align}
\label{eqn:bke}
\mathcal{L}(q) &:=\sum_{i=1}^d\left(-\gamma_i^{-1}\frac{\partial V(\Bx)}{\partial x_i} \frac{\partial q(\Bx)}{\partial x_i}+\gamma_i^{-1}\beta^{-1} \frac{\partial^2 q(\Bx)}{\partial x_i^2}\right)=0,\ \ \\ \Bx &\in \Omega\setminus(A\cup B), \quad 
q(\Bx)=0, \ \Bx\in A, \quad q(\Bx)=1,\ \Bx\in B. \nonumber 
\end{align}
When dimension of $\Omega$ is low (2 or 3), this partial differential equation can be solved numerically by standard methods such as the finite difference or finite element method. However, when the dimension of $\Omega$ is high, it is not practical to use these methods to solve for $\qx$. We will discuss an alternative approach that uses a feed-forward neural network to solve for $\qx$ in Section~\ref{sec:committnn}.

\textbf{Probability of being reactive}. We use $\rho(\Bx)$ to denote the probability of observing a reactive trajectory crossing $\Bx$. It follows from the Markovian property of the underlying overdamped Langevin dynamics that $\rho(\Bx)$ can be written as the product of $1-q(\Bx)$, which describes the probability of the trajectory arriving at $\Bx$ from $A$ without going through $B$ (under the assumption of time reversibility at the statistical equilibrium \cite{metzner2006illustration}), $p(\Bx)$, which represents the probability of being at $\Bx$, and $q(\Bx)$, which describes the probability of reaching $B$ first after leaving $\Bx$, i.e.,
\begin{align}
\rho(\Bx)=(1-q(\Bx))q(\Bx)p(\Bx).
\label{eqn:reactiveprob}
\end{align}
A configuration $\Bx$ that has a high reactive density $\rho(\Bx)$ is of particular interest because it marks a transition region along a reactive trajectory associated with 
an overdamped Langevin dynamics.
However, the definition of $q(\Bx)$ given in \eqref{eqn:reactiveprob} requires the committor function $\qx$ to be known in advance.  Because $\qx$ is generally difficult to calculate for an arbitrary $\Bx$, it is not easy to calculate $\rho(\Bx)$ in practice.


\textbf{Reaction rate}. 
The committor function is used in Ref.~\cite{vanden2010transition} to introduce the notion of the \textit{current} or \textit{flux} associated with a reactive trajectory. At a configuration $\Bx$ along a reactive trajectory, the flux $\BJ(\Bx)$ across $\Bx$ is defined as
\begin{equation}
J_i(\Bx)=Z^{-1} e^{-\beta V(\Bx)} \gamma_i^{-1}\beta^{-1}\frac{\partial{q(\Bx)}}{\partial x_i},
\label{eqn:flux}
\end{equation}
where $\BJ(\Bx)=(J_1(\Bx),\cdots,J_d(\Bx))$ and $Z$ is again the partition function.

If $\BJ(\Bx)$ is known for all $\Bx$ along a dividing surface $S$ that separates $A$ and $B$, we can use it to evaluate the reaction rate constant $\kappa$ via
\begin{align}
\kappa=\int_S n_S(\Bx) \BJ(\Bx) d \sigma_S(\Bx),
\label{eqn:rate}
\end{align}
where $n_S(\Bx)$ is normal vector of $S$. 

Note that the main contribution to the integral \eqref{eqn:rate} comes from configurations $\Bx$ along the dividing surface that has a relatively large magnitude of the flux $\BJ(\Bx)$. Because these configurations typically occupy a small area $S'$ on $S$ for rare events~\cite{vanden2010transition,metzner2006illustration},  we can focus on this area and approximate~\eqref{eqn:rate} by  
\begin{equation}
\kappa \approx \int_{S'} n_{S'}(\Bx) \BJ(\Bx) d \sigma_S(\Bx).
\label{eqn:rate2}
\end{equation}
The area $S'$ can be defined by the intersection of $S$ and the so-called reaction channel to be defined in Sec.~\ref{sec:method}. 



We should note that as long as $S'$ can be easily identified and $\BJ(\Bx)$ can be efficiently evaluated for all $\Bx \in S'$, the formula~\eqref{eqn:rate2} provides a more practical way to compute the rate constant compared to a brute force approach in which the rate constant is computed according to the alternative definition
\begin{equation}
\kappa=\lim_{T\to \infty}\frac{N_T}{T},
\end{equation}
where $N_T$ is the number of trajectory segments that leave $A$ and enter $B$ within the time interval $[0,T]$~\cite{vanden2010transition}. In the latter approach, $\kappa$ can be approximated by a \textit{direct simulation}, i.e., we can generate a sufficiently long Langevin trajectory from a random starting point and count $N_T$. However, when the system contains a high barrier, it becomes extremely rare to observe a reactive trajectory even with a long time simulation. 

If the dividing surface is not accessible, an alternative method is to calculate the integral of the flux on the transition region, which can be done as follows:
\begin{align}
\kappa=\int_{\Omega\setminus(A\cup B)} Z^{-1} e^{-\beta V(\Bx)} \beta^{-1}\sum_{i=1}^d\gamma_i^{-1}\left(\frac{\partial{q(\Bx)}}{\partial x_i}\right)^2 d \Bx \, ,
\label{eqn:rate_sq}
\end{align}
for the systems considered here.

\subsection{Solving committor function via deep learning}
\label{sec:committnn}
In recent year, deep learning has demonstrated remarkable progress in various domains of scientific exploration~\cite{zhang2022identifying,nan2020deep,lin2022data}, thanks to the exceptional approximation and generalization capabilities of neural networks (NNs)~\cite{Shen4}. In particular, deep learning has emerged as a powerful tool for solving a wide range of partial differential equations (PDEs), including those formulated in high-dimensional spaces where conventional solvers like finite difference and finite element methods suffer from the curse of dimensionality~\cite{weinan2021algorithms}. Furthermore, NN-based solvers can be easily adapted to solve PDEs defined on an irregular domain. These two key attributes of an NN-based PDE solver make it highly advantageous for the specific problem we aim to address in this study. As we will show in Section~\ref{sec:num}, an NN-based PDE solver enables us to solve a 66-dimensional BKE associated with a 22-atom molecule within a reaction channel that consists of an ensemble of configurations not uniformly distributed or organized in a regular domain in the configuration space. The NN-based PDE solver overcomes these complexities and enables us to effectively handle this challenging scenario.

Typically in an NN-based PDE solver, the approximation to the solution of the PDE is represented as an NN $q(\Bx;\Btheta)$ parameterized by a set of weights and biases denoted by a vector $\Btheta$. The network takes $\Bx$ as the input and generates $q(\Bx;\Btheta)$ as the output. The NN parameters are determined in an iterative training procedure that minimizes a loss function with respect to $\Btheta$ for different choices of the input $\Bx$. 
To use deep learning to solve the BKE~\eqref{eqn:bke} on $\Omega$, we define the loss function as
\begin{align}
     \|\mathcal{L}(q(\Bx; \Btheta))\|^2_{L^2(\Omega)}+\ell\|q(\Bx; \Btheta)\|^2_{L^2(A)}+\ell\|q(\Bx; \Btheta)-1\|^2_{L^2(B)},
    \label{eqn:nnloss}
\end{align}
where $\ell$ is a penalty coefficient used to impose the boundary constraints. In practice, the $L^2$-norm in~\eqref{eqn:nnloss} is evaluated by summing the loss on the data points sampled randomly and uniformly in $\Omega$ and $\mathcal{L}$ is computed by auto-differentiation using advanced deep learning frameworks (e.g., Pytorch~\cite{paszke2019pytorch}). The NN-based optimization~\eqref{eqn:nnloss} can be conducted by stochastic gradient descent (SGD), such as Adam~\cite{kingma2014adam}, a variant of SGD based on momentum. Similarly, when solving BKE~\eqref{eqn:bke} on $K$ sub-domains $\Omega_1, \cdots, \Omega_K\subset \Omega$, the PDE becomes $\mathcal{L}(q)=0$ for $\Bx \in \cup_{s=1}^K\Omega_s$ along with the boundary conditions, which is referred to as \textit{restricted BKE}. We can define the NN-optimization problem by
\begin{align}
    \min_{\Btheta} \sum_{s=1}^K\|\mathcal{L}(q(\Bx; \Btheta))\|^2_{L^2(\Omega_s)}+\ell\|q(\Bx; \Btheta)\|^2_{L^2(A)}+\ell\|q(\Bx; \Btheta)-1\|^2_{L^2(B)}. 
    \label{eqn:nn-channels}
\end{align}
When we have data points $\{\Bx^{i,s}\}_{i=1}^{N_s}\subset \Omega_s$, $s=1,\cdots, K$, $\{\hat{\Bx}^{i}\}_{i=1}^{N_A}\subset A$ and $\{\tilde{\Bx}^{i}\}_{i=1}^{N_B}\subset B$, the loss function in~\eqref{eqn:nn-channels} can be evaluated as
\begin{align}
\frac{1}{\sum_{s=1}^K N_s}\sum_{s=1}^K\sum_{i=1}^{N_s}(\mathcal{L}q(\Bx^{i,s}; \Btheta))^2+\frac{\ell}{N_A}\sum_{i=1}^{N_A}q(\hat{\Bx}^{i}; \Btheta)^2+\frac{\ell}{N_B}\sum_{i=1}^{N_B}(q(\tilde{\Bx}^{i}; \Btheta)-1)^2.
\label{eqn:discrete}
\end{align}

\section{Methodology}
\label{sec:method}
As we indicated in Sec.~\ref{sec:preTPT}, configurations $\Bx$ with high reactive density $\rho(\Bx)$ are of interest because they mark a transition region in which reactive trajectories are more likely to be observed.  Intuitively, if we shoot a trajectory from a configuration $\Bx$ with a high reactive density $\rho(\Bx)$ and initiate it with a random momentum, it is likely that the trajectory will stay within a region where reactive trajectories pass through. Even though such a trajectory may not be part of a reactive trajectory, the configurations along such a trajectory occupy a small subspace that is likely to contain several reactive trajectories. We will refer to the subspace formed by these configurations as a \textit{reactive channel}.  Because configurations within such a subspace are likely to have high reactive flux $\BJ(\Bx)$, we will focus on these configurations, and solve a restricted BKE within a reactive channel using the neural network technique discussed in Sec.~\ref{sec:committnn} to obtain an approximate committor function and and its gradient at configurations within the channel. With these, one can approximately calculate the reaction rate constant by evaluating \eqref{eqn:rate2}.

We should note that the concept of reaction channel introduced here is similar in spirit to the notion of \textit{transition tube} introduced in Ref.~\cite{vanden2010transition}. A transition tube is defined to be an ensemble of regions on non-intersection dividing surfaces between two metastable regions $A$ and $B$ that have localized flux. Because a transition tube is characterized by configurations with relatively high reactive flux which depends on the unknown committor function, it is not easy to identify directly. Although the central curve within the transition tube can be approximated by the minimum energy path which can be computed by the string method~\cite{weinan2002string}, defining the region of the tube is still not trivial.


Because a reaction channel is generated by shooting trajectories from a single configuration, it is relatively easy to produce as long as we can select a proper configuration to shoot from.  Ideally, that configuration should be the one that has a high reactive density $\rho(\Bx)$. However, because $\rho(\Bx)$ is defined in terms of the committor function, it is not easy to identify configurations with high $\rho(\Bx)$ directly because that would require solving the original BKE \eqref{eqn:bke}.  In the following, we will present a reinforcement learning-based technique to identify configurations that are likely to have a high $\rho(\Bx)$, and we will refer to these configurations as \textit{connective configurations}.


To create reaction channels, we start by performing a shooting procedure from connective configurations. Within each reaction channel, we can determine the committor function on each configuration by solving a restricted BKE using a neural network. Using the NN solution and its gradient, we can then calculate the reactive flux for every configuration within the reaction channels. As we will see in the next section, the reaction channel generated by shooting trajectories from connective configurations is likely to contain configurations with relatively high reactive flux. This is sufficient to provide a good estimation of statistics, such as rate, even if not all configurations within the channel have high reactive flux.

\subsection{Seeking connective configurations via reinforcement learning}
\label{sec:rlenv}
In this section, we show how to use a reinforcement learning (RL) method to identify connective configurations. Our basic strategy is to treat each configuration as a state $\Bx^t$ and train an agent to take a sequence of actions $\{\Ba^0, \Ba^1, \cdots, \Ba^n\}$ to move from an arbitrary state $\Bx^0$ to $\Bx^1$, $\Bx^2, \cdots$ successively through the operation $\Bx^{t+1} = \Bx^{t} + \Ba^{t}$, for $t=0,1,\cdots,n-1$, until it ultimately reaches a desired state $\Bx^n$ which corresponds to a connective configuration. The sequence of state-action pairs $\{(\Bx^t,\Ba^t)\}$, with $t=0,1,2,..,n-1$ is an instance of a policy $\pi$ the agent follows, which is initially not optimal.  However, over a multi-episode learning process, the policy is gradually improved based on the feedback the agent receives from the environment, which consists of configurations not being visited, through a policy gradient. An optimal policy allows the agent to move from an arbitrary state to the desired state efficiently.

In an RL algorithm, an action that an agent takes at a particular state $\Bx$ is associated with a reward $r(\Bx,\Ba)$ that measures the effectiveness of that state action pair.  The policy an agent follows at a particular state $\Bx$ is often designed to maximize not just the reward $r(\Bx,\Ba)$, but the expectation of a sequence of discounted future rewards, i.e.,
\begin{equation}
\mathbb{E}_{\tau\sim\pi} \left[R(\tau) | \Bx^0=\Bx, \Ba^0 = \Ba \right],
\end{equation}
where $\tau$ is an instance of a policy $\pi$ that is specified by a sequence of state action pairs $\tau:=\{(\Bx^0, \Ba^0),\cdots, (\Bx^t, \Ba^t), \cdots\}$, and 
\[
R(\tau):=\sum_{t=1}^\infty \eta^t r(\Bx^t, \Ba^t),
\]
for some discount factor $0<\eta\leq 1$. Such expectation of discounted future rewards is often referred to as a $Q$-value function or $Q$-function in short and denoted by $Q^{\pi}(\Bx,\Ba)$.

We use $A(\Bx)$ to denote the action that maximizes $Q^{\pi}(\Bx,\Ba)$ for a given policy $\pi$, i.e.,
\begin{align}
    A(\Bx)=\arg \max _{\Ba} Q^\pi(\Bx, \Ba).
    \label{eqn:action}
\end{align}

It is well known that the optimal $Q$-function $Q^*(\Bx, \Ba):=\max_{\pi}Q^{\pi}(\Bx,\Ba)$ satisfies the Bellman equation~\cite{sutton2018reinforcement}
\begin{equation}
Q^*(\Bx, \Ba)={\mathbb{E}}_{\Bx^{\prime}}\left[r(\Bx, \Ba)+\eta \max _{\Ba^{\prime}} Q^*\left(\Bx^{\prime}, \Ba^{\prime}\right)\right],
\label{eqn:bellman}
\end{equation}
where the expectation is taken with respect to the conditional probability of the new state $\Bx^{\prime}$ given $\Bx$ and $\Ba$. 
Associated with this optimal $Q$-function is the optimal action
\begin{align}
    A^*(\Bx)=\arg \max _{\Ba} Q^*(\Bx, \Ba).
    \label{eqn:optimalaction}
\end{align}

Because the state space that contains $\Bx$ and the action space that contains $\Ba$ in our problem are not discrete and because the analytical form of the optimal $Q$-function is generally unknown, it is difficult to solve the Bellman equation~\eqref{eqn:bellman} or the optimization problem~\eqref{eqn:optimalaction} directly. Finding the optimal $Q$-function is often an intractable problem, and approximate solutions are commonly used instead. Several methods such as the deep deterministic policy gradient~\cite{silver2014deterministic} (DDPG) method, Twin Delayed DDPG~\cite{fujimoto2018addressing} (TD3) method have been developed to solve the problem approximately. In these methods, $Q(\Bx,\Ba)$ and $A(\Bx)$ are represented by neural networks parameterized by $\Psi$ and $\Phi$ respectively. These neural networks are trained by a set of data $\mathbb{P}$ that contains a collection of $(\Bx, \Ba, r(\Bx, \Ba), \Bx^{\prime})$, where $\Bx^{\prime}$ denotes a new state reached by the agent after taking the action $\Ba$. Roughly speaking, the parameters $\Psi$ and $\Phi$ are optimized in an alternate fashion by maximizing the objectives derived from~\eqref{eqn:optimalaction} and the Bellman equation. We have included a diagram in Appendix~\ref{Qlearningdiagram} to provide a visual illustration of the Q-learning training process.  To be specific, the training process is used to solve the following optimization problems alternately
\begin{equation}
    \max_{\Phi}\underset{\left(\Bx, \Ba, r, \Bx^{\prime}\right) \sim \mathbb{P}}{\mathbb{E}} Q(\Bx, A(\Bx;\Phi); \Psi)
    \label{eqn:qobj1}
\end{equation}
and 
\begin{equation}
    \min_{\Psi}\underset{\left(\Bx, \Ba, r, \Bx^{\prime}\right) \sim \mathbb{P}}{\mathbb{E}}\left[Q(\Bx, \Ba; \Psi)-\left(r(\Bx,\Ba)+\eta Q\left(\Bx^{\prime}, A(\Bx^{\prime};\Phi); \Psi\right)\right)\right]^2.
    \label{eqn:qobj2}
\end{equation}

\textbf{Reward.} Note that the solution of the second problem~\eqref{eqn:qobj2} depends on how the reward function $r(\Bx,\Ba)$ is defined. Because our ultimate goal is to identify connective configurations that are expected to have a high reactive density $\rho(\Bx)$, ideally, we would like to use $\rho(\Bx')$, where $\Bx' = \Bx + \Ba$, as the reward function. 
The difficulty is that, as we indicated earlier, $\rho(\Bx)$ is defined in terms of the committor function $q(\Bx)$ which is unknown in general. Therefore, setting $r(\Bx,\Ba)$ to the exact $\rho(\Bx')$ is not practical.


However, as $\qx$ is defined as the probability of a trajectory starting from a point $\Bx(0)=\Bx$ and reaching $B$ before reaching $A$, we can estimate this probability numerically by shooting trajectories from $\Bx$, and performing statistical analysis of these trajectories. To be specific, we propose to use a shooting procedure to estimate $q(\Bx)$ by counting the number of trajectories originating from $\Bx$ with a random momentum and terminating in one of the metastable regions $A$ or $B$ within a fixed number of time steps.

We shoot $N$ trajectories from $\Bx$ with a random momentum or force.  
Let $N_A$ be the number of trajectories reaching $A$ first rather than $B$ within a fixed number of time steps $T$, and $N_B$ be the number of trajectories reaching $B$ first rather than $A$. Typically, $T$ is far smaller than the time scale required to observe a reactive trajectory.
Clearly, $N_A+N_B\leq N$ since some trajectories may hover around the starting configuration for a long time and never reach either $A$ or $B$ within $T$ time interval. We can view $N_A/(N_A+N_B)$ as an approximation to $\qx$ and $N_B/(N_A+N_B)$ as an approximation to $1-\qx$. Consequently, we can use \begin{equation}
\hat{\rho}(\Bx)=\frac{N_AN_B}{(N_A+N_B)^2}p(\Bx)
\end{equation}
as a proxy for $\rho(\Bx)$.  As a result, the reward function associated with the state action pair $(\Bx,\Ba)$ can be defined as 
\begin{align}
r(\Bx,\Ba):=\frac{N_AN_B}{(N_A+N_B)^2}\cdot \frac{\exp(-\beta V(\Bx'))}{Z},
\label{eqn:reward}
\end{align}
where $\Bx'=\Bx+\Ba$ is a new state. In practice, we can ignore the constant $Z$ in \eqref{eqn:reward}. 


Once the neural networks are properly trained, the optimal action to be taken at each $\Bx$ satisfies
\[
Q^*(\Bx, A(\Bx))=\max_{\Ba} Q^*(\Bx, \Ba).
\]

\begin{figure}[htbp]
\centering
\includegraphics[width=0.8\linewidth]{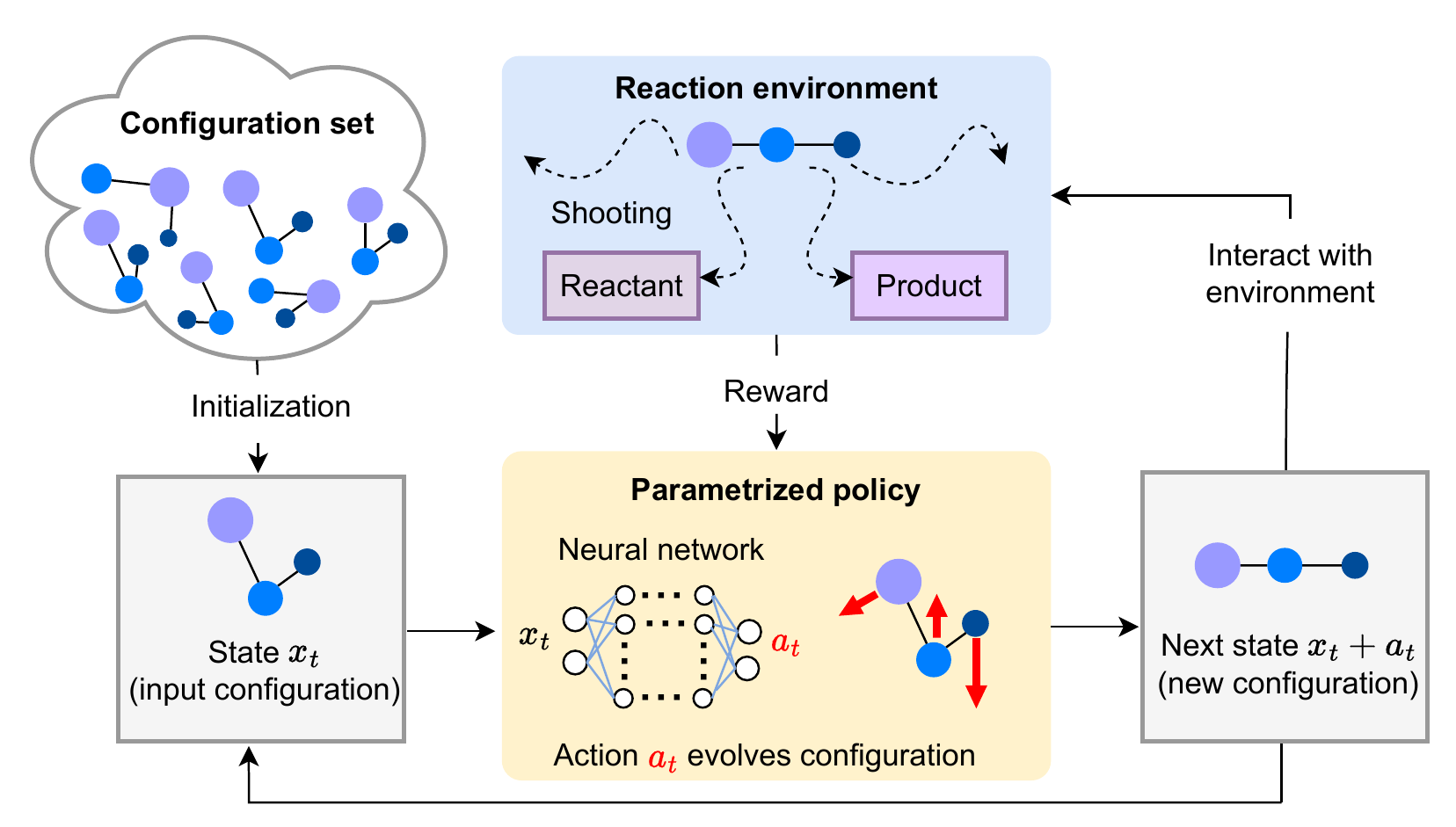}
\caption{\label{fig:framework} A schematic description of the proposed RL action-reward feedback loop. In each episode, a sequence of states (configurations) $\{\Bx^0, \cdots,\Bx^t, \cdots\}$ is generated iteratively. Each iteration consists of the following stages.
 \textcircled{1} \textbf{Initialization:} a configuration $\Bx^0$ is randomly drawn from the configuration set; \textcircled{2} \textbf{Evolution:} the parameterized policy yields the action $\Ba^t$ after observing the state $\Bx^{t}$. After taking the action, the agent moves to a new configuration $\Bx^{t+1}{=}\Bx^{t}+\Ba^t$; \textcircled{3} \textbf{Interaction:} we shoot multiple trajectories from $\Bx^{t+1}$, and evaluate its reward $r(\Bx^{t}, \Ba^{t})$. The quadruples $(\Bx^t, \Ba^t, r(\Bx^{t}, \Ba^{t}), \Bx^{t+1})$ is used in Q-learning to update the policy.}
\end{figure}

RL is a multi-episode iterative learning process. In each episode, a random configuration $\Bx^0$ is chosen to start the learning process.  A neural network that represents the action function $A(\Bx)$ takes the configuration as the input and generates an action $\Ba^0$ as the output. Taking such an action yields a new configuration $\Bx^1$ for which a reward $r$ can be obtained by shooting several trajectories from $\Bx^1$ and evaluating \eqref{eqn:reward}.  This process can be repeated several times until we generate a sequence of state action pairs $\{(\Bx^t,\Ba^t)\}$.
Figure~\ref{fig:framework} gives a schematic illustration of the process of generating a sequence of state action pairs within a single episode.

The generated sequence of state action pairs, together with the rewards evaluated for states form the training data set $\mathbb{P}$ that we use to optimize the neural network representions of the Q-function $Q(\Bx,\Ba)$ and action $A(\Bx)$ by solving the minimization problems \eqref{eqn:qobj1} and \eqref{eqn:qobj2} in an alternate fashion.


We use the TD3~\cite{} method to perform the optimization of $A(\Bx)$ and $Q(\Bx,\Ba)$. 
This variant of the DDPG method uses a variety of techniques to improve the stability of the training process and mitigate the risk of potentially over-estimating the $Q$-value function. In particular, TD3 uses the moving average of parameters associated with multiple neural networks to solve ~\eqref{eqn:qobj1} and ~\eqref{eqn:qobj2}.  Furthermore, TD3 uses a ``delayed'' policy update, where the policy network (which is used to determine the optimal action) is only updated after a certain number of Q-network updates. This delayed updating scheme helps to stabilize the training process and reduces the likelihood of the policy network training being stuck in an undesirable local minimum.



The main steps of a multi-episode RL algorithm for seeking the optimal policy that allows us to quickly identify connection configurations from an arbitrary starting configuration is shown in Algorithm~\ref{alg:ean}. 

\begin{algorithm}[t]
    \caption{An RL algorithm for seeking connective configurations}\label{alg:ean}   
    \textbf{Input:} Random parameter initialization $\Psi_1$ and $\Psi_2$ of critic networks and $\Phi$ of actor network. 
    
    \textbf{Output:} Action network $A(\cdot; \Phi)$. 
        
    \begin{algorithmic}[1]

    \State Initialize target networks $\Psi'_1\leftarrow\Psi_1$, $\Psi'_2\leftarrow\Psi_2$ and $\Phi'\leftarrow\Phi$
    \State Replay buffer $\mathbb{P}\leftarrow \emptyset$
    \For{episode from 1 to the maximal number of episodes} 
        \State $\Bx^0\sim {\rm Uniform}(\Omega)$  \algorithmiccomment{Start a new episode}
        \For{$t$ from $0$ to $L-1$}
        \State Obtain a new action $\Ba^t = A(\Bx^t; \Phi)$ and a new state $\Bx^{t+1}=\Bx^{t}+\Ba^t$ \algorithmiccomment{New state update rule}
            \If{$p(\Bx^{t+1})$ is smaller than a threshold}{ Break}
            \EndIf
        \State Compute the reward $r(\Bx^{t}, \Ba^t)$ using~\eqref{eqn:reward} \algorithmiccomment{Compute the reward by shooting}
        \State Append $(\Bx^t, \Ba^t, r(\Bx^{t}, \Ba^{t}), \Bx^{t+1})$ to $\mathbb{P}$
        \EndFor
    \For{$tt$ from $0$ to $t$}
    \State Sample a mini-batch $(\Bx, \Ba, r, \Bx^{\prime})$ of size $B$ from $\mathbb{P}$
    \State $y\leftarrow r + \eta \min\{Q(\Bx^{\prime}, A(\Bx^{\prime};\Phi'); \cdot)):\Psi_1', \Psi_2'\}$
    \State Update $\Psi_i$ with the loss $\frac{1}{B}\sum (Q(\Bx, \Ba; \Psi_i)-y)^2$ for $i=1,2$
    \If{$tt$ mod $policy\_delay=0$}
    \State Update $\Phi$ with the loss $\frac{1}{B}\sum-Q(s, A(s;\Phi); \Psi_1)$
    \State Update target networks: $\Phi'\leftarrow\tau\Phi+(1-\tau)\Phi'$ and  $\Psi'_i\leftarrow\tau\Psi_i+(1-\tau)\Psi_i'$ for $i=1,2$
    \EndIf
    \EndFor
    \EndFor  
    \end{algorithmic} 
\label{alg:TD3}
\end{algorithm}  

\subsection{Generating reaction channels and computing reaction rate constant}
\label{sec:reactchannelrate}
After performing several episodes of RL using Algorithm~\ref{alg:ean}, 
we obtain an optimal action function $A(\cdot; \Phi^*)$ parameterized by $\Phi^*$. Such a function yields the policy we follow at each configuration $\Bx$ to quickly move towards a connective configuration. Figure~\ref{fig:tri-action} shows how such a policy (represented as the vector field) looks like for a simple potential energy surface. The value of $A(\Bx)$, which is a vector with two components, is plotted as an arrow for each $\Bx$ uniformly sampled in $[-2.0,2.0]\times [-1.5,2.0]$. We see the arrows point to two configurations marked by crosses. These correspond to two connective configurations. 

Once we identify a connective configuration, we perform additional shooting operations from that configuration to generate multiple trajectories originating from the connective configuration. 
The configurations generated along these trajectories are considered as samples within a reactive channel between $A$ and $B$. 
If multiple connective configurations are identified, each one of them can be used to generate a distinct reactive channel.

The configurations generated within reactive channels can be used to compute the committor function $\qx$ by solving a restricted BKE on these configurations using a deep neural network as we described in Section~\ref{sec:committnn}. The utilization of NN-based PDE solver is often associated with an implicit bias towards fitting smooth functions that exhibit fast decay in the frequency domain.~\cite{FP} Consequently, This bias can make it challenging for NN models to capture drastic changes in the committor function. However, by choosing an appropriate training dataset, one can mitigate this bias.~\cite{liang2022stiffness} In Appendix~\ref{sec:dataset}, we demonstrate the advantage of using configurations within  reaction channels to train the NN designed to solve the restricted BKE.

The neural network not only returns $\qx$ for each $\Bx$ within all reactive channels, but also its gradient $\nabla \qx$.
This will allow us to compute the reactive flux at each $\Bx$ within the reaction channel.  As a result, by using \eqref{eqn:rate2}, we can calculate the rate constant.  

%

  \section{Numerical results}
\label{sec:num}
In this section, we present several numerical experiments that demonstrate the effectiveness of the RL method introduced in Sec.~\ref{sec:method} for identifying connective configurations and using them to generate configurations that characterize reactive channels. Our experiments were performed on model potentials (triple-well and rugged Muller potentials) as well as the Alanine dipeptide molecule in {vacuum}. While we demonstrate the RL results with initializations uniformly sampled from $\Omega$ in this section, we have the flexibility to relax this constraint by considering initializations from metastable states (see Appendix~\ref{sec:stable}).


\subsection{Potential with multiple reaction channels}
\begin{figure*}[ht]
\centering
\subfigure[Learned action function (vector field) and the connective configurations it reveals (crosses).]{\label{fig:tri-action}\includegraphics[width=0.45\linewidth ]{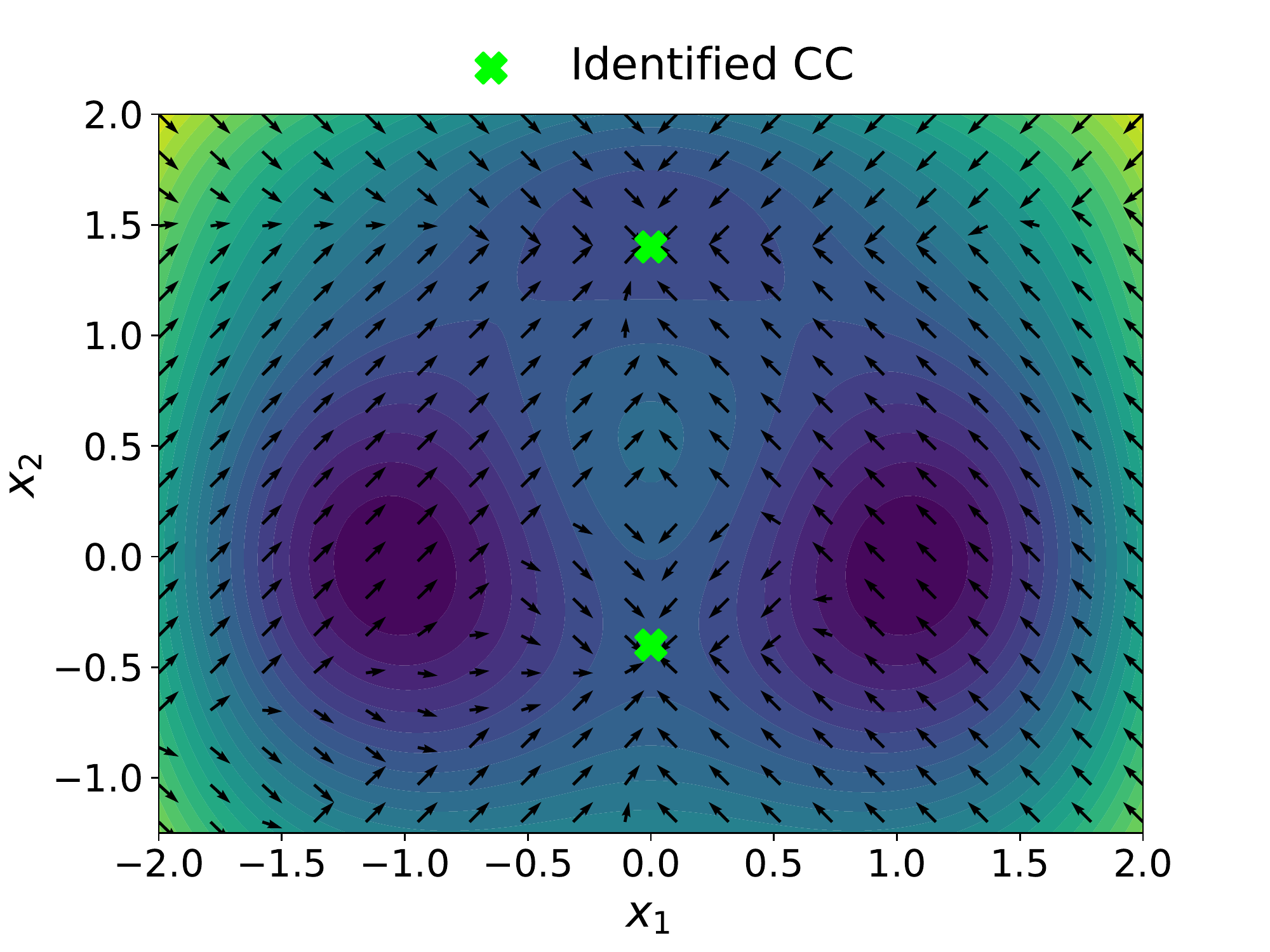}}
\hfill
\subfigure[Sampled configurations  generated by shooting from two identified connective configurations.]{\label{fig:tri-pts}\includegraphics[width=0.45\linewidth ]{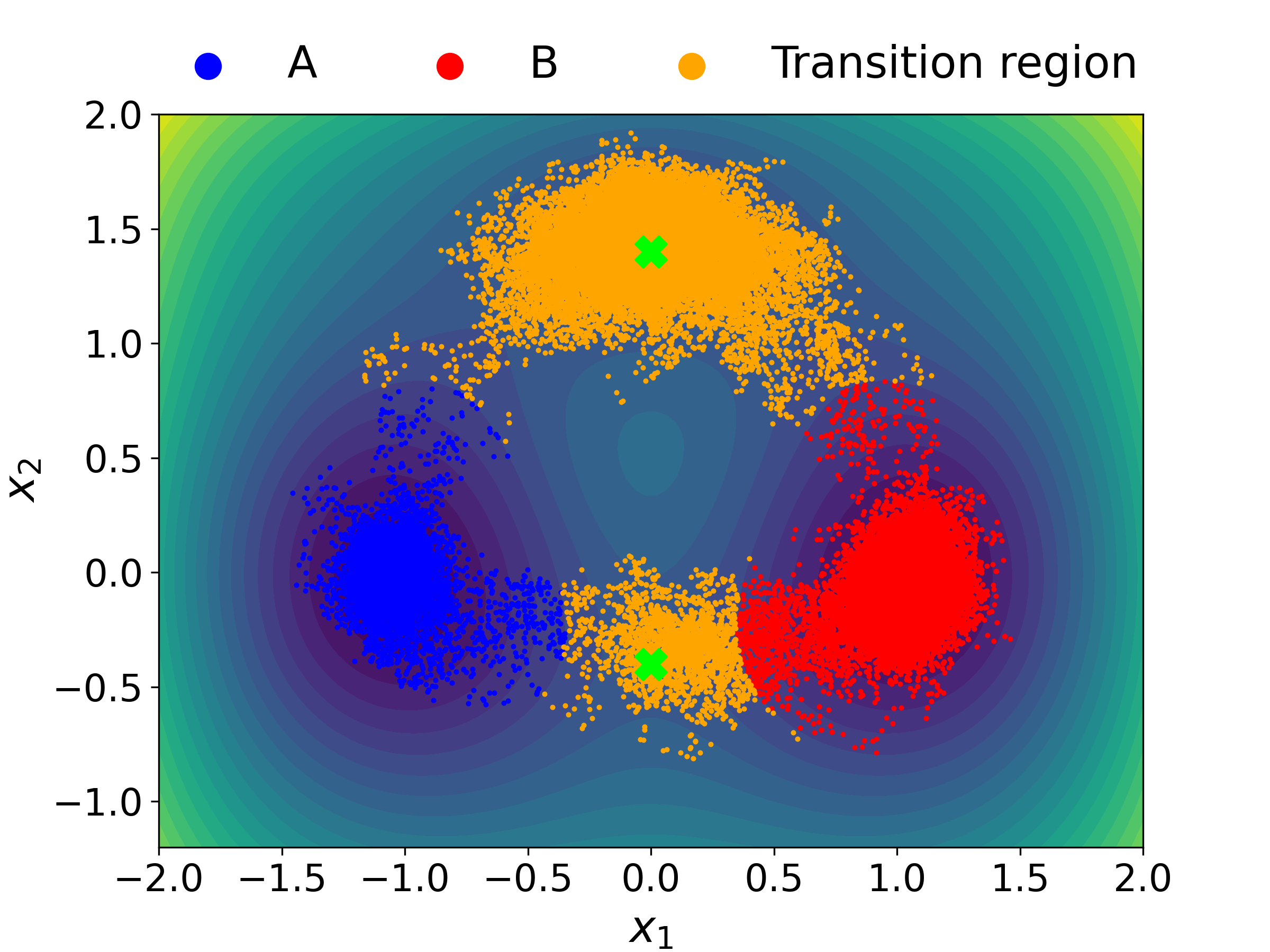}}
\subfigure[The FDM solution of \eqref{eqn:bke} on $\Omega$.]{\label{fig:tri-fem}\includegraphics[width=0.32\linewidth ]{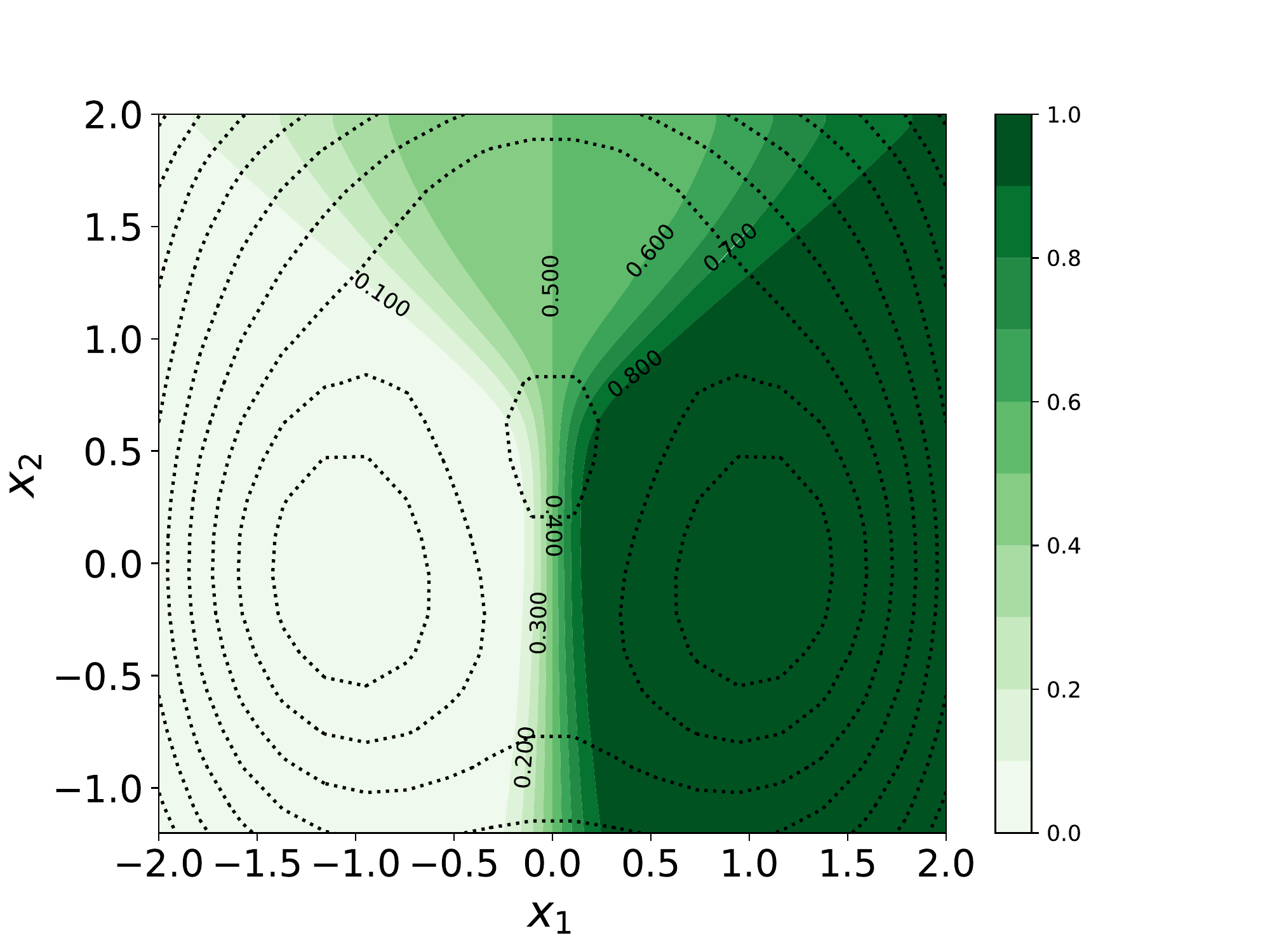}}
\subfigure[The NN solution of \eqref{eqn:bke}.]{\label{fig:tri-nn}\includegraphics[width=0.32\linewidth ]{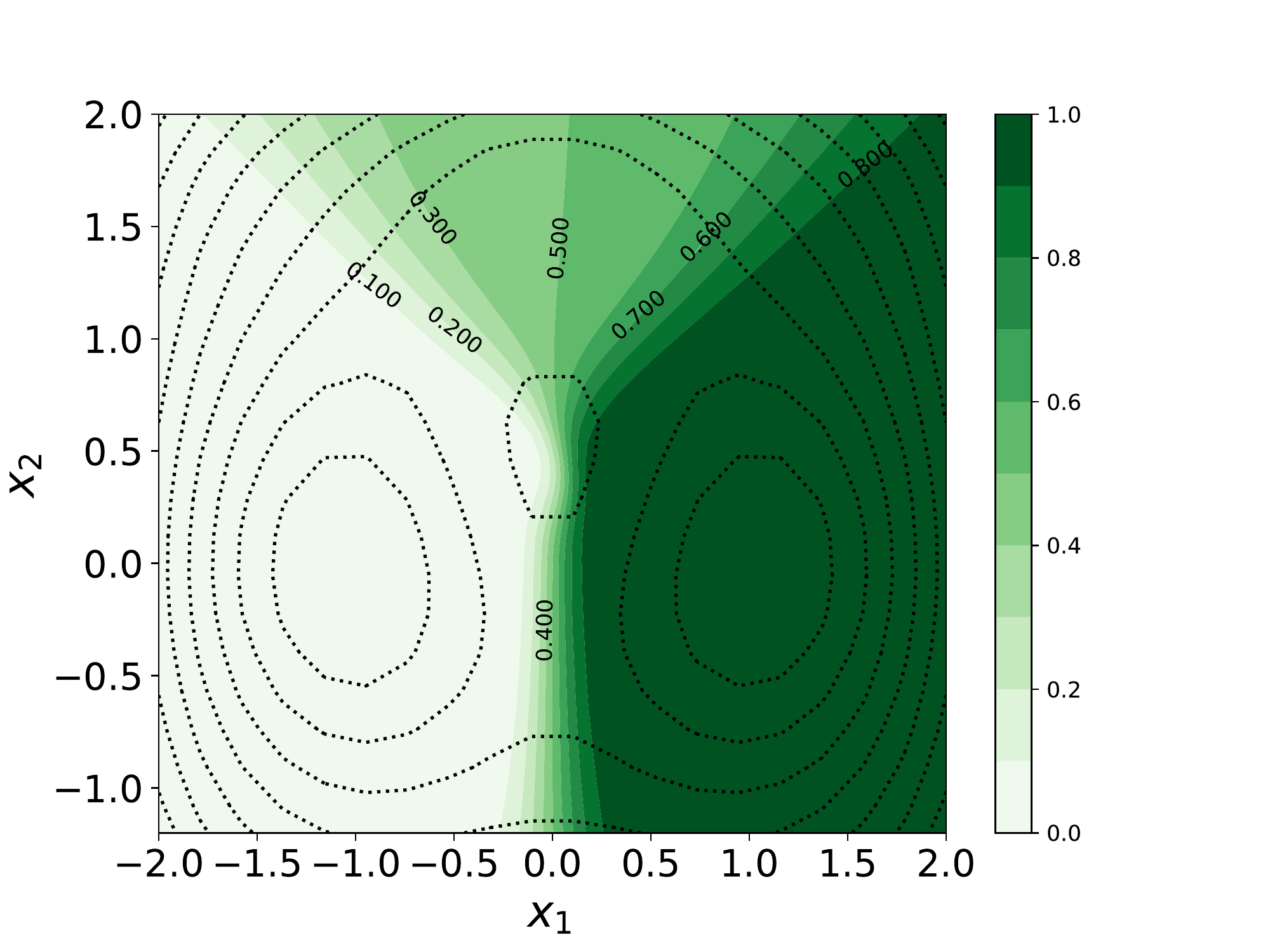}}
\subfigure[The difference between (c) and (d).]{\label{fig:tri-error}\includegraphics[width=0.32\linewidth ]{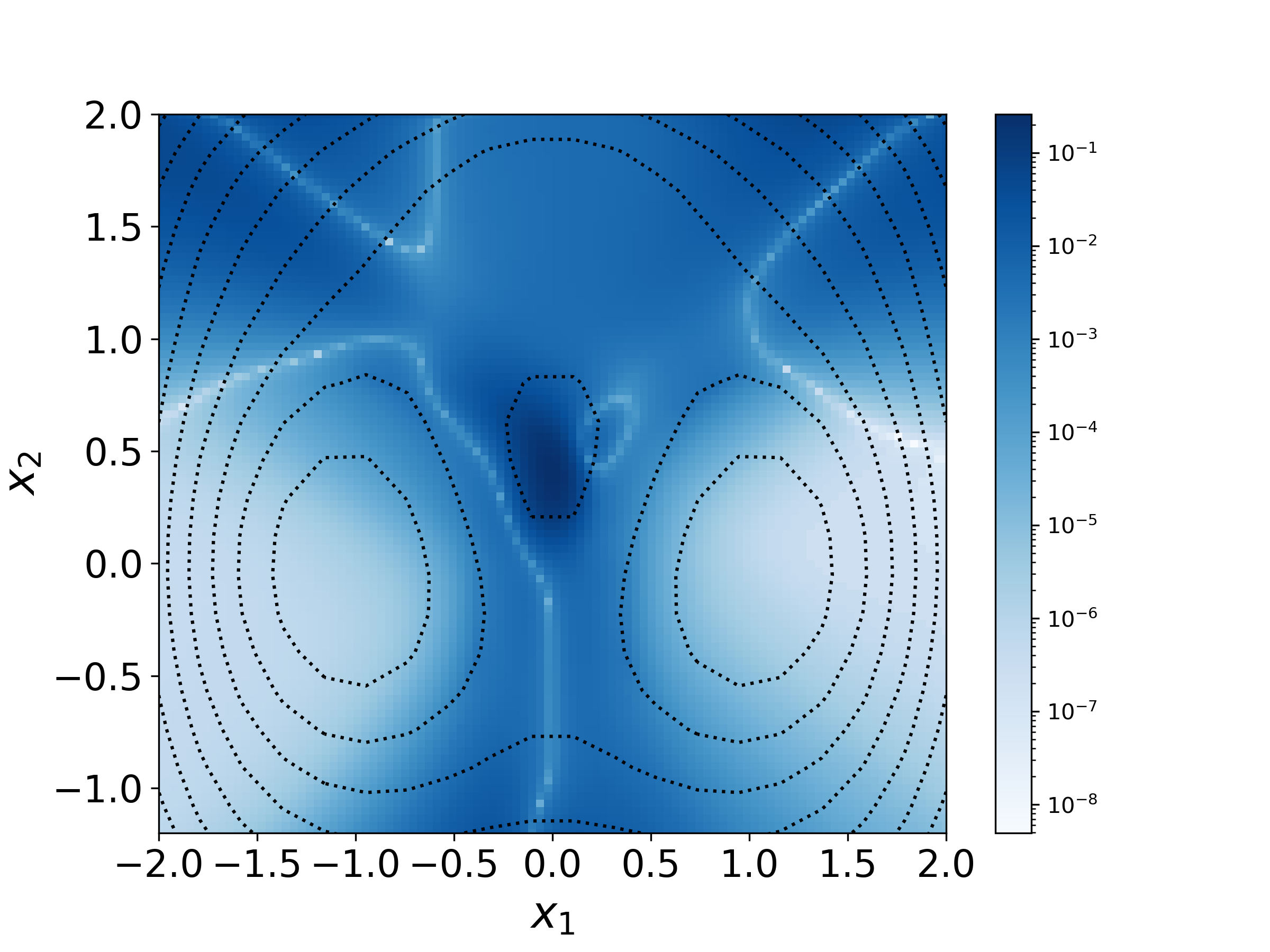}}
\caption{
The results obtained from running the RL algorithm for Triple-well potential at inverse temperature $\beta=6.67$. (a) The action function (vector field) learned by our proposed RL. The action function reveals two connective configurations (crosses). (b) The configurations generated by shooting trajectories initiated at the identified connective configurations. (c-e) The FDM (c) and NN (d) solutions of the BKE \eqref{eqn:bke} and their difference (e).}
\label{fig:tri}
\end{figure*}

We consider the triple-well potential defined by
\begin{align}
\begin{split}
V(x_1, x_2)= 3 e^{-x_1^2-\left(x_2-\frac{1}{3}\right)^2}-3 e^{-x_1^2-\left(x_2-\frac{5}{3}\right)^2}-5 e^{-(x-1)^2-x_2^2}-5 e^{-(x_1+1)^2-x_2^2}+0.2 x_1^4+0.2(x_2-\frac{1}{3})^4.
\end{split}
\end{align}
We focus on the domain $\Omega= [-2,2]\times[-1.2,2]$. 
Figure~\ref{fig:tri-action} shows this potential as a color-mapped image.
The two meta-stable regions $A$ and $B$ are defined by 
\begin{align*}
A=\{\Bx: V(x_1,x_2)<-2 \text{ and } x_1\leq -0.1\}, \quad B=\{\Bx: V(x_1,x_2)<-2 \text{ and } x_1\geq 0.1\}.
\end{align*}
We can see from Figure~\ref{fig:tri-action} that there are two transition paths from $A$ to $B$. The top transition path goes through the third well in the top part of the image and two transition states between the third well and $A$ ($B$). The bottom transition path goes directly from $A$ to $B$ and crosses the transition state. Because the potential function is symmetric with respect to $x_1=0$, all configurations along $\{\Bx: x_1=0\}$ have an equal probability of reaching either $A$ or $B$. As a result, $\{\Bx: x_1=0\}$ represents the half-isocommittor surface $\{\Bx: q(\Bx)=0.5\}$. We experimented with both a relatively low-temperature regime $\beta=6.67$ and a relatively high-temperature regime $\beta=1.67$. We discuss the results for $\beta=6.67$ here, which is more challenging for studying rare events. We report a similar observation for the $\beta=1.67$ case in Appendix~\ref{sec:triwell-appdx}. By default, the friction coefficient is set as 1. 

\textbf{Identifying reaction channels. }In our experiments,  the initial configuration for each RL episode is randomly sampled from a uniform distribution of configurations in $\Omega$. The reward $r(\Bx, \Ba)$~\eqref{eqn:reward} is obtained by shooting $N=50$ trajectories from $\Bx$. The maximal number of evolution steps is set to $L=15$. The maximum number of episodes used in RL is set to 1000. While we initially set a large number of episodes for the RL training process, it is worth noting that the convergence of RL does not require an extensive number of episodes. Additional discussion can be found in Appendix~\ref{sec:trainingcurve}.

Figure~\ref{fig:tri-action} shows the learned action $A(\Bx;\Phi^*)$ as a vector field that represents an optimal policy. 
Two attractors can be seen from this policy field. They correspond to two connective configurations located in two different transition paths.

From each of the identified connective configuration, we shoot $50$ trajectories by simulating the overdamped Langevin dynamics~\eqref{eqn:overdamplangvin} using the Euler-Maruyama scheme. We choose a uniform time step size $\Delta t = 5\times 10^{-3}$, and propagate the solution from $T=0.0$ to $T=2.0$. The total number of configurations generated along these trajectories is 40,000. These configurations lie either in the metastable region $A$ or $B$ or two transition regions between $A$ and $B$ as shown in Fig.~\ref{fig:tri-pts}.  These two transition regions correspond to the two reactive channels associated with this potential energy surface.

\textbf{Solving BKE.} We then solve the restricted BKE in the identified reaction channels by using the NN-based solver discussed in Sec.~\ref{sec:committnn}. The loss function~\eqref{eqn:discrete} is optimized by the Adam optimizer~\cite{kingma2014adam}. The hyperparameters used in the NN are listed in Appendix~\ref{sec:hyperNN}. Figure~\ref{fig:tri-nn} shows the contour plots of the NN solution (colored contour lines) and the potential (dotted contour lines).
The $0.5$-level set of NN solution marked in the figure almost coincides with the true half-isocommittor $\{\Bx: x_1=0\}$. As a reference, we also used the finite difference method (FDM) to solve the BKE on the entire domain $\Omega$. The absolute difference  between the NN and FDM solutions on a $100\times 100$ uniform mesh is shown in Figure~\ref{fig:tri-error}. We observe small errors in the NN solution in the regions that contain a large number of sampled configurations (such as the top transition path) and relatively large errors in regions where configurations are sparsely sampled (such as the region near $(0.0, 0.5)$). 
 
\textbf{Rate estimation.} The NN approximation to the committor function and its gradient on configurations with two reactive channels are used to calculate the reaction rate by integrating~\eqref{eqn:rate} over the dividing sub-surfaces identified with each reaction channel based on the data. 
We compare the computed rate with the one obtained from a direct dynamics simulation (see Sec.~\ref{sec:preTPT}) and that computed from the committor function obtained from the finite difference solution of the KBE on the entire domain $\Omega$ in Table~\ref{tab:triplewell}. We list the computed rates for both $\beta=1.67$ and $\beta=6.67$.  In the direct dynamics simulation, we generate a long trajectory by time evolving the solution to the overdampled Langevin equation to $T=2\times 10^{7}$. In the FDM calculation, the rate is computed with numerical integration of \eqref{eqn:rate} on the entire line segment $\{\Bx: x_1=0, -1.2\leq x_2\leq 2\}$. From these numerical experiments, we find that the generated configurations mainly cross the line segment $\{\Bx: x_1=0, -1.0\leq x_2\leq 2.2\}$ for $\beta=1.67$ and line segments $\{\Bx: x_1=0, -0.8\leq x_2\leq 0.0, 0.8\leq x_2\leq 1.8\}$ for $\beta=6.67$. We then compute the rates using the NN solutions on these segments. As we can see, the NN solution on reaction channels gives comparable rates as the one obtained by the FDM and a direct simulation. Finally, we validate the rate calculation with our NN solution using the formula~\eqref{eqn:rate_sq}. 

\begin{table}[ht]
  \centering
  \caption{Reaction rate of the triple-well potential under various temperatures. ``N/A'' means that no reactive trajectory is observed in the long trajectory of time $T=2\times 10^7$. }
    \begin{tabular}{lllll}
    \toprule
    \toprule
    \multirow{2}[4]{*}{Methods} &       & \multicolumn{3}{c}{$\beta$} \\
\cmidrule{3-5}          &       & 1.67     &       & 6.67 \\
    \midrule
    Direct simulation &       & $2.18\times 10^{-2}$     &       & N/A \\
    \midrule
    Finite difference (Eqn.~\eqref{eqn:rate2}) &       & $2.16\times 10^{-2}$     &       & $7.43\times 10^{-8}$ \\
    \midrule
    Our NN (Eqn.~\eqref{eqn:rate2}) &       & $2.00\times 10^{-2} $    &       &  $7.11\times 10^{-8}$\\
    \midrule
    Our NN (Eqn.~\eqref{eqn:rate_sq}) &       & $1.99\times 10^{-2}$    &       &  $7.32\times 10^{-8}$\\
    \bottomrule
    \bottomrule
    \end{tabular}%
  \label{tab:triplewell}%
\end{table}%

\begin{figure*}[ht]
\centering
\subfigure[Learned action function (vector field) and the connective configurations it reveals (crosses).]{\label{fig:RMlow-action}\includegraphics[width=0.45\linewidth ]{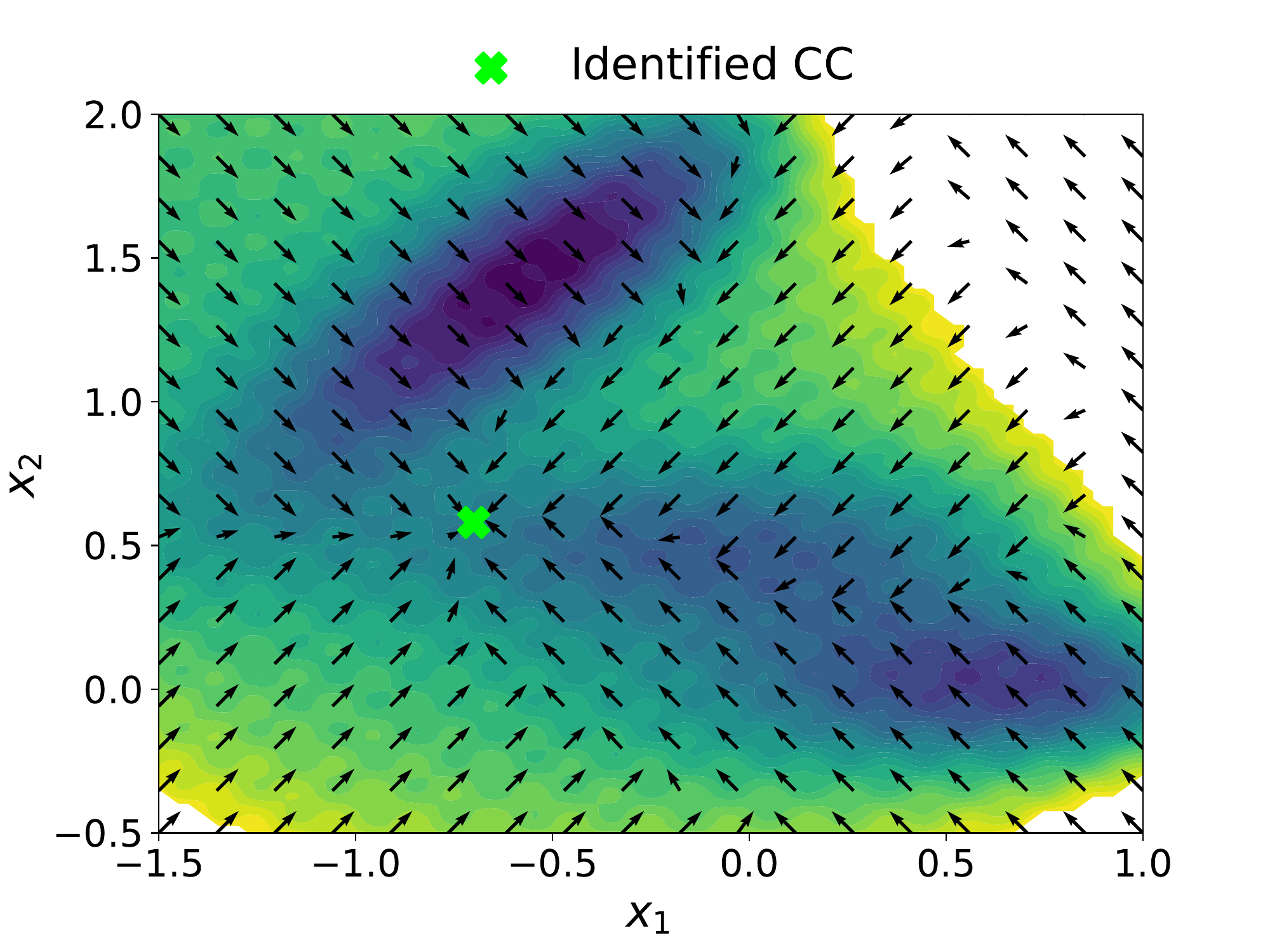}}
\hfill
\subfigure[Sampled configurations generated by shooting trajectories from the  identified connective configuration.]{\label{fig:RMlow-pts}\includegraphics[width=0.445\linewidth ]{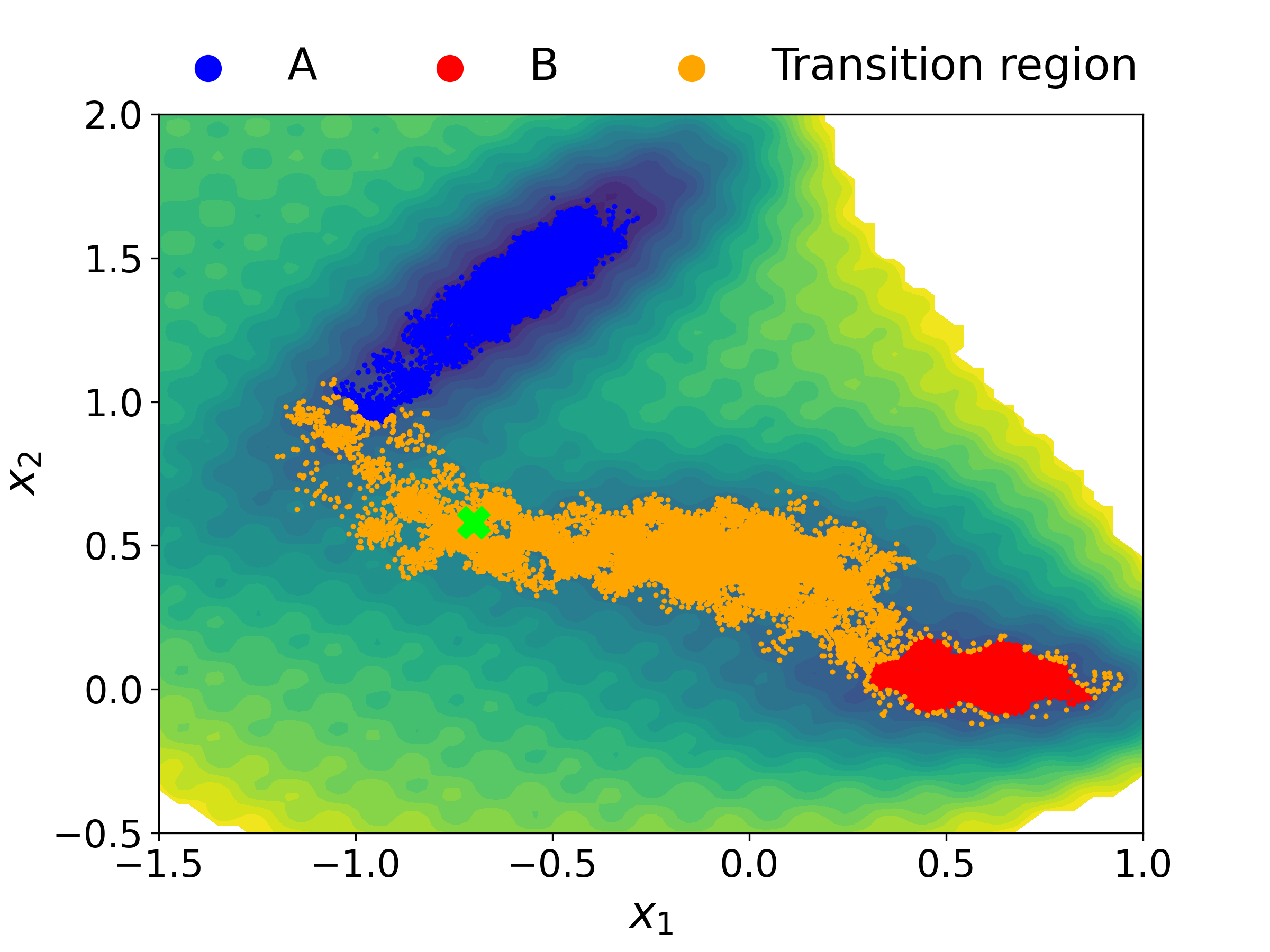}}
\subfigure[The FDM solution of \eqref{eqn:bke} on $\Omega$.]{\label{fig:RMlow-fdm}\includegraphics[width=0.32\linewidth ]{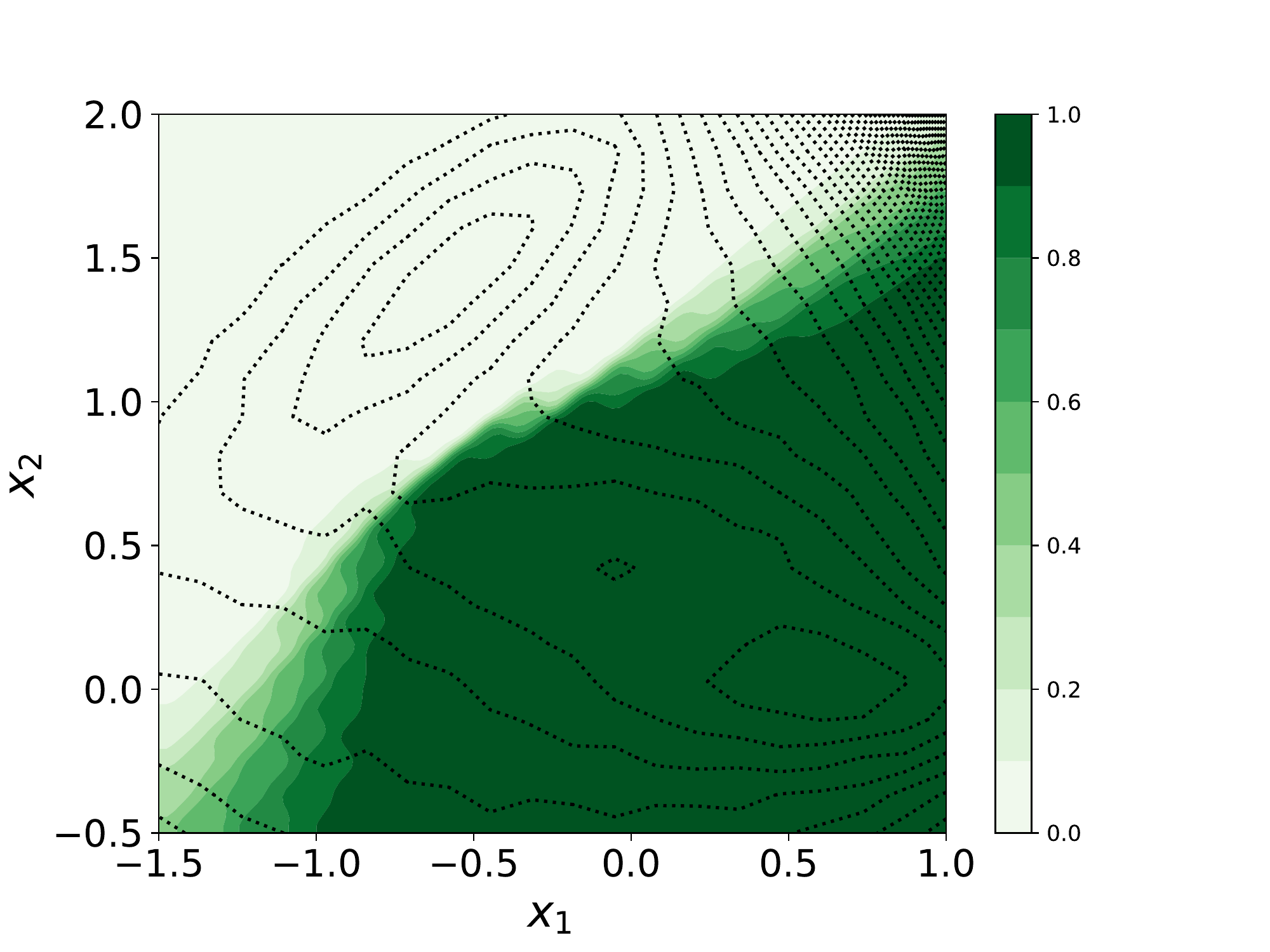}}
\subfigure[The NN solution of \eqref{eqn:bke}.]{\label{fig:RMlow-nn}\includegraphics[width=0.32\linewidth ]{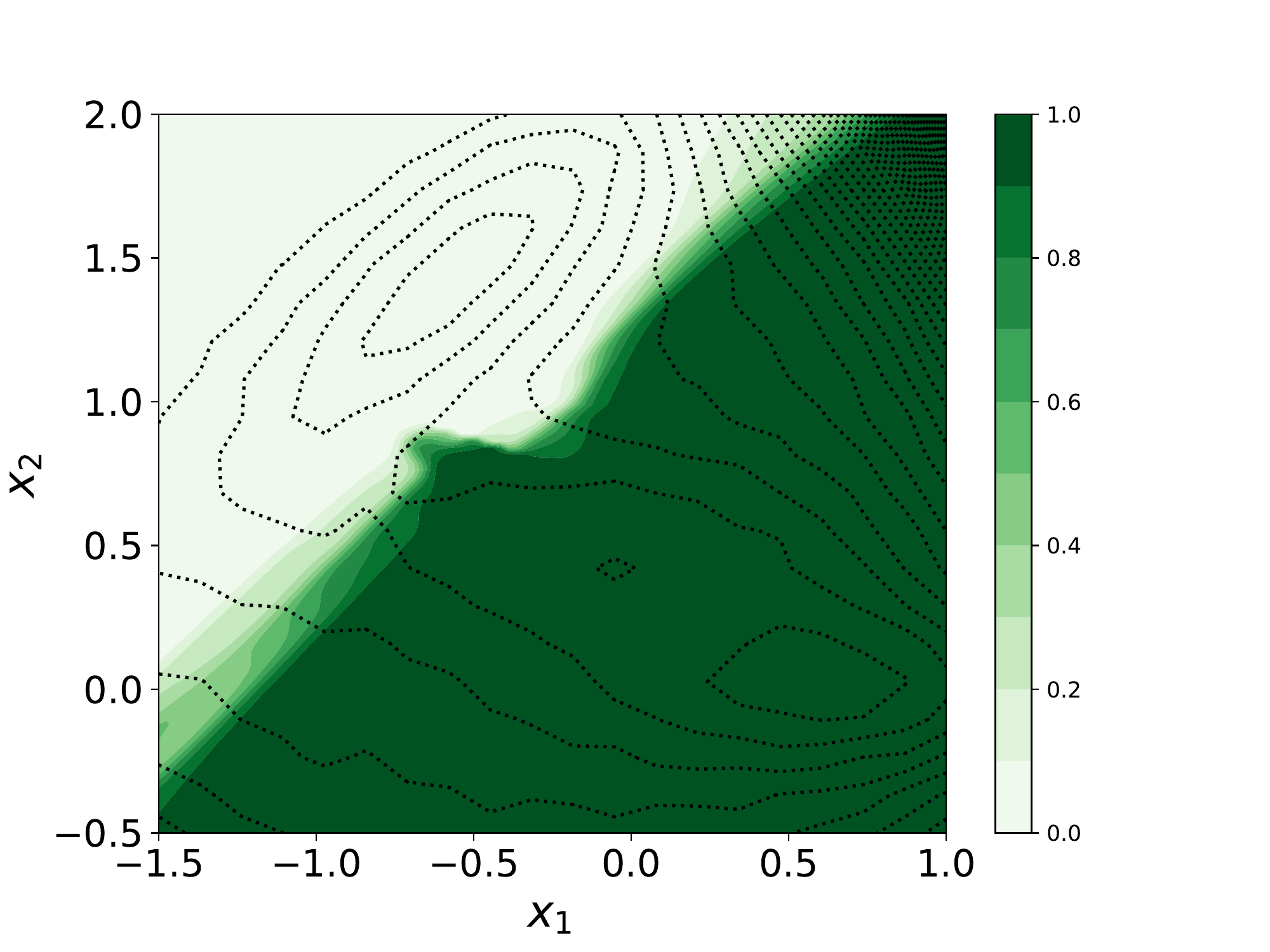}}
\subfigure[The difference between (c) and (d).]{\label{fig:RMlow-error}\includegraphics[width=0.32\linewidth ]{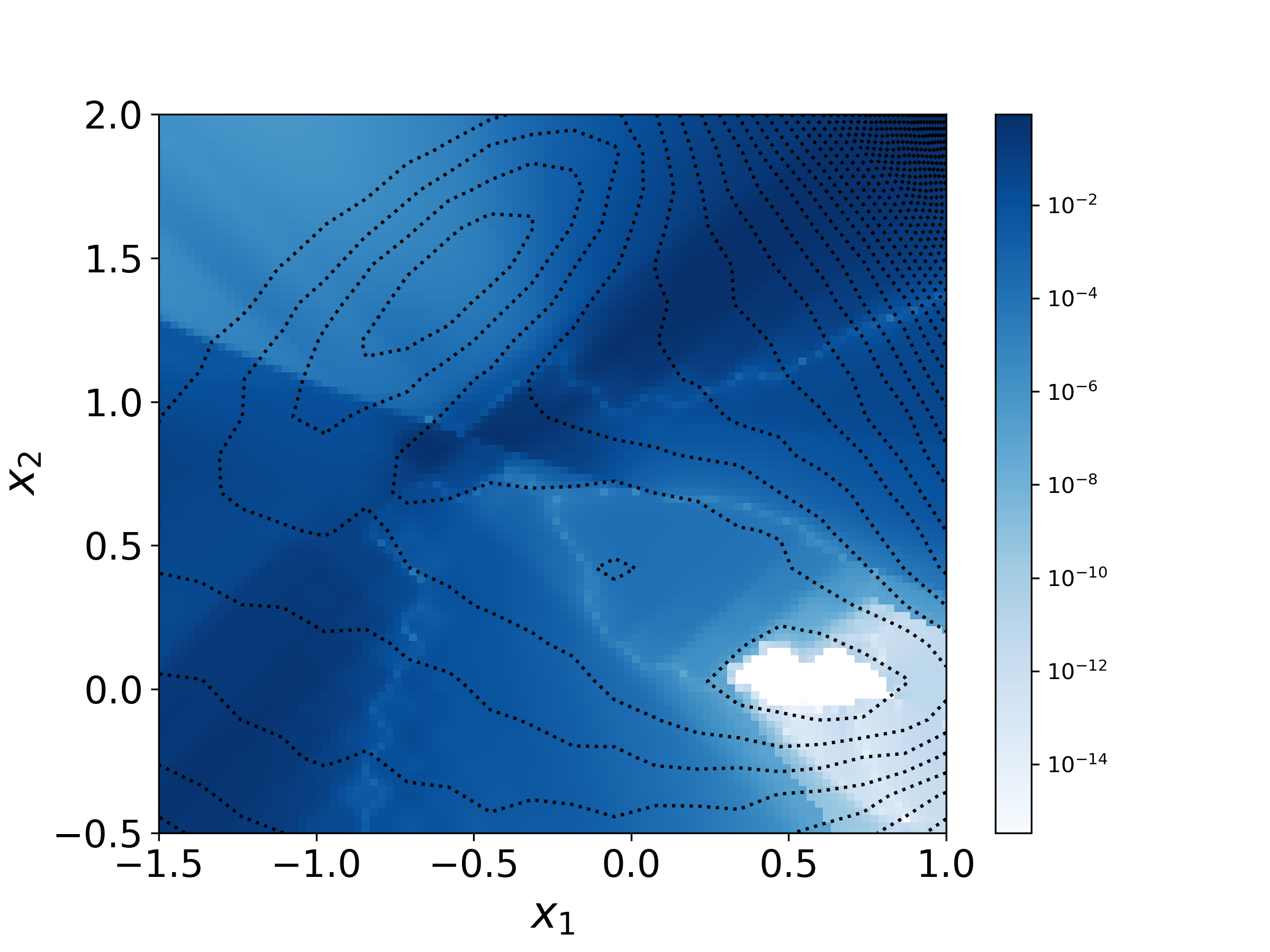}}
\caption{
The results obtained from running the RL algorithm for the Rugged Muller potential at the inverse temperature $\beta=0.25$. (a) The action function (vector field) learned by the RL algorithm. It reveals one connective configuration (cross). (b) The configurations generated by shooting trajectories from the identified connective configurations. (c-e) The FDM (c) and NN (d) solutions of the BKE \eqref{eqn:bke} and their difference (e).}
\label{fig:RMlow}
\end{figure*}

\subsection{Potential with rough landscape}
In the second example, we consider the rugged Muller potential on the domain $\Omega=[-1.5,1]\times[-0.5,2]$. The potential function is defined by
\begin{align}
V\left(x_1, x_2\right)=& \sum_{i=1}^4 D_i \exp \left[a_i\left(x_1-X_i\right)^2+b_i\left(x_1-X_i\right)\left(x_2-Y_i\right)+c_i\left(x_2-Y_i\right)^2\right]+\gamma \sin \left(2 k \pi x_1\right) \sin \left(2 k \pi x_2\right).
\end{align}
Here the parameters $\gamma$ and $k$ control the roughness of the landscape, which are set to $9$ and $5$, respectively. Other model parameters ($a_i,b_i,c_i,X_i,Y_i, D_i$, $i=1,\cdots,4$) are exactly the same as the ones used in Refs.~\cite{li2019computing,khoo2019solving}. By default, the friction coefficient is set as 1. 

\textbf{Identifying reaction channels.} We discuss the numerical results in the low temperature regime $\beta=0.25$ (Fig.~\ref{fig:RMlow}) here and refer readers to  Appendix~\ref{sec:rugmul-appdx} for results obtained for $\beta=0.1$. In this example, the reward in the RL algorithm  is calculated by shooting $N=20$ trajectories up to $T = 0.25$. The step size of used in each trajectory is set to $5\times 10^{-5}$. Each RL episode consists of $L=20$ steps (actions). We ran the RL algorithm for $1000$ episodes. Figure~\ref{fig:RMlow-action} shows that the learned policy points to a single connective configuration. From that configuration, we shoot $50$ trajectories using the Euler-Maruyama scheme. These trajectories contain 100,000 configurations that lie in the metastable regions $A$ and $B$ as well as the transition region in between (Fig.~\ref{fig:RMlow-pts}). The latter is viewed as the reactive channel for this particular potential energy surface. The hyperparameter setting for the NNs used in the RL algorithm is listed in Appendix~\ref{sec:hyperNN}. 

\textbf{Solving BKE.} 
 We use the configurations contained in the reactive channel to solve the BKE via a NN. Figures~\ref{fig:RMlow-fdm} and \ref{fig:RMlow-nn} show the contour plots of the solution to the BKE obtained from both the FDM and the NN, respectively. The difference between the two is also shown in Figure~\ref{fig:RMlow-error}. From these plots, we observe that the NN solution agrees well with the FDM solution. In particular, the NN solution captures drastical changes near the point $(-0.8, 0.6)$. 
 
\textbf{Rate estimation}. The computed committor functions obtained by the FDM and the NN approach are used to compute the reaction rate under different $\beta$ values. The computed rates are compared with those obtained from direct numerical simulations of the corresponding overdamped Langevin dynamics in Table~\ref{tab:ruggedMuller}. In direction numerical simulations, we ran a long trajectory until $T=1.5\times 10^4$ with a time step size of $10^{-5}$. When $\beta=0.25$, no reactive trajectory is observed. When the committor function is obtained from the FDM, the rate is computed from the numerical integration of \eqref{eqn:rate} on the dividing surface $\{x_1=0.0, -0.5\leq x_2\leq 2.0\}$.  When using an NN to compute the commitor function within the reactive channel,  
 the reaction rate is computed from the numerical integration of \eqref{eqn:rate} along the line segments $\{x_1=-0.5, 0.0\leq x_2\leq 0.8\}$ for $\beta=0.1$.  When $\beta=0.25$, we calculate the rate by integrating \eqref{eqn:rate} along the line segment $\{x_1=-0.5, 0.3\leq x_2\leq 0.65\}$. We observe that the rates computed by all three methods are comparable when $\beta=0.1$. When $\beta=0.25$, the rates obtained from the NN and FDM approximation of the committor function have the same magnitude. 
\begin{table}[htbp]
  \centering
  \caption{Reaction rate of the rugged Muller potential under various temperatures. ``N/A'' means that no reactive trajectory is observed in the long trajectory of time $T=1.5\times 10^4$.}
    \begin{tabular}{lllll}
    \toprule
    \toprule
    \multirow{2}[4]{*}{Methods} &       & \multicolumn{3}{c}{$\beta$} \\
\cmidrule{3-5}          &       & 0.1     &       & 0.25 \\
    \midrule
    Direct simulation &       & $4.75\times 10^{-3}$     &       & N/A \\
    \midrule
    Finite difference (Eqn.~\eqref{eqn:rate2}) &       & $4.31\times 10^{-3}$     &       & $1.78\times 10^{-10}$ \\
    \midrule
    Our NN (Eqn.~\eqref{eqn:rate2})&       & $4.36\times 10^{-3}$    &       &  $6.27\times 10^{-10}$\\
    \bottomrule
    \bottomrule
    \end{tabular}%
  \label{tab:ruggedMuller}%
\end{table}%
\subsection{Alanine dipeptide in vacuum}
\begin{figure*}[ht]
\centering
\subfigure[Potential energy in $(\phi,\psi)$ and two metastable regions.]{\label{fig:AD-potential}\includegraphics[width=0.32\linewidth ]{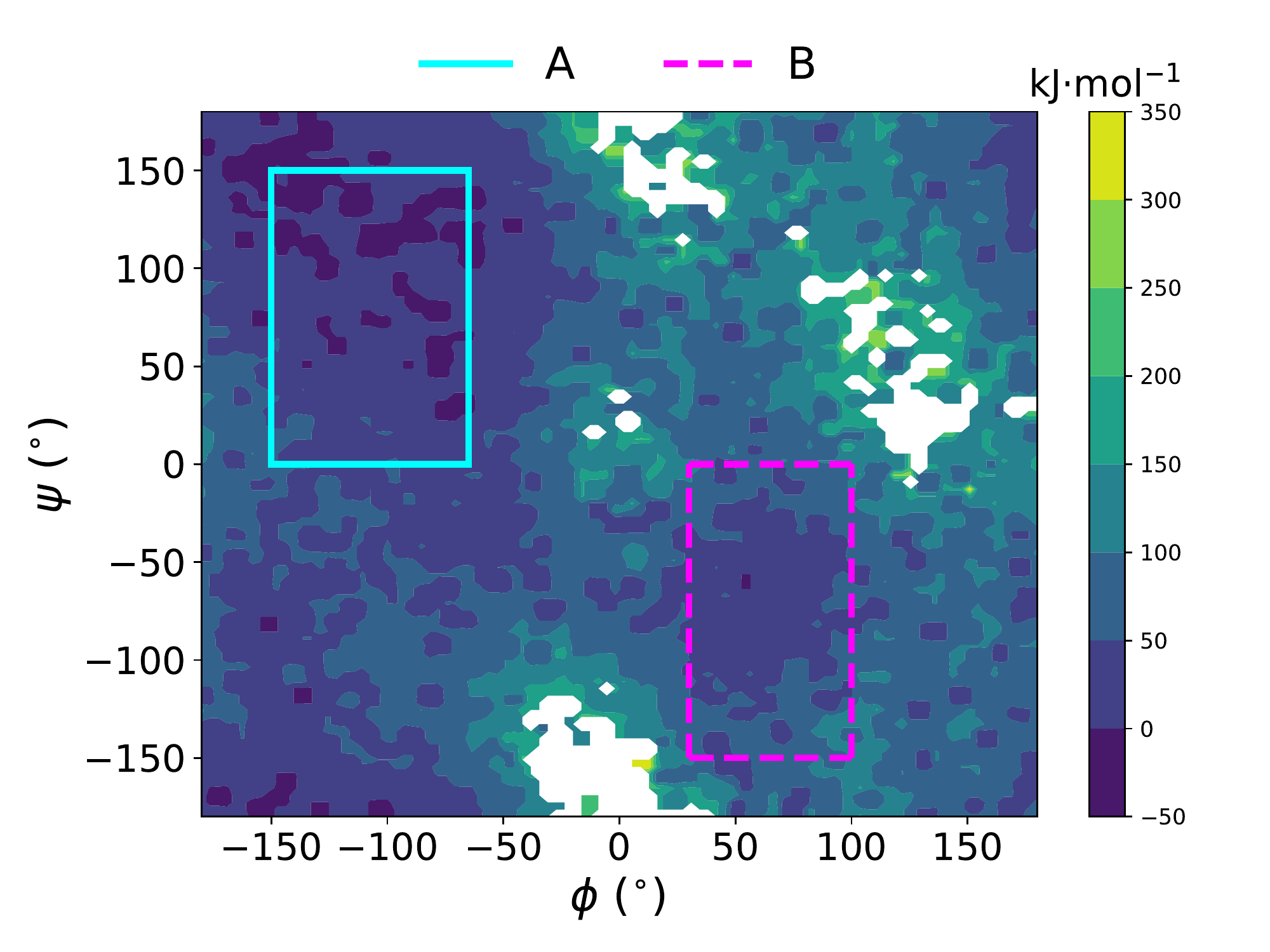}}
\subfigure[Configurations generated from Langevin dynamics at temperature $300$K]{\label{fig:AD-pts1}\includegraphics[width=0.32\linewidth ]{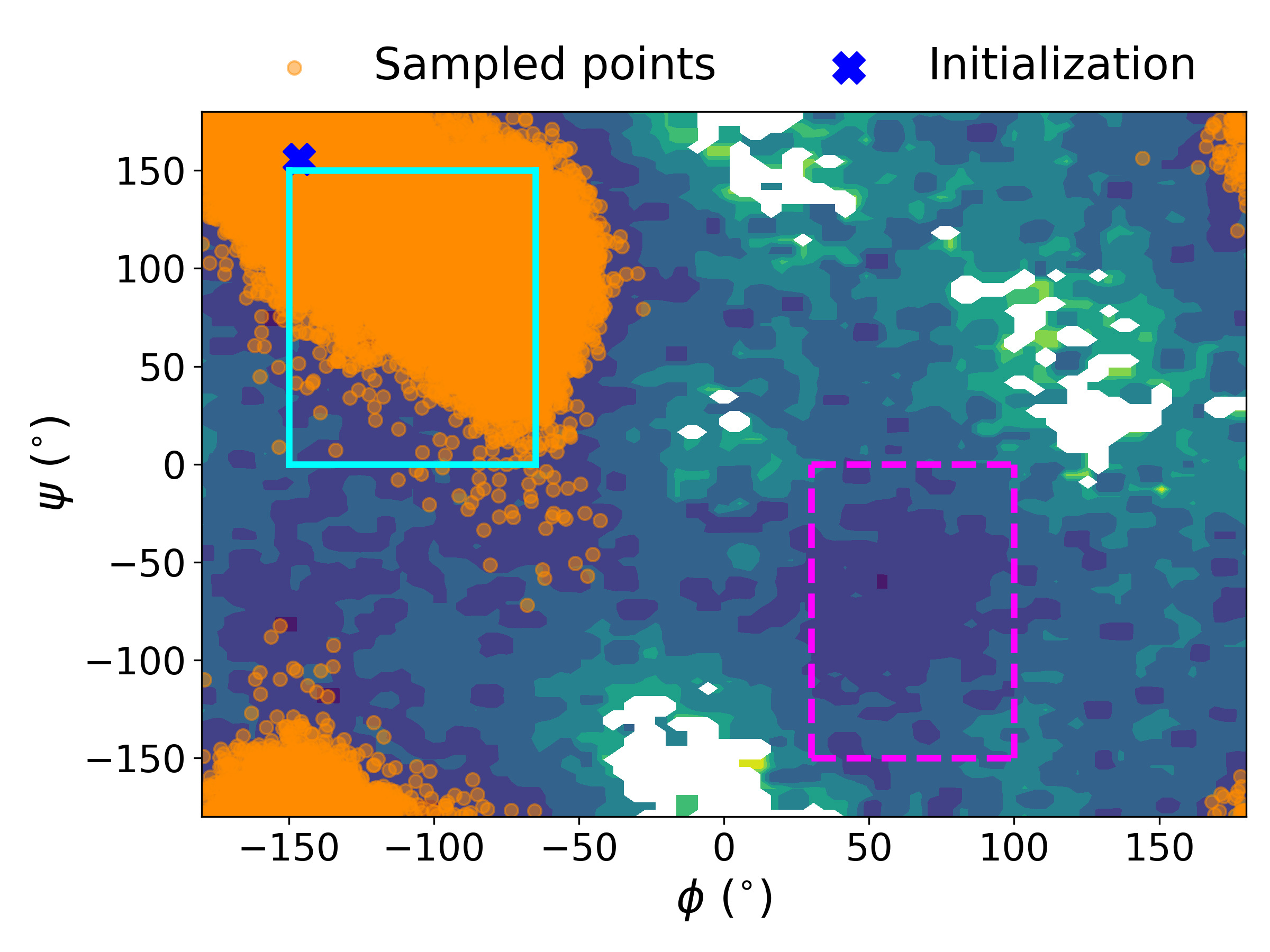}}
\subfigure[Learned action and identified CCs]{\label{fig:AD-action}\includegraphics[width=0.32\linewidth ]{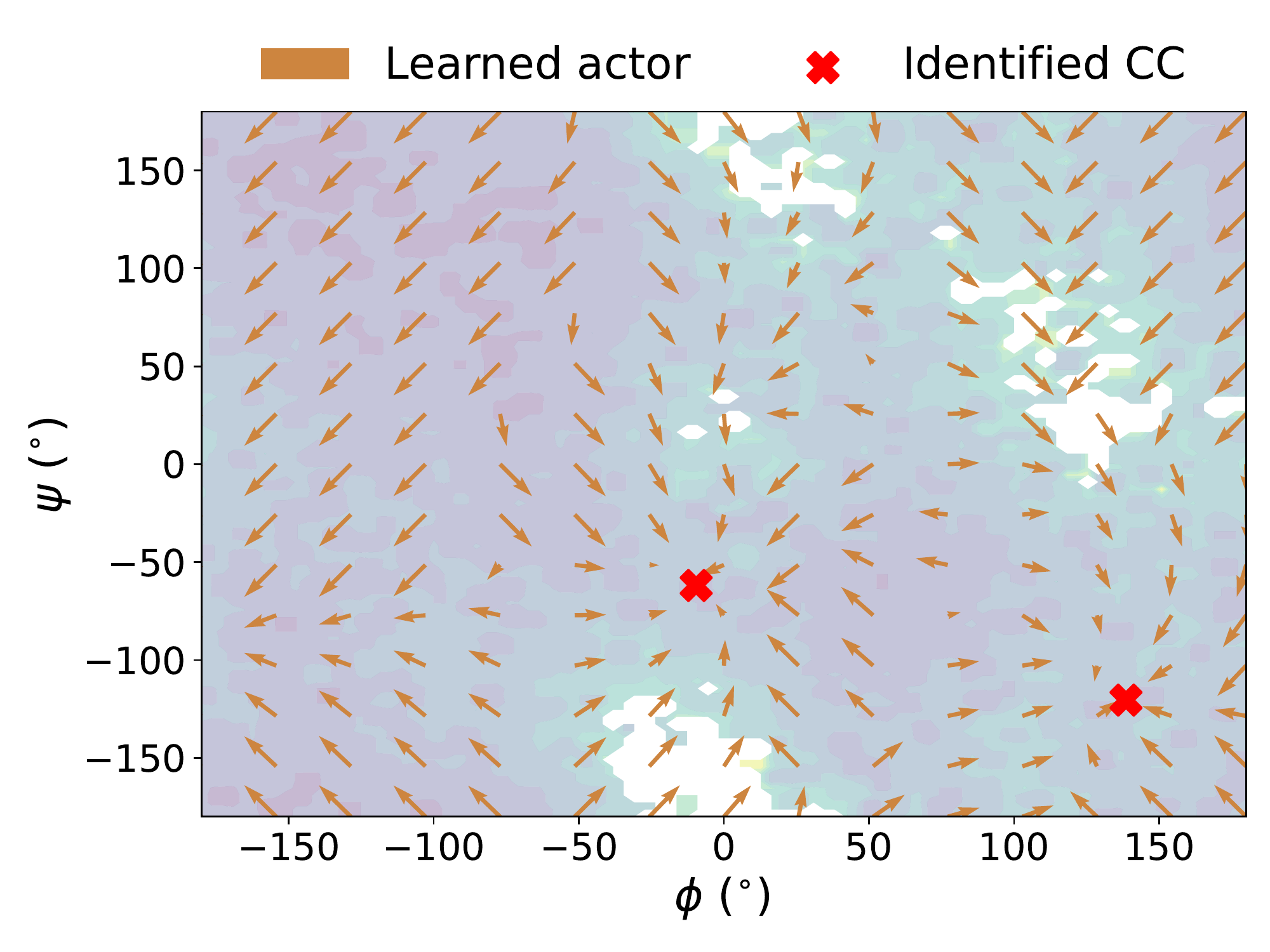}}
\subfigure[Reactive channel generated from ${\rm CC}_1$]{\label{fig:AD-pts-CC1}\includegraphics[width=0.32\linewidth ]{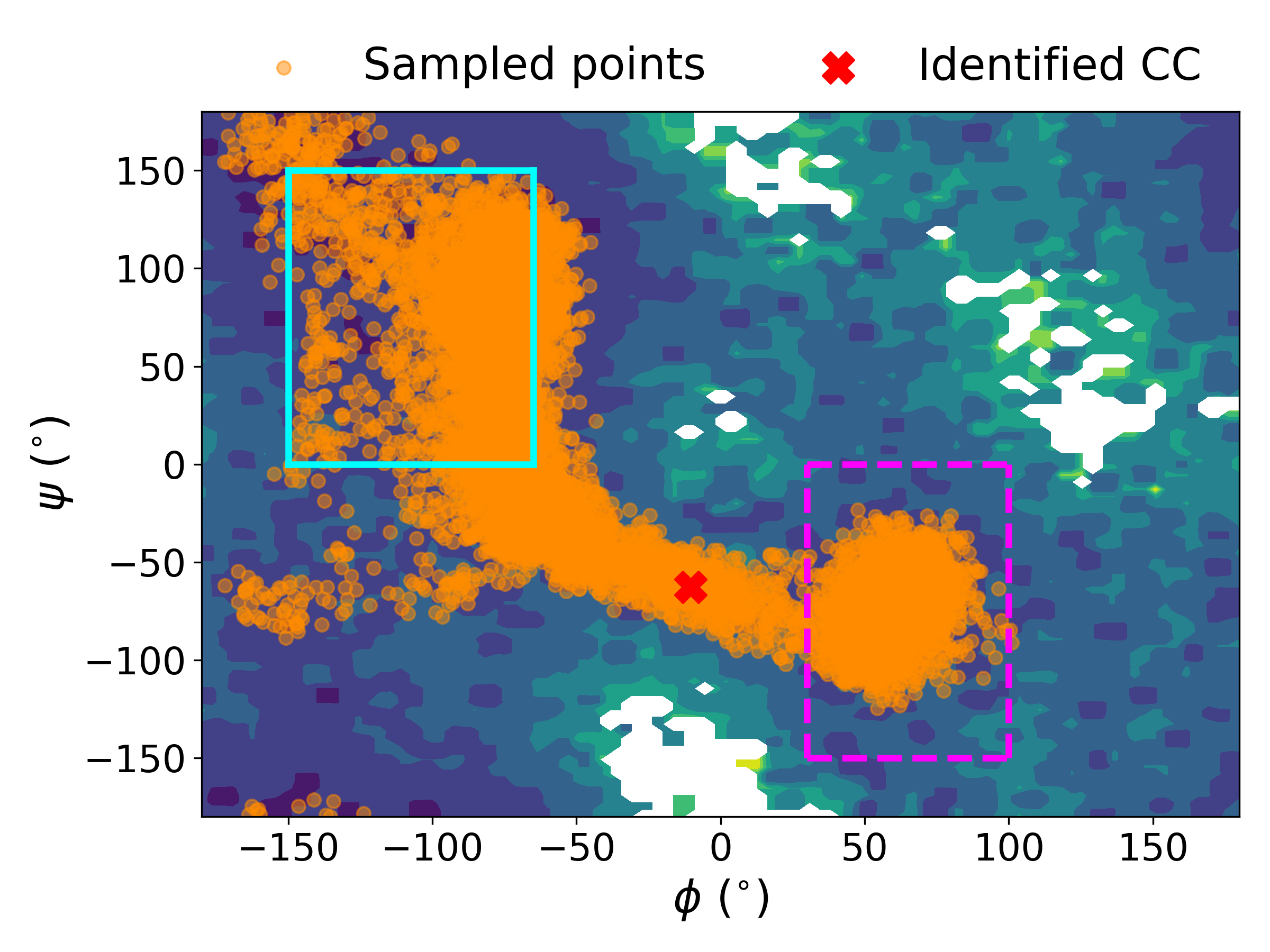}}
\subfigure[Change of configuration from  from ${\rm CC}_1$ to metastable states.]{\label{fig:AD-pts-CC1-tworegions}\includegraphics[width=0.55\linewidth ]{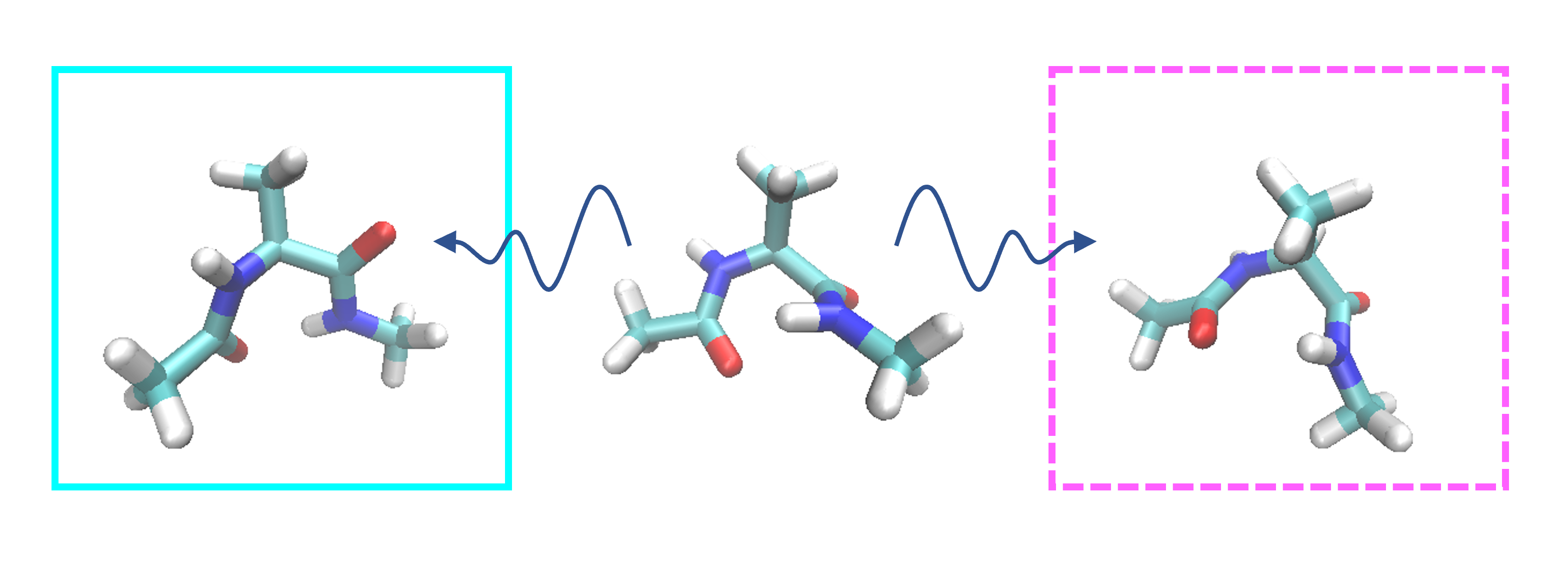}}
\caption{Alanine dipeptide in vacuum under temperature $300$ K. (a) The potential energy landscape visualized in $(\phi,\psi)$-space and two metastable states (solid and dotted rectangle). (b) configurations generated from overdamped Langevin dynamis at temperature 300 K.  (c) The action function learned by the proposed RL algorithm. The action function reveals two connective configurations (CC's). (d) Configurations generated by shooting from the identified connective configurations. (e) The change of molecule configuration from the identified connective configuration to metasable states. }
\label{fig:AD}
\end{figure*}

In this example, we show how the RL algorithm introduced above can be used to identify a reaction channel of an Alanine dipeptide (ADP) molecule in vacuum that corresponds to its isomerization process. In our numerical experiment, we set the temperature to $300$ K. The ADP molecule contains $22$ atoms (see Fig.~\ref{fig:AD-pts-CC1-tworegions}). Therefore, the dimension of the configuration space is 66. The isomerization process is principally described by two the dihedral angles $(\phi, \psi)\in [-180^{\circ},180^{\circ}]^2$ of a subset of atoms (indexed by $4,6,8,14$ and $6,8,14, 16$, respectively.) With a slight abuse of notation, we use $\phi(\Bx)$ and $\psi(\Bx)$ to donate the mapping from a configuration $\Bx$ to the two specific torsion angles. Figure~\ref{fig:AD-potential} shows the potential energy landscape in the $(\phi, \psi)$-space. The plot is constructed as follows. We first generate a long trajectory at a relatively high temperature ($1200$ K). We denote the set of configurations along this trajectory by $\mathbb{S}$. The potential energy for each configuration in $\mathbb{S}$ is stored. Next, we discretize $(\phi, \psi)$ in $(-180^{\circ},180^{\circ}]\times (-180^{\circ},180^{\circ}]$ by generating a $100\times 100$ uniform grid. For each $(\phi_i,\psi_j)$ pair on the grid, we define a neighborhood 
\[
C(\phi_i,\psi_j) = \{\Bx \in \mathbb{S} : |\phi(\Bx)-\phi_i| \leq 5^\circ, |\psi(\Bx)-\psi_j| \leq 5^\circ\}.
\] 
If $C(\phi_i,\psi_j) \neq \emptyset$, we set the potential energy associated with $(\phi_i,\psi_j)$ to  $V_{\min}$, where 
\[
V_{\min} = \min_{\Bx \in C(\phi_i,\psi_j)} V(\Bx).
\]
Otherwise, the potential energy of $(\phi_i,\psi_j)$ is undefined and colored by white in Figure~\ref{fig:AD}. The two metastable regions $A$ (solid box in Fig.~\ref{fig:AD-potential}) and $B$ (dotted box) are defined by $A = \{\Bx:-150^{\circ}\leq \phi(\Bx)\leq -65^{\circ}, 0^{\circ}\leq\psi(\Bx)\leq 150^{\circ}\}, B= \{\Bx:30^{\circ}\leq \phi(\Bx)\leq 100^{\circ}, -150^{\circ}\leq\psi(\Bx)\leq 0^{\circ}\}$. Figure~\ref{fig:AD-pts1} shows the snapshots of one trajectory initiated at a configuration near $A$ of $T=1\times 10^8$ fs when the temperature is $300$ K and we can see that the no reactive trajectory is observed. 

\textbf{Identifying reaction channels.} Our RL algorithm aims to find a connective configuration in the $(\phi, \psi)$-space.  Subsequently, we use this configuration to generate additional configurations that bridge two metastable states. We should note that, for the ADP system, the actions taken in the RL algorithm are defined in a low-dimensional space specified by $(\phi,\psi)$ whereas the shooting procedure used to evaluate the reward takes place in the 66-dimensional configuration space. To address this disparity in dimensionality, it is necessary to establish a one-to-one mapping between each $(\phi,\psi)$ pair and a configuration in the phase space. To this end, we first construct a configuration set $\mathbb{P}$ by generating trajectories at high temperature $1200$ K. We then map a given torsion coordinate in $(\phi, \psi)$ back to the configuration space by choosing a configuration from $\mathbb{P}$ with lowest potential energy. We simulate the Langevin dynamics with a step size of $2$ fs and friction $10 \text{ ps}^{-1}$ using the package Openmm Python API~\cite{eastman2017openmm}. The reward for each action is computed by shooting 10 trajectories of $T=2\times 10^3$ fs with kinetic initialization randomly sampled from the Boltzmann-Gibbs distribution. Each RL episode consists of $L=10$ steps (actions). Figure~\ref{fig:AD-action} shows the learned action function reveals $2$ different connective configurations, i.e., $(-10^\circ, -62^\circ)$ and $(139^\circ, -120^\circ)$. {However, our primary interest is in $(-10^\circ, -62^\circ)$ configuration, as it has been extensively studied in the existing literature~\cite{kikutsuji2022explaining,li2019computing,singh2023variational}. We shoot 100 trajectories of $T=2\times 10^3$ fs from this connective configuration to generate the additional configurations that bridge two metastable states as shown in Fig.~\ref{fig:AD-pts-CC1}. } We also visualize 
the change of molecular structures from the connective configuration to two metastable states in Fig.~\ref{fig:AD-pts-CC1-tworegions}. 

\textbf{Solving BKE.} Our next objective is to solve the BKE~\eqref{eqn:bke} in a 66-dimensional space. The obtained numerical solution is used to generate the plot of the approximate committor function in $(\phi,\psi)$. Such an approximation is then compared with the approximation obtained from the diffusion map (DM)~\cite{coifman2008diffusion,ko2023using} method. 
Note that it is not possible to use traditional PDE solvers, such as finite difference and finite element to solve \eqref{eqn:bke} because they suffer from the curse of dimensionality, i.e., their computational cost increase exponentially with respect to dimension of the problem~\cite{weinan2021algorithms}.  The DM method is another sample-based method that allows for the solution of the BKE to be approximated on an arbitrary set of configurations $\{\Bx^i\}$. However, it is important to note that DM may not always produce an accurate approximation of the derivatives at configurations near the boundary~\cite{liang2021solving, jiang2023ghost}. This can result in less accurate DM solutions to BKE near the boundaries of a fixed domain.

\begin{figure*}[ht]
\centering
\subfigure[Diffusion map.]{\label{fig:ADDM}\includegraphics[width=0.45\linewidth ]{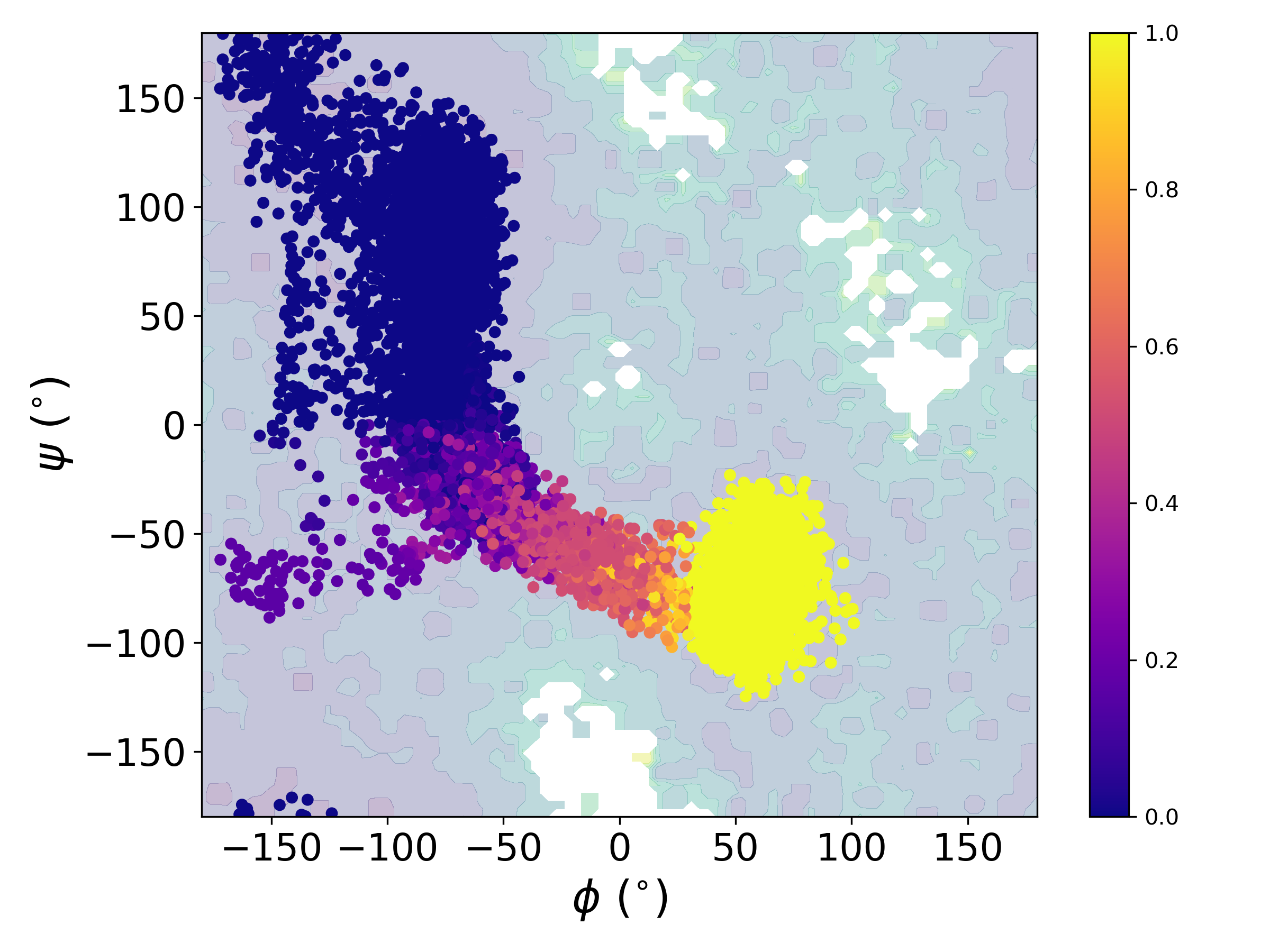}}
\subfigure[NN.]{\label{fig:ADNN}\includegraphics[width=0.45\linewidth ]{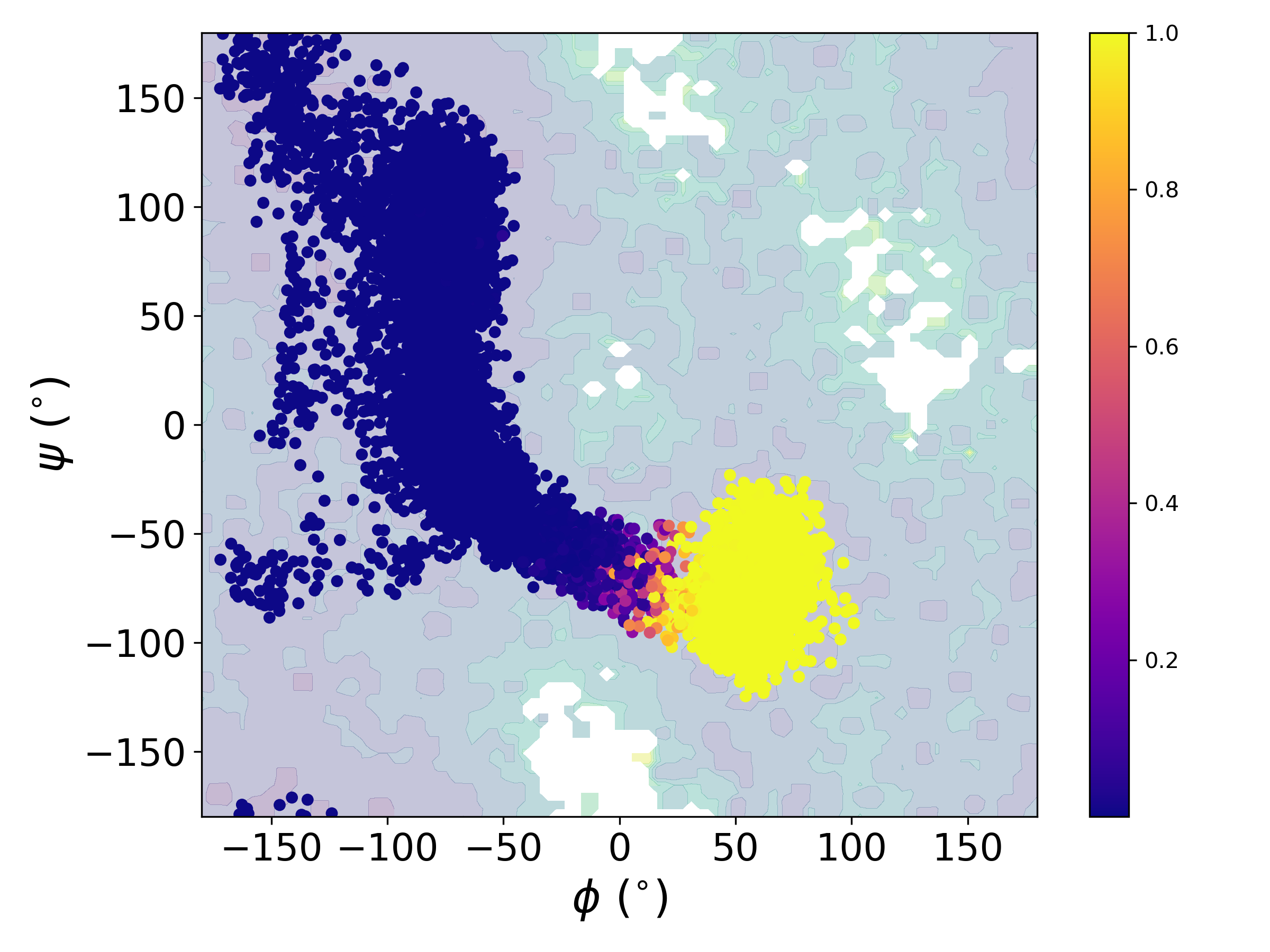}}
\caption{A comparison of solutions of the BKE~\eqref{eqn:bke} obtained from a diffusion map and NN respectively at temperature 300K. }
\label{fig:ADbke}
\end{figure*}

We use the dataset $\{\Bx^i\}$ identified by our RL method as shown in Fig.~\ref{fig:AD-pts-CC1} and apply DM and NN method to get the solutions of the BKE. The solutions are presented in Fig.~\ref{fig:ADbke}. We observe that the half-isocommittor region of the solution, i.e., the set of configurations on which the committor function value is close to 0.5, obtained from the DM method occupies a relatively large area in the $(\phi,\psi)$ plane, whereas the half-isocommittor region defined by the NN solution is confined to a small area defined by $[-25^{\circ},25^{\circ}]\times[-80^{\circ},25^{\circ}]$. Our solution is found to be more consistent with the results presented in Fig. 1 Panel 2 of Ref.~\cite{kikutsuji2022explaining}. 

\textbf{Rate estimation.} To estimate the reaction rate, we use formula \eqref{eqn:rate_sq} and approximate the evaluation of the integral using the generated configurations $\{\Bx^i\}$ obtained through the shooting method. This approach yields the approximation formula:
\begin{align}
\kappa=\frac{\int_{\Omega} e^{-\beta V(\Bx)} \beta^{-1}\left\|\nabla q(\Bx)\right\|^2 d \Bx}{\int_{\Omega} e^{-\beta V(\Bx)} d \Bx}\approx \frac{\sum_{i=1}^N e^{-\beta V(\Bx_i)} \beta^{-1}\left\|\nabla q(\Bx_i)\right\|^2 }{\sum_{i=1}^N e^{-\beta V(\Bx_i)}}.
\label{eqn:rate_sq_est}
\end{align}
Finally, evaluating \eqref{eqn:rate_sq_est} using the approximated solution of KBE, we obtain a rate of $1.58\times 10^{-5}\text{ ps}^{-1}$. This calculated estimate is comparable to the approximated value $4.54\times 10^{-5}\text{ ps}^{-1}$ (reported in Figure 5 of Ref.~\cite{singh2023variational}). 

\section{Conclusion}
\label{sec:conclu}
We presented a novel RL-based approach for identifying and characterizing an ensemble of configurations where reactive trajectories are likely to be found.  The optimized action function returned from the RL algorithm reveals connective configurations that have high reactive probabilities. One of the key elements of the RL algorithm is the proper construction of a reward function that serves as a surrogate for measuring the reactive probability of a configuration, which is normally defined in terms of the value of it committor function. Because the exact committor function is generally unknown in advance, we employ a randomized shooting procedure to estimate its value at an arbitrary configuration. Our results demonstrate the effectiveness of this approach across multiple model problems of different sizes.  Using the identified connective configurations, we generate trajectories directed towards metastable regions. The configurations along these trajectories are utilized to define reactive channels on which a restricted BKE is solved by a NN-based PDE solver. The solution yields an approximate committor function evaluated within these channels. This committor function can then be used to estimate reaction rates.

\section*{Acknowledgement}
This material is based upon work supported by the U.S. Department of Energy, Office of Science, Office of Advanced Scientific Computing Research and Office of Basic Energy Science, Scientific Discovery through Advanced Computing (SciDAC) program under Contract No. DE-AC02-05CH11231. This work used the computational resources of the National Energy Research Scientific Computing (NERSC) center under NERSC Award ASCR-ERCAP m1027 for 2023, which is supported by the Office of Science of the U.S. Department of Energy under Contract No. DE-AC02-05CH11231.



\bibliography{sample}

\appendix
\section{Q-learning algorithm}
\label{Qlearningdiagram}
\begin{figure*}
\centering
\includegraphics[width=1.0\linewidth]{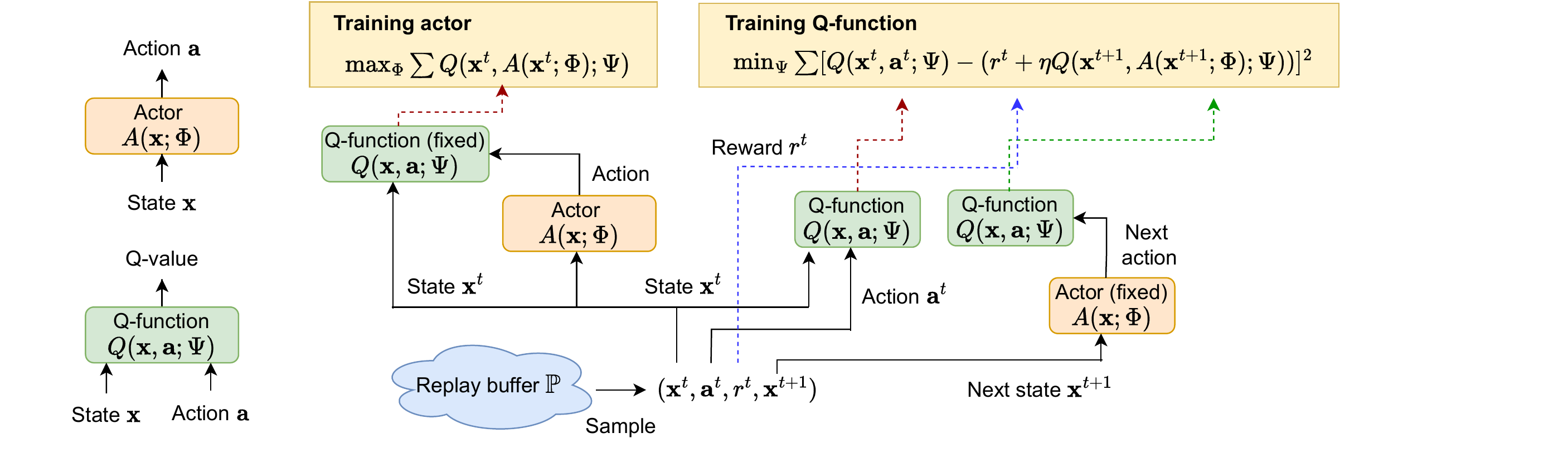}
\caption{\label{fig:qlearning} A schematic illustration of how the Q-learning is performed for a continuous action.}
\end{figure*}

Figure~\ref{fig:qlearning} gives a schematic description of how the action function and the $Q$-function are optimized in the RL algorithm used to identify connective configurations. 

\section{Numerical results for high temperature regime}
\label{sec:hightemresults}
\subsection{Triple-well potential}
\label{sec:triwell-appdx}
This section presents the the results obtained from running the RL algorithm to identify connective configurations for a triple-well potential with an inverse temperature of $\beta=1.67$. The reward function defined in Equation~\eqref{eqn:reward} was evaluated by shooting 50 trajectories of that evolve up to time $T=0.75$. The maximum number of time steps taken in each trajectory is set to $L=20$. The RL procedure was run for a total of 1000 episodes.

\begin{figure*}[ht]
\centering
\subfigure[Learned action function and the connective configurations it reveals.]{\label{fig:trihigh-action}\includegraphics[width=0.45\linewidth ]{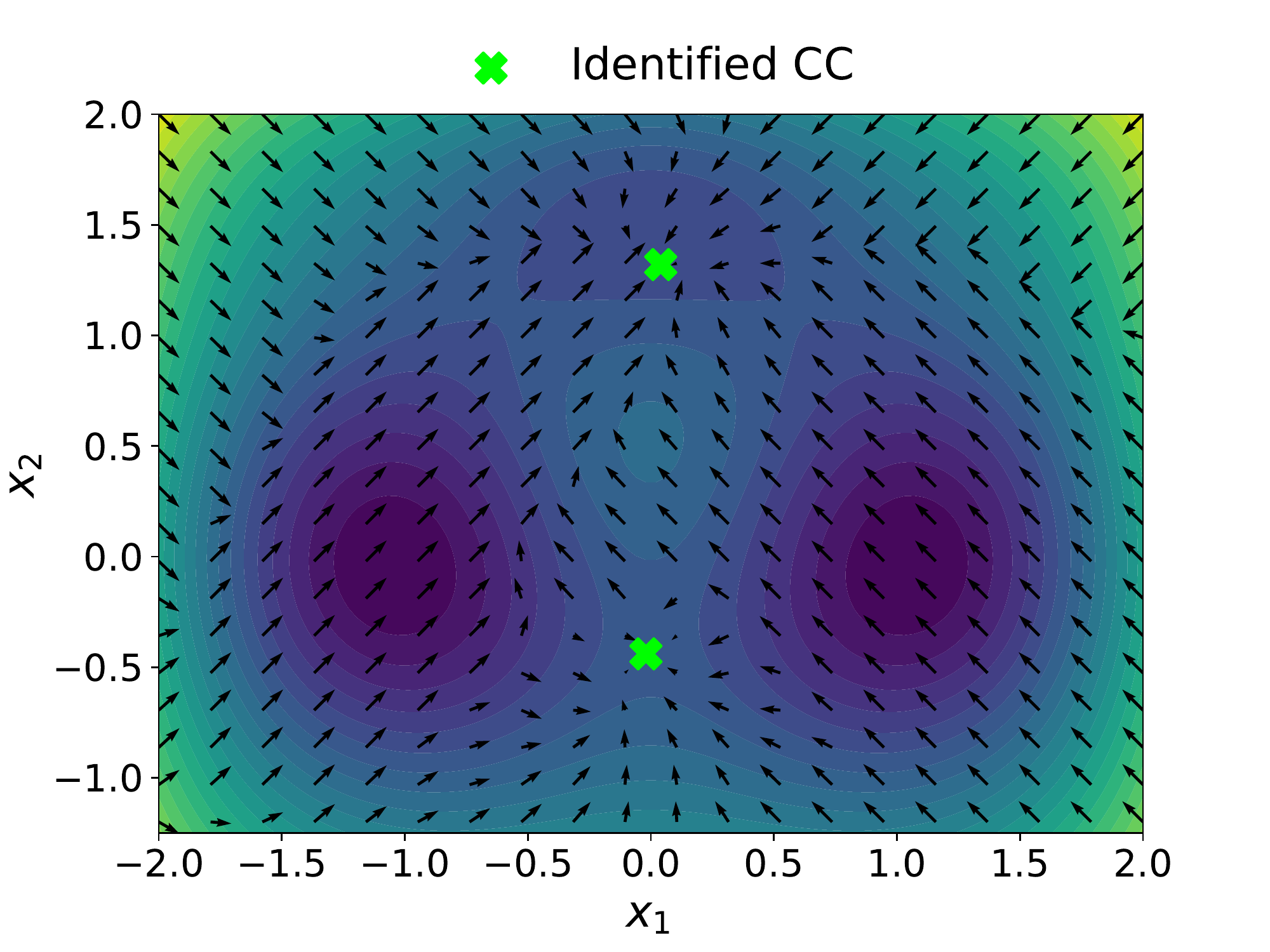}}
\hfill
\subfigure[Sampled configurations generated by shooting trajectories from the two identified connective configurations.]{\label{fig:trihigh-pts}\includegraphics[width=0.45\linewidth ]{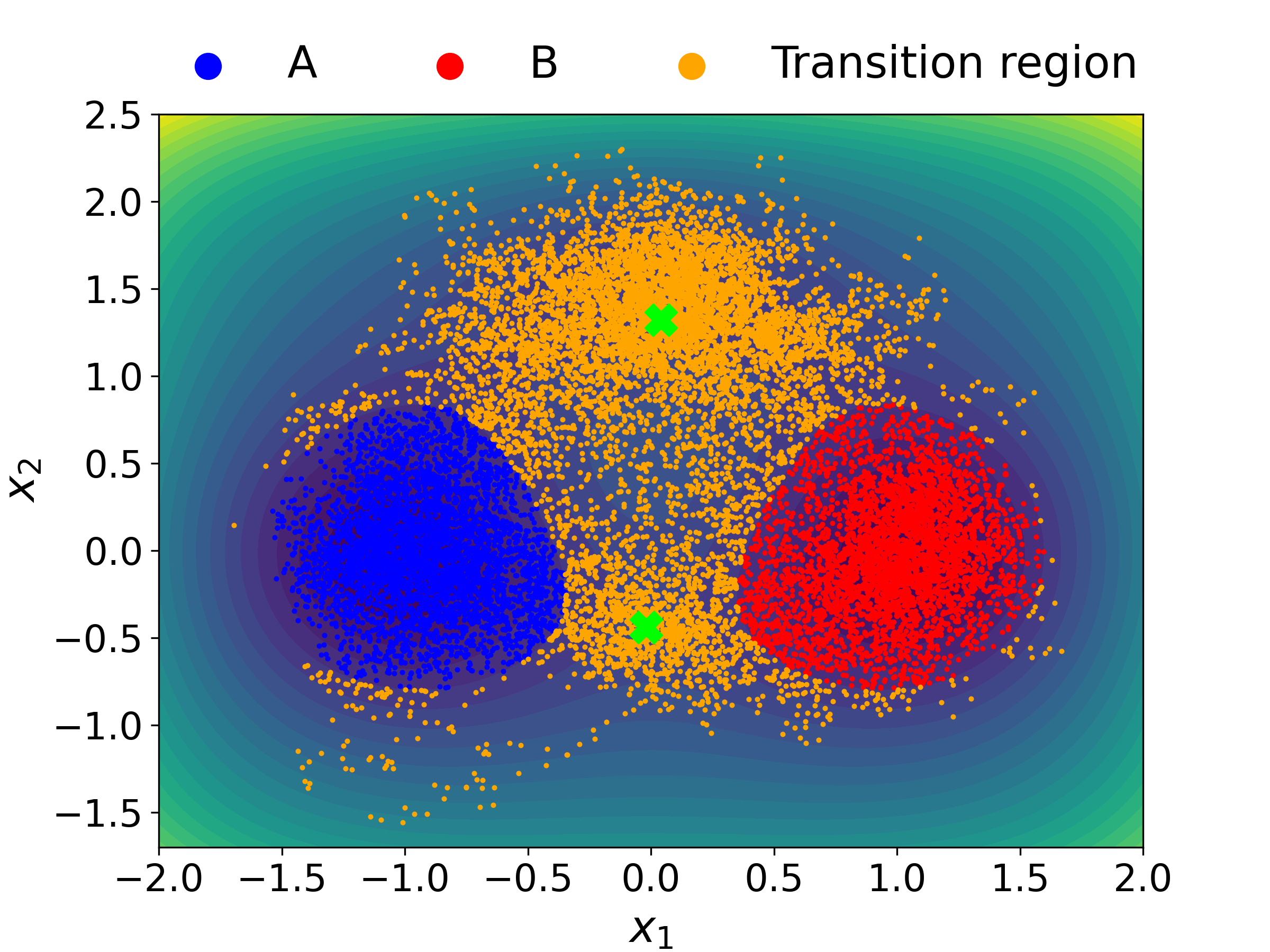}}
\subfigure[The FDM solution of \eqref{eqn:bke} on $\Omega$]{\label{fig:trihigh-fem}\includegraphics[width=0.32\linewidth ]{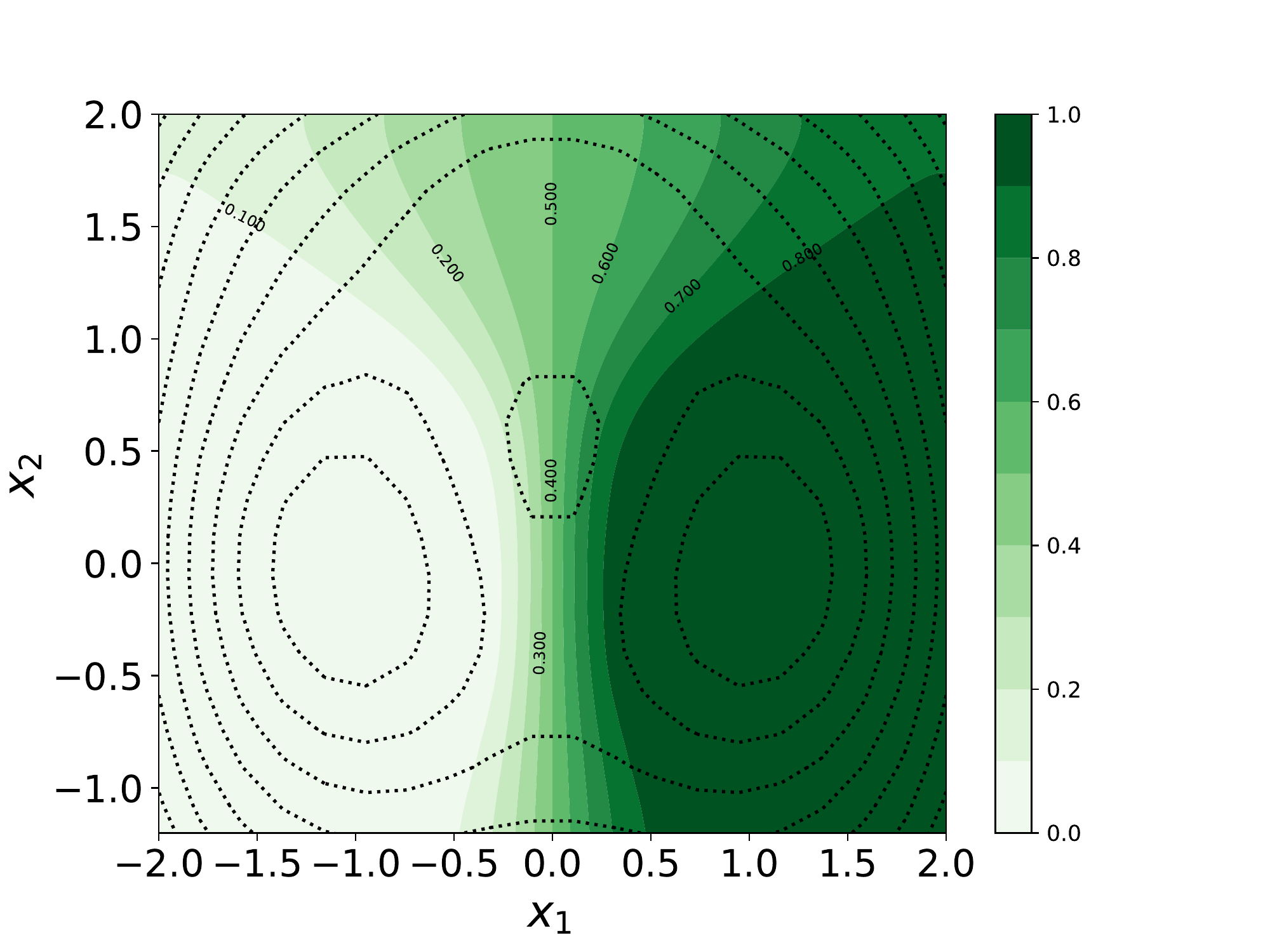}}
\subfigure[The NN solution of \eqref{eqn:bke}.]{\label{fig:trihigh-nn}\includegraphics[width=0.32\linewidth ]{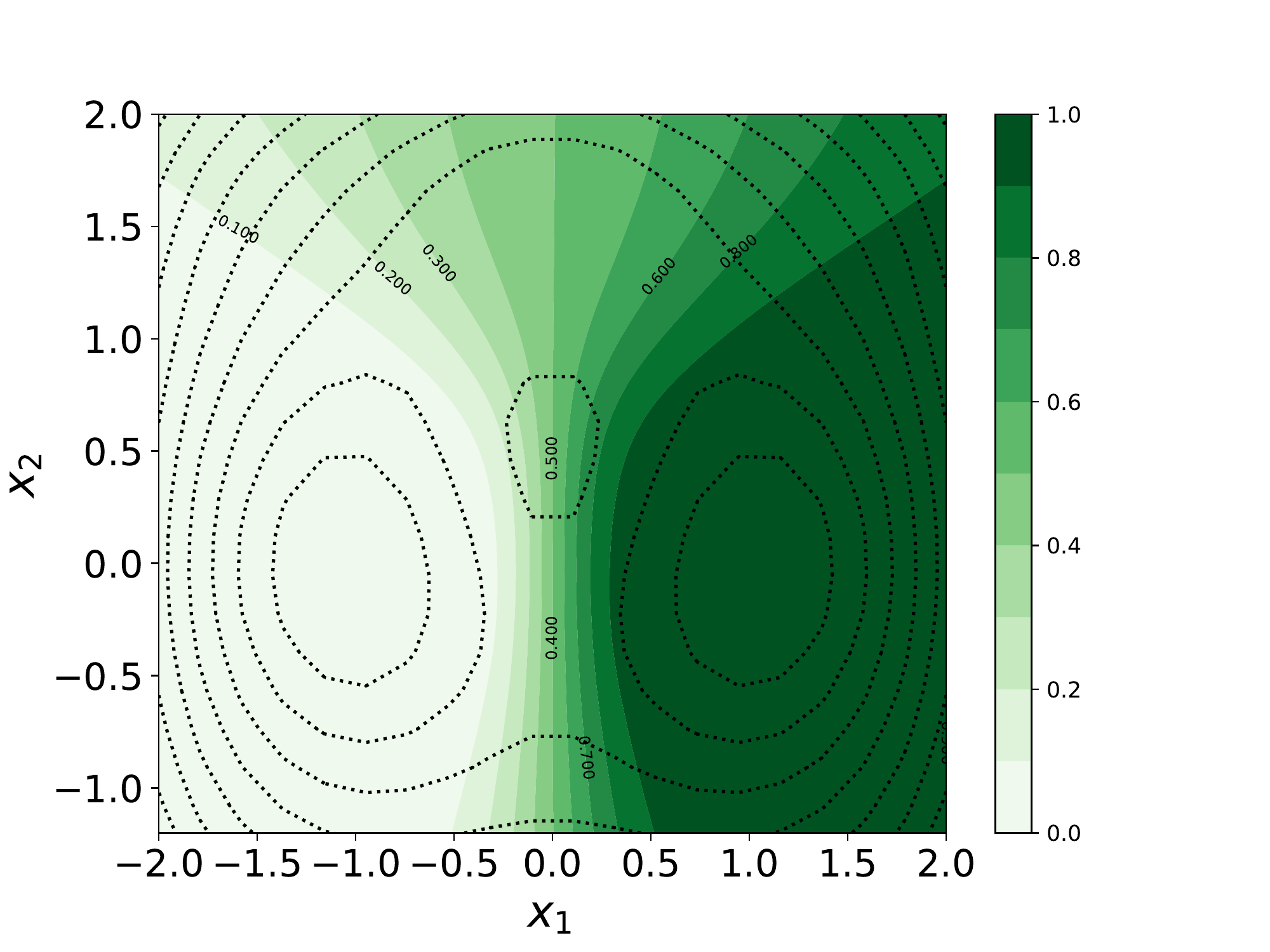}}
\subfigure[The difference between (c) and (d).]{\label{fig:trihigh-error}\includegraphics[width=0.32\linewidth ]{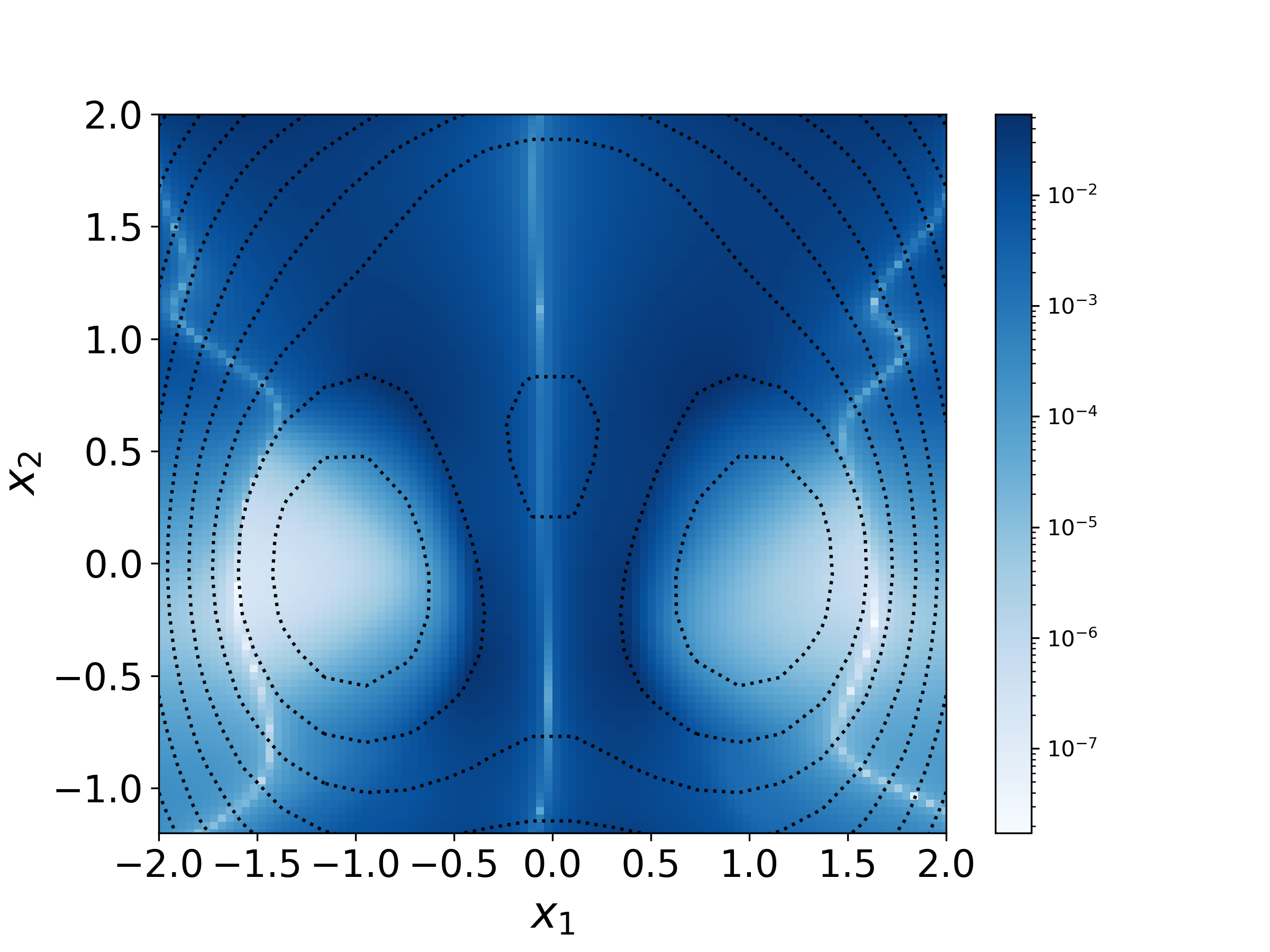}}
\caption{The results obtained from running the RL algorithm for Triple-well potential at inverse temperature $\beta=1.67$. (a) The action function learned by our proposed RL. The action function reveals two connective configurations (crosses). (b) The configurations generated by shooting trajectories initiated at the identified connective configurations. (c-e) The FDM (c) and NN (d) solutions of the BKE \eqref{eqn:bke} and their difference (e). }
\label{fig:trihigh}
\end{figure*}

Figure~\ref{fig:trihigh-action} shows the action function $A(x;\Phi)$ produced at the end of the RL run. We mark two distinct connectivity configurations identified by crosses. They are nearly identical to the connective configurations we found when the trajectories were generated using $\beta = 6.67$. In Figure~\ref{fig:trihigh-pts}, we plot configurations generated by shooting trajectories from two connective configurations.  We observe that the configurations generated from running trajectories using $\beta=1.67$ cover a wider region of the configuration space that includes the local maximum near $(0.0, 0.5)$, as well as some configurations outside the pre-defined domain $\Omega=[-2,2]\times[-1.2,2]$. Figures~\ref{fig:trihigh}(c-e) show that the neural network solution to BKE is comparable to solution obtained from the finite difference method (FDM).

\subsection{Rugged Muller potential}
\label{sec:rugmul-appdx}
\begin{figure*}[ht]
\centering
\subfigure[Learned action function and the connective configurations it reveals.]{\label{fig:RMhigh-action}\includegraphics[width=0.45\linewidth ]{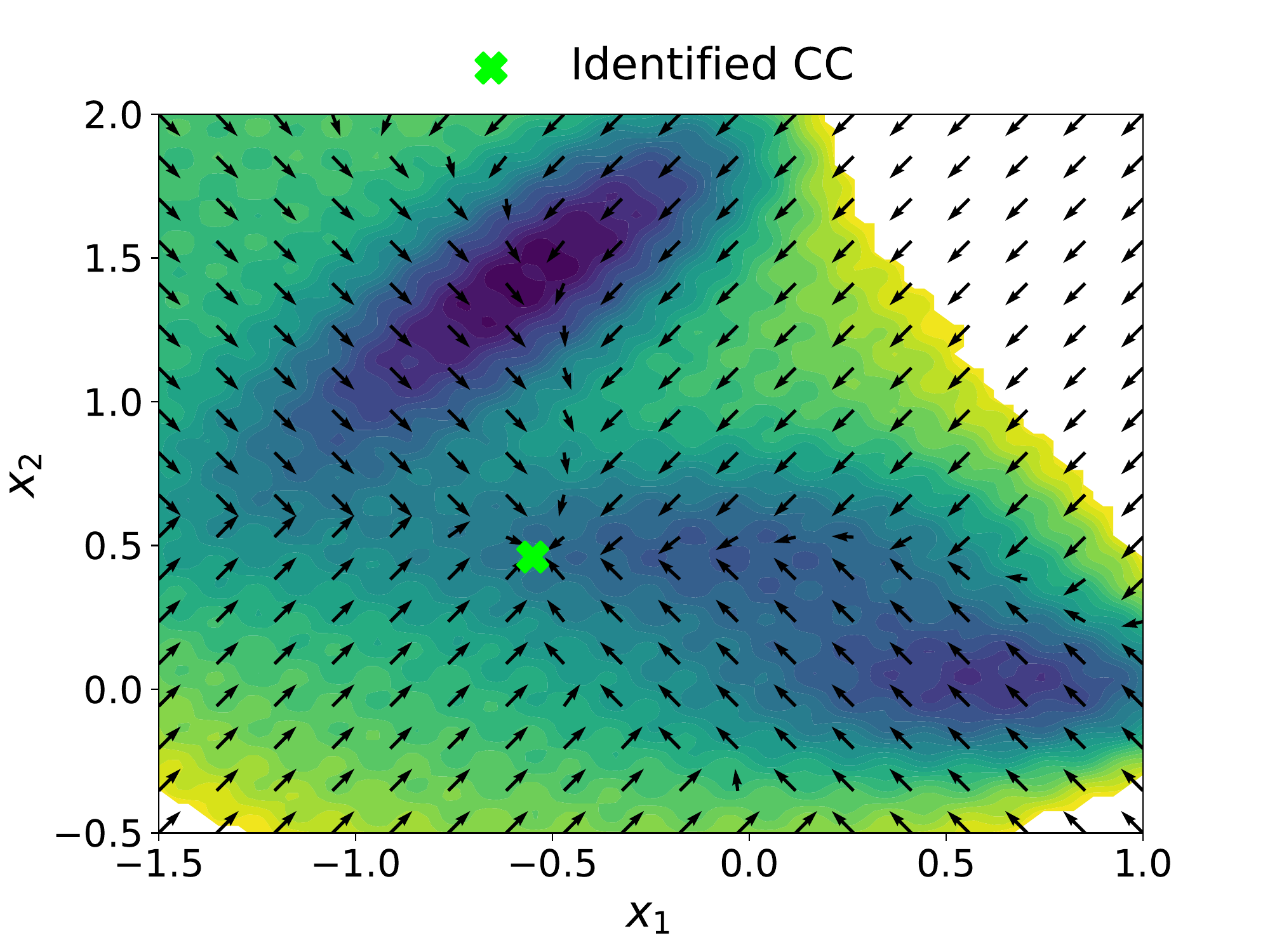}}
\hfill
\subfigure[Sampled configurations generated by shooting from identified connective configurations.]{\label{fig:RMhigh-pts}\includegraphics[width=0.45\linewidth ]{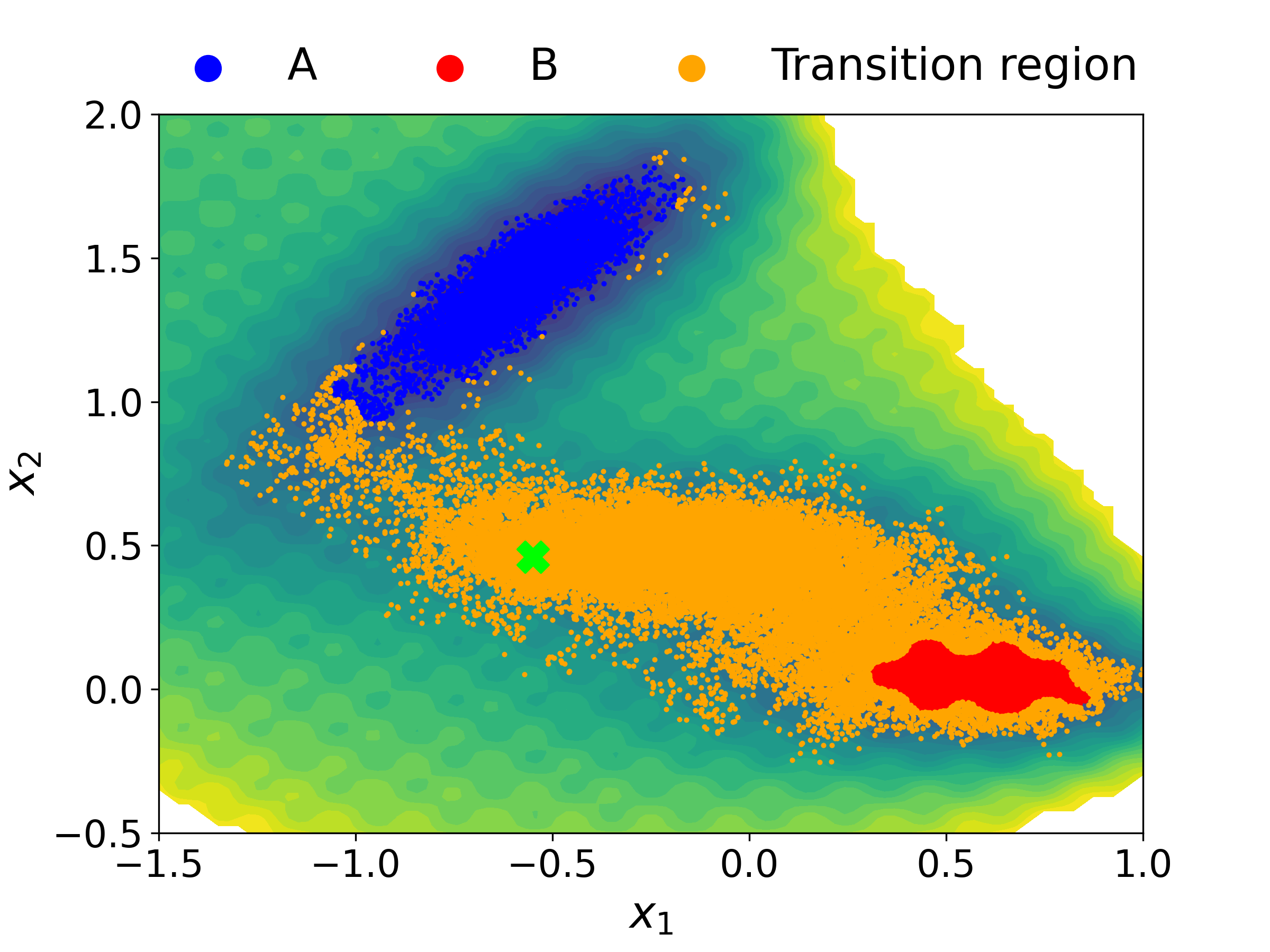}}
\subfigure[The FDM solution of \eqref{eqn:bke}.]{\label{fig:RMhigh-fdm}\includegraphics[width=0.32\linewidth ]{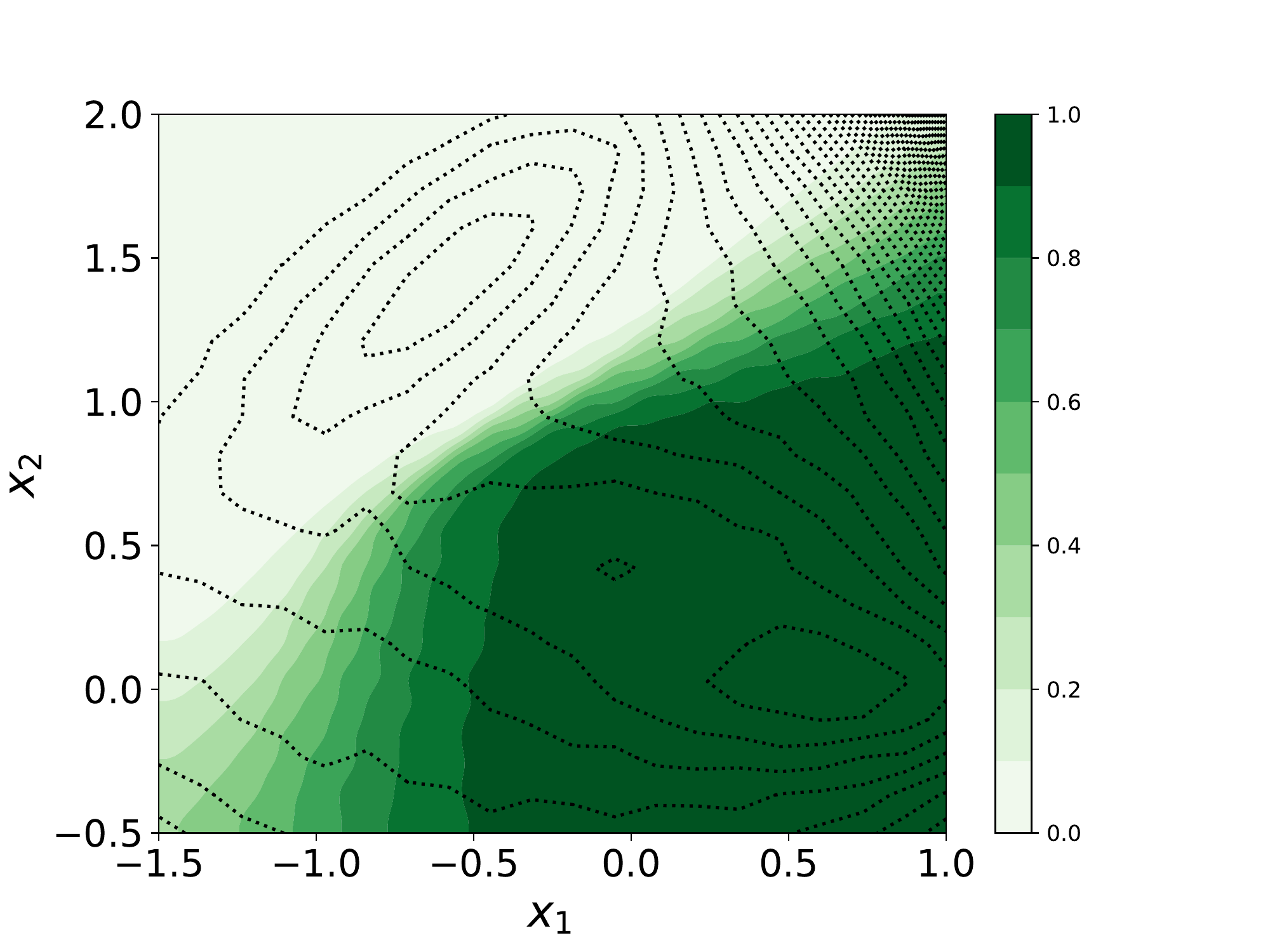}}
\subfigure[The NN solution of \eqref{eqn:bke}.]{\label{fig:RMhigh-nn}\includegraphics[width=0.32\linewidth ]{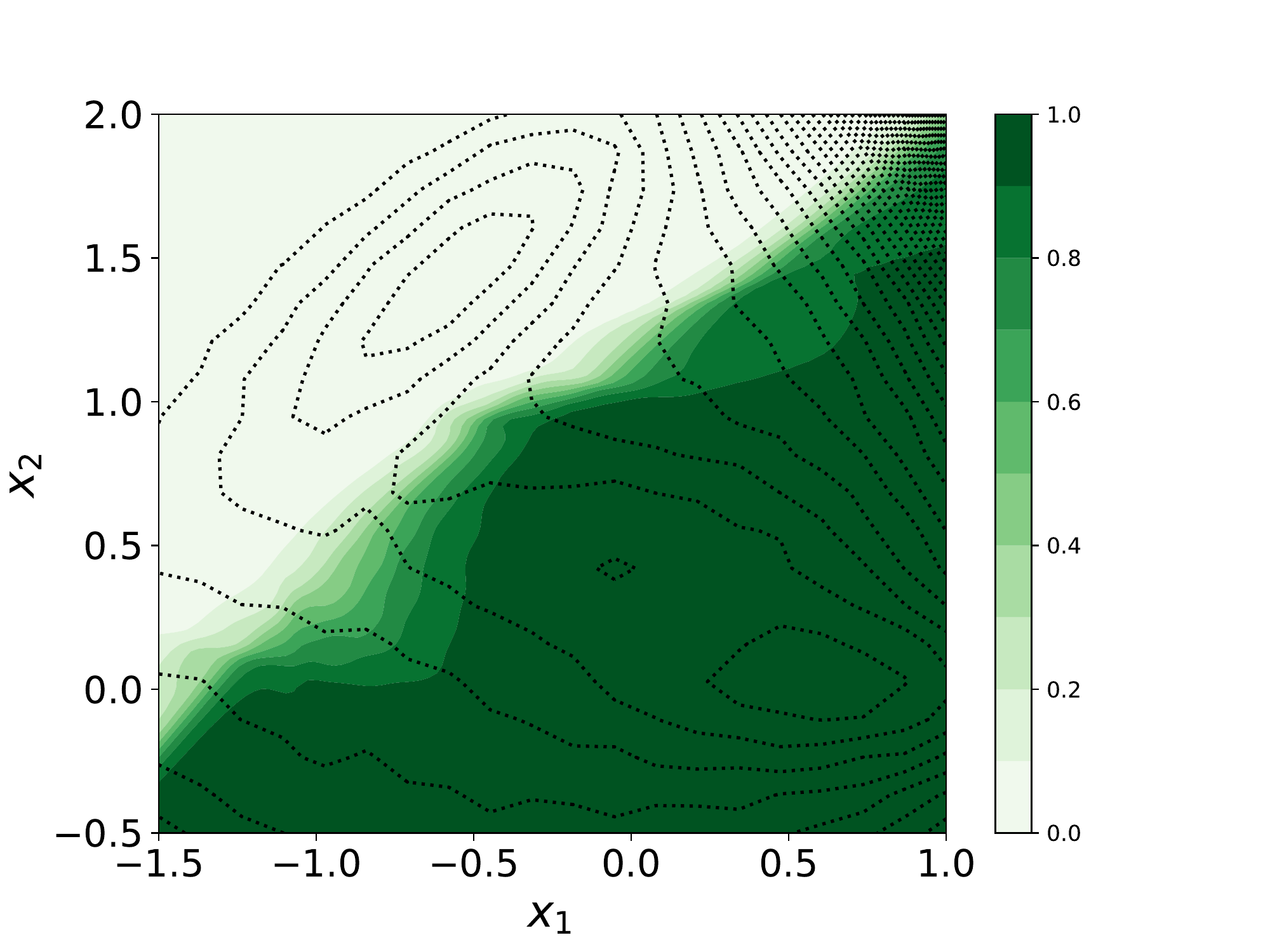}}
\subfigure[The difference between (d) and (d).]{\label{fig:RMhigh-error}\includegraphics[width=0.32\linewidth ]{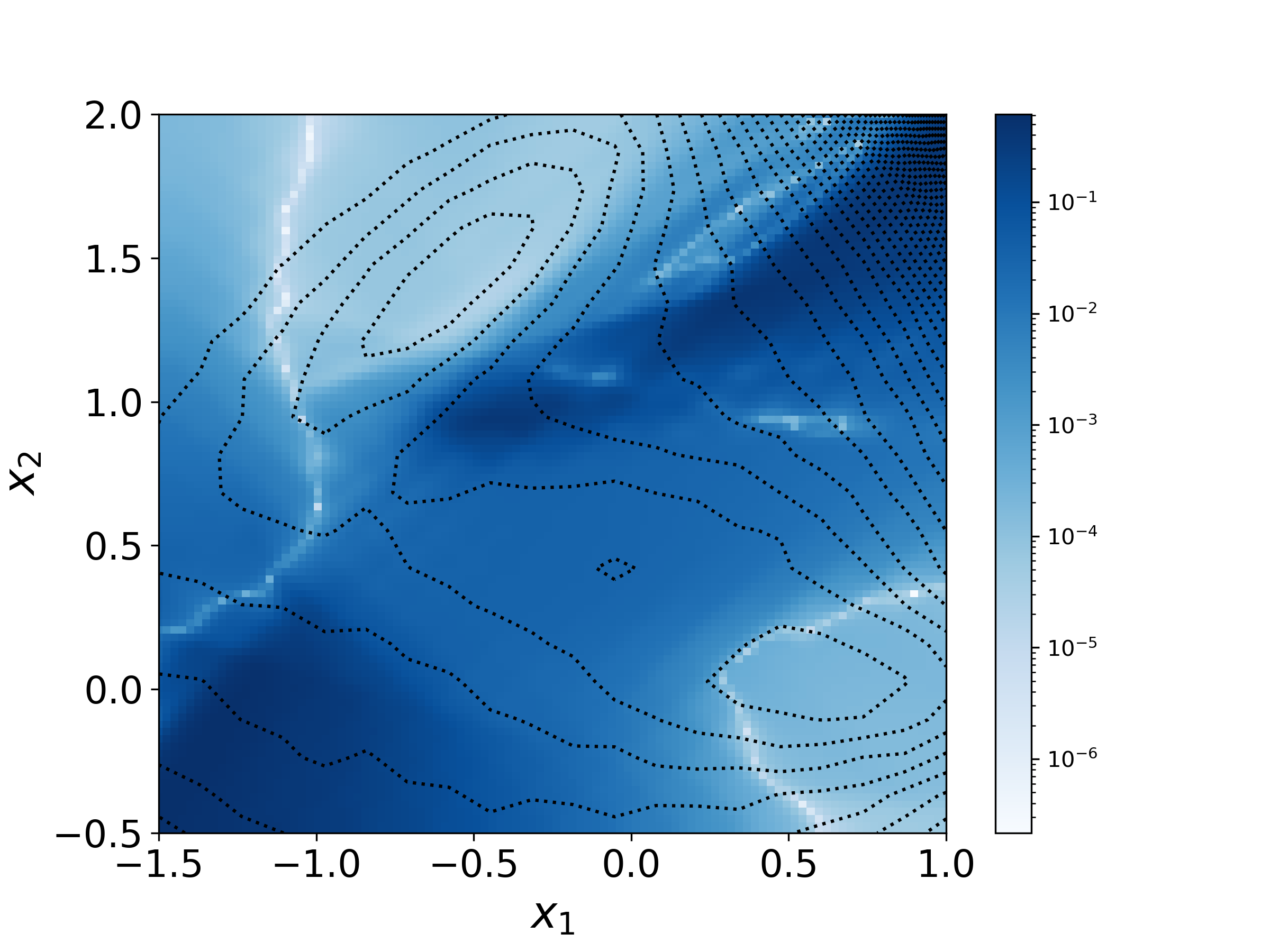}}
\caption{The results obtained from running the RL algorithm for the Rugged Muller potential at the inverse temperature $\beta=0.1$. (a) The action function learned by the RL algorithm. It reveals two connective configurations (crosses). (b) The configurations generated by shooting trajectories from the identified connective configurations. (c-e) The FDM (c) and NN (d) solutions of the BKE \eqref{eqn:bke} and their difference (e).}
\label{fig:RMhigh}
\end{figure*}

Figure~\ref{fig:RMhigh} shows the results obtained from applying the previously presented RL algorithm to the Rugged Muller potential at an inverse temperature of $\beta=0.1$. The reward function defined in Equation~\eqref{eqn:reward} was computed based on shooting $N=10$ trajectories that evolve up to time $T=0.05$. The maximum number of time steps taken in each trajectory was set to $L=20$. The RL algorithm was run for a total of 1000 episodes. We can observe that the configurations generated by shooting trajectories from the identified connective configurations with $\beta=0.1$ (see Fig.~\ref{fig:RMhigh-pts}) cover a wider area compared to that obtained from performing RL and generating trajectories at $\beta=0.25$. The committor function appears to be more stable near the region around $(-0.8, 0.6)$. The error is relatively small in the region covered by the sampled configurations.

\begin{figure*}[ht]
\centering
\subfigure[Muller: $\beta=0.1$]{\label{fig:rw-rm-0p1}\includegraphics[width=0.23\linewidth ]{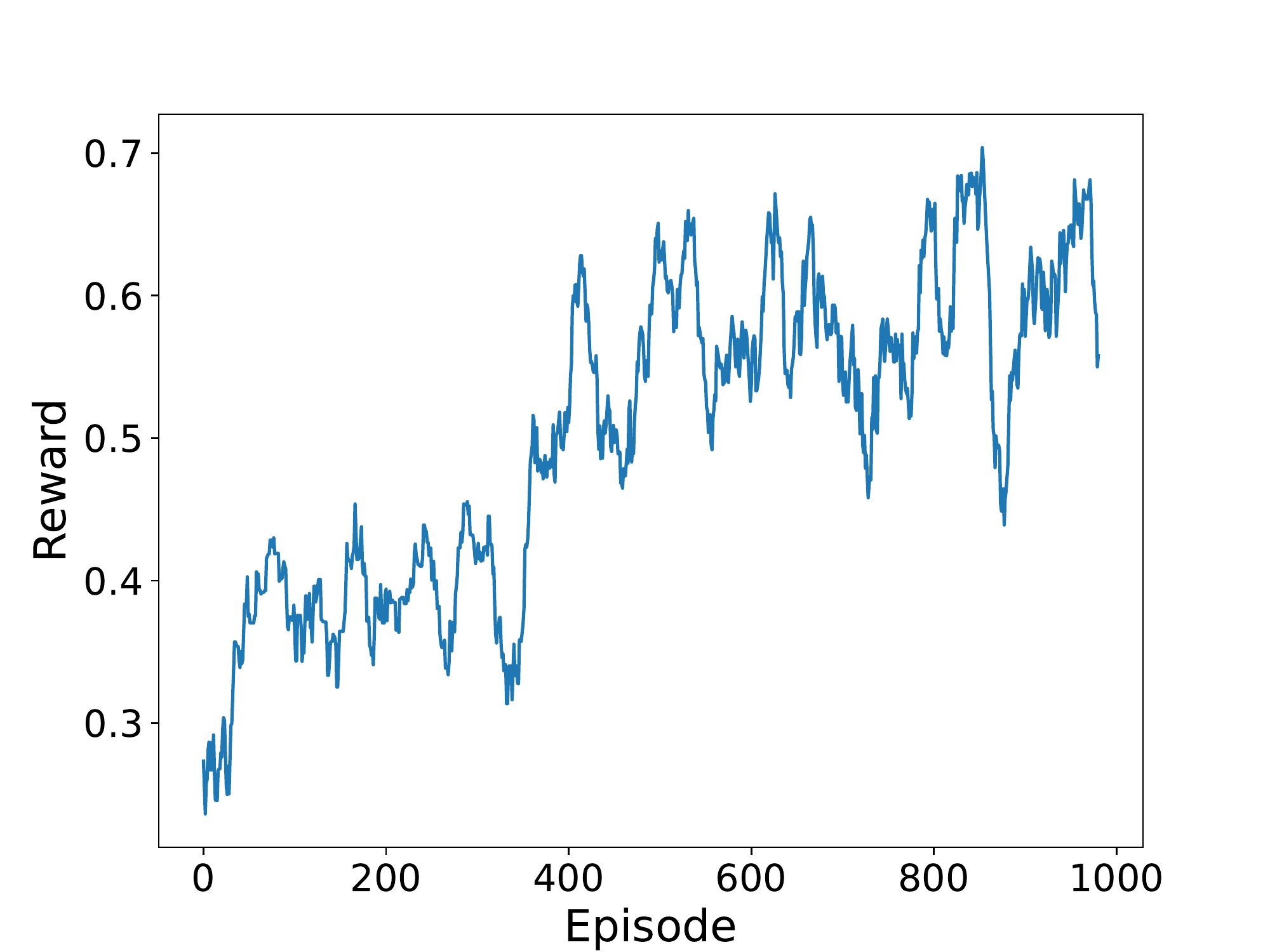}}
\subfigure[Muller: $\beta=0.25$]{\label{fig:rw-rm-025}\includegraphics[width=0.23\linewidth ]{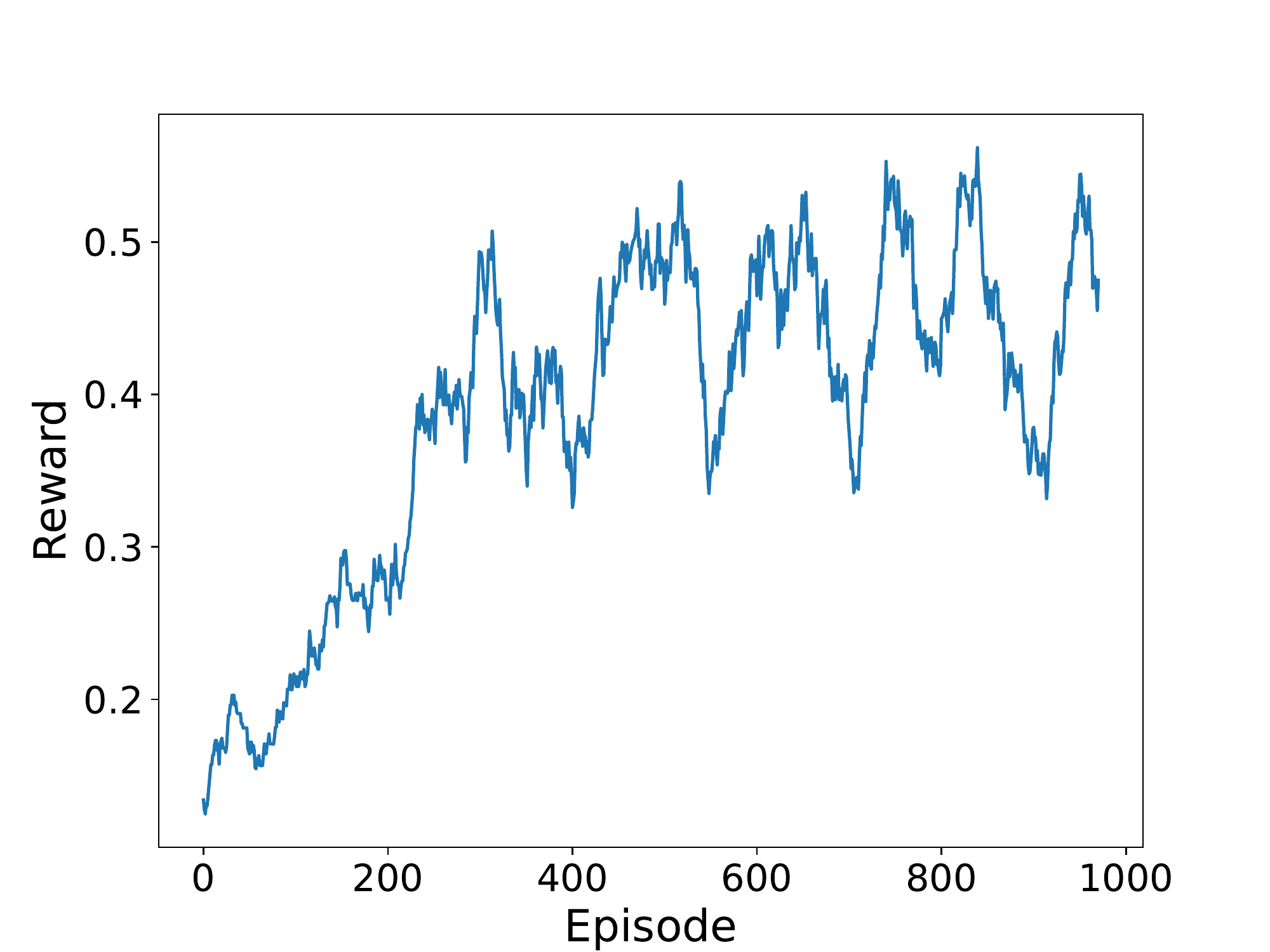}}
\subfigure[Tri-well: $\beta=1.67$]{\label{fig:rw-tw-1p67}\includegraphics[width=0.23\linewidth ]{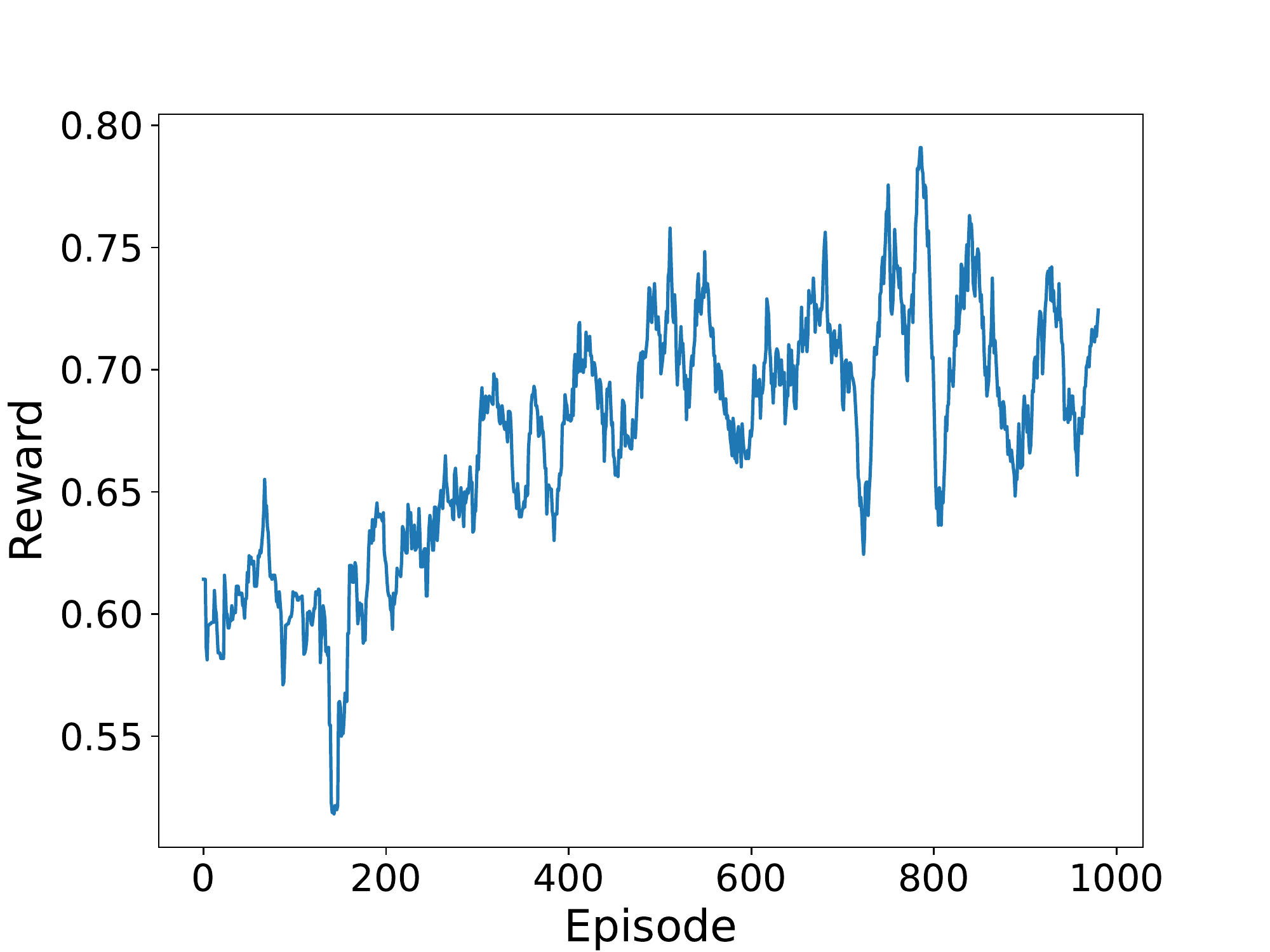}}
\subfigure[Tri-well: $\beta=6.67$]{\label{fig:rw-tw-6p67}\includegraphics[width=0.23\linewidth ]{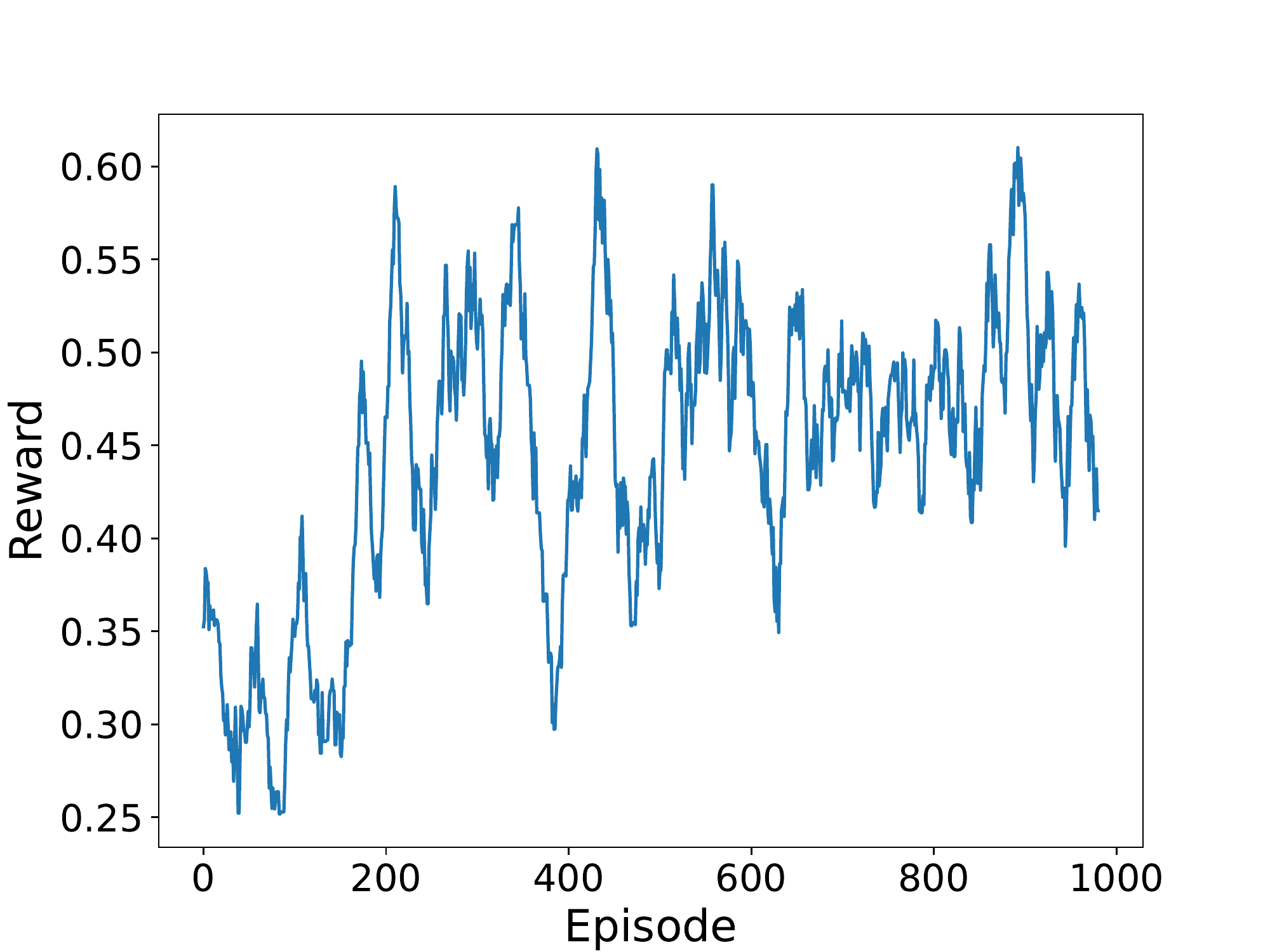}}
\caption{The moving average of the reward as a function of the episode number during the RL procedure. Here the value of the reward is linearly scaled to the interval $[0,1]$ by min-max normalization. }
\label{fig:reward}
\end{figure*}

\subsection{Monitoring RL progress}
\label{sec:trainingcurve}
Figure~\ref{fig:reward} shows the reward as a function of the episode in the RL algorithm. A maximum of 1000 episodes are allowed. However, the largest rewards are achieved around episode 500. This suggests that the RL process is efficient and reaches a high level of performance before reaching the limit on the allowed number of episodes.

\section{Implementation details of NN solutions of the BKE}
\label{sec:hyperNN}
In this paper, we approximate the solution of the BKE by a fully connected neural network (FNN), which can be viewed as a composition of $L$ simple nonlinear functions, i.e.,
$\phi(\mathbf{x};\bm{\theta}):=\sigma_2\circ\mathbf{a}^\top \mathbf{h}_L \circ \mathbf{h}_{L-1} \circ \cdots \circ \mathbf{h}_{1}(\mathbf{x})$. Here, $\mathbf{h}_{\ell}(\mathbf{x})=\sigma\left(\mathbf{W}_\ell \mathbf{x} + \mathbf{b}_\ell \right)$ with $\mathbf{W}_\ell \in \mathbb{R}^{N_{\ell}\times N_{\ell-1}}$, $\mathbf{b}_\ell \in \mathbb{R}^{N_\ell}$ for $\ell=1,\dots,L$, $\mathbf{a}\in \mathbb{R}^{N_L}$, $\sigma$ is the tanh function, and $\sigma_2$ is a sigmoid function such that the range of output is $[0,1]$. We use an FNN with $L=2$ and a uniform width $m$, i.e., $N_\ell = m$ for all $\ell\neq 0$. 
 
The FNN is optimized by Adam~\cite{kingma2014adam}. In Adam, we use an initial learning rate of 0.001 for $T$ iterations. The learning rate is then adjusted in each iteration by a factor of $0.5( \cos(\frac{\pi t}{T}) + 1)$, where $t$ is the current iteration number. We set the batch size to be the total number of training points. The hyperparameter setting for each numerical example is listed in Table~\ref{tab:hypercommittor}.

\begin{table}[htbp]
  \centering
  \caption{Hyperparameter setting for solving BKE with NN. Note that we use two different boundary coefficients for boundary conditions of $A$ and $B$ in the Alanine dipeptide case. }
    \begin{tabular}{lrrrrrrr}
    \toprule
        Example  & \multicolumn{2}{c}{Triple-well} &       & \multicolumn{2}{c}{Rugged Muller} &       & {Alanine dipeptide} \\
    \midrule
    $\beta$  & 1.67  & 6.67  &       & 0.1   & 0.25   &  & 1/2.5 kJ$^{-1}$mol  \\
    \midrule
    Width $m$ & 50    & 50    &       & 100   & 100 &  &  200  \\
    Boundary coefficient $\ell$ & 10    & 100   &       & 10,000 & 100,000 &  & 1 million \& 2 million\\
    Iteration & 30,000 & 30,000 &       & 100,000 & 30,000 &  & 100,000\\
    \bottomrule
    \end{tabular}%
  \label{tab:hypercommittor}%
\end{table}%

\section{Additional notes on the RL method}
\subsection{RL initialization from stable states}
\label{sec:stable}
\begin{figure*}[ht]
\centering
\subfigure[Triple-well $\beta=1.67$]{\includegraphics[width=0.4\linewidth ]{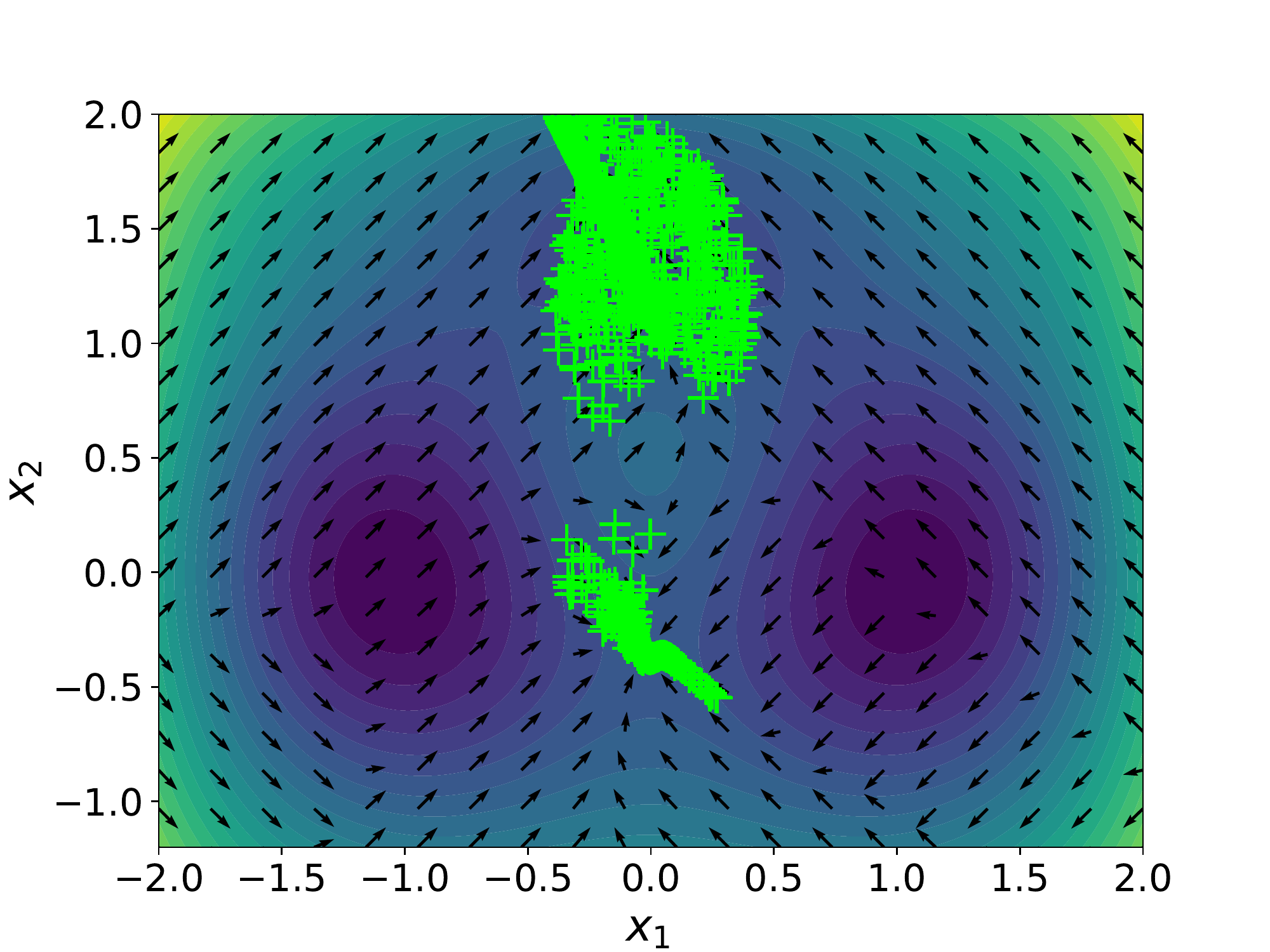}}
\subfigure[Triple-well $\beta=6.67$]{\includegraphics[width=0.4\linewidth ]{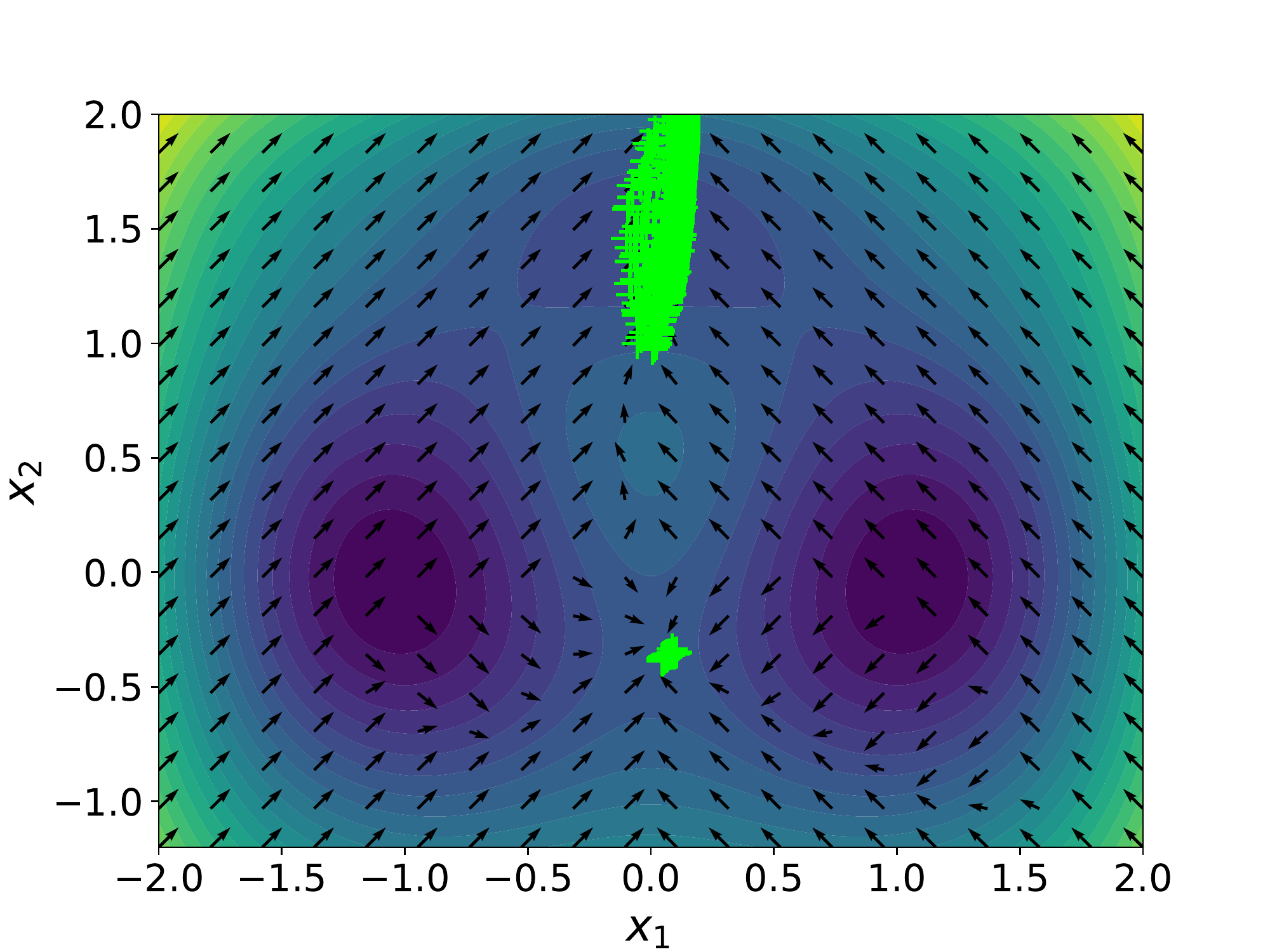}}
\subfigure[Rugged Muller $\beta=0.1$]{\includegraphics[width=0.4\linewidth ]{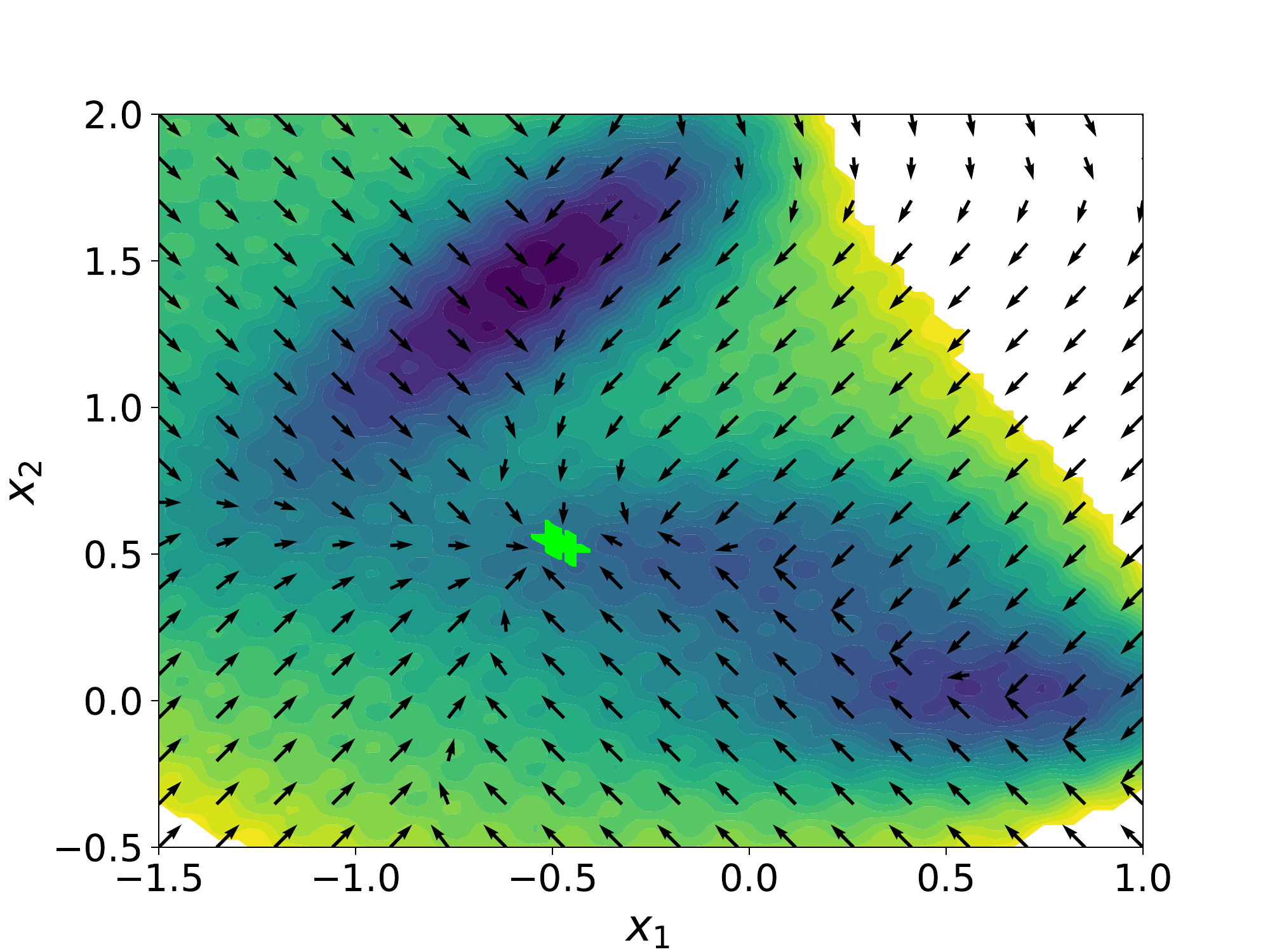}}
\subfigure[Rugged Muller $\beta=0.25$]{\includegraphics[width=0.4\linewidth ]{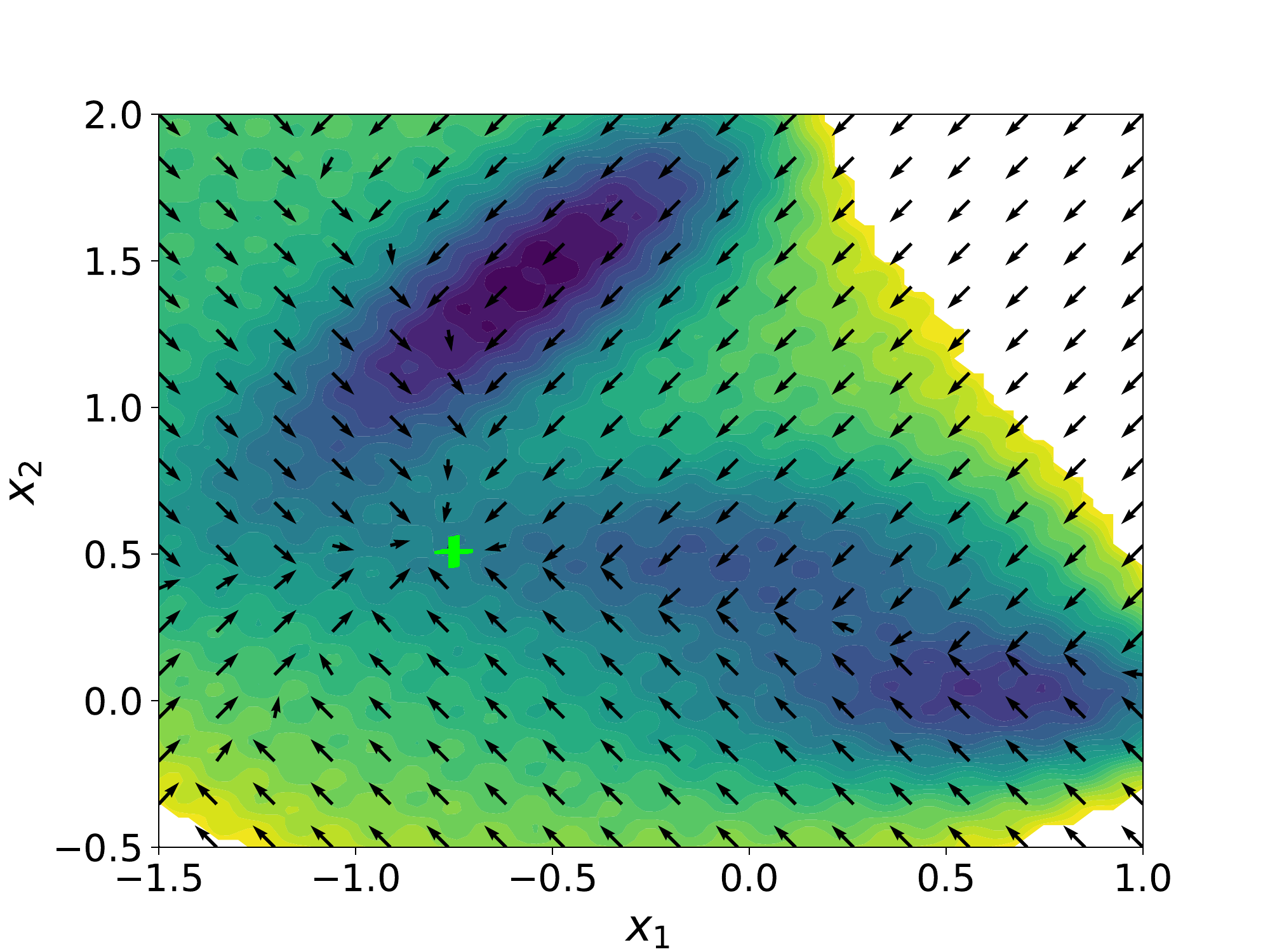}}
\subfigure[Alanine dipeptide]{\includegraphics[width=0.5\linewidth ]{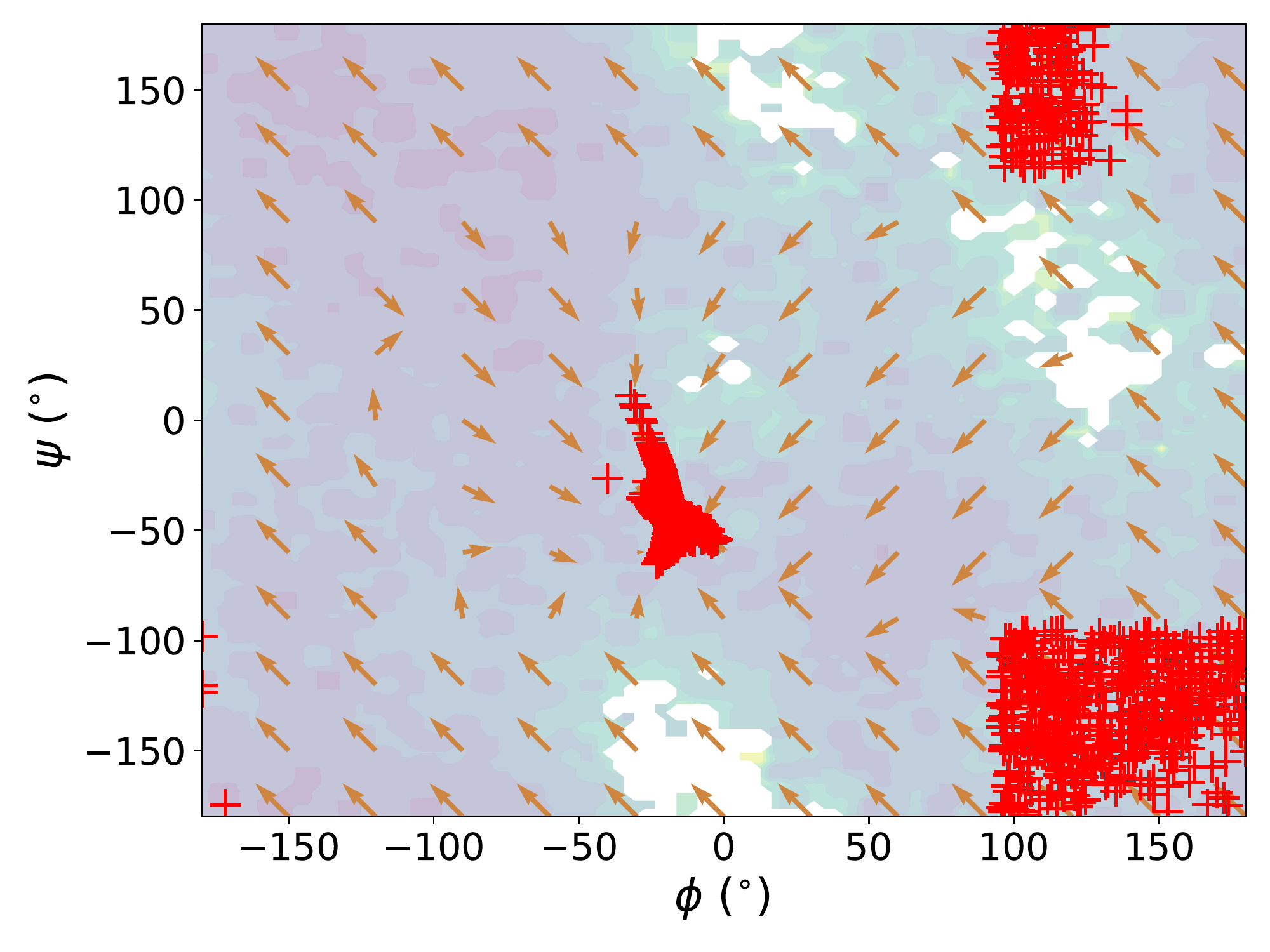}}
\caption{The learned action produced by the RL algorithm where the initial configurations are randomly sampled from the meta-stable region uniformly. Each ``+'' represents the configuration generated in the final configuration of RL episode.}
\label{fig:stable}
\end{figure*}

In the numerical results presented in section~\ref{sec:hightemresults}, the initialization of the reinforcement learning (RL) algorithm is performed by randomly sampling from $\Omega$ uniformly. In a  revised approach, we revise the initialization scheme by sampling the initial configurations from the meta-stable regions. To show the impact of this change, we retrain the RL model and plot the final configurations obtained from RL procedure using 1,000 different initializations sampled from the meta-stable region in Figures~\ref{fig:stable}.

With this initialization scheme, the RL method still successfully identifies two regions with a high reactive probability for the triple-well potential. Similarly, in the case of the Rugged Muller potential, the RL algorithm converges closely to the configurations depicted in Figures~\ref{fig:RMlow-action} and \ref{fig:RMhigh-action}. Furthermore, in the Alanine dipeptide case, the RL approach reveals reactive regions similar to those shown in Figure~\ref{fig:AD-action}. These findings demonstrate the robustness of the RL model with respect to different initialization strategies. 

\subsection{Solving the BKE using NN with different training datasets}
\label{sec:dataset}
\begin{figure*}[ht]
\centering
\subfigure[Uniformly sampled in $\Omega$.]{\includegraphics[width=0.32\linewidth ]{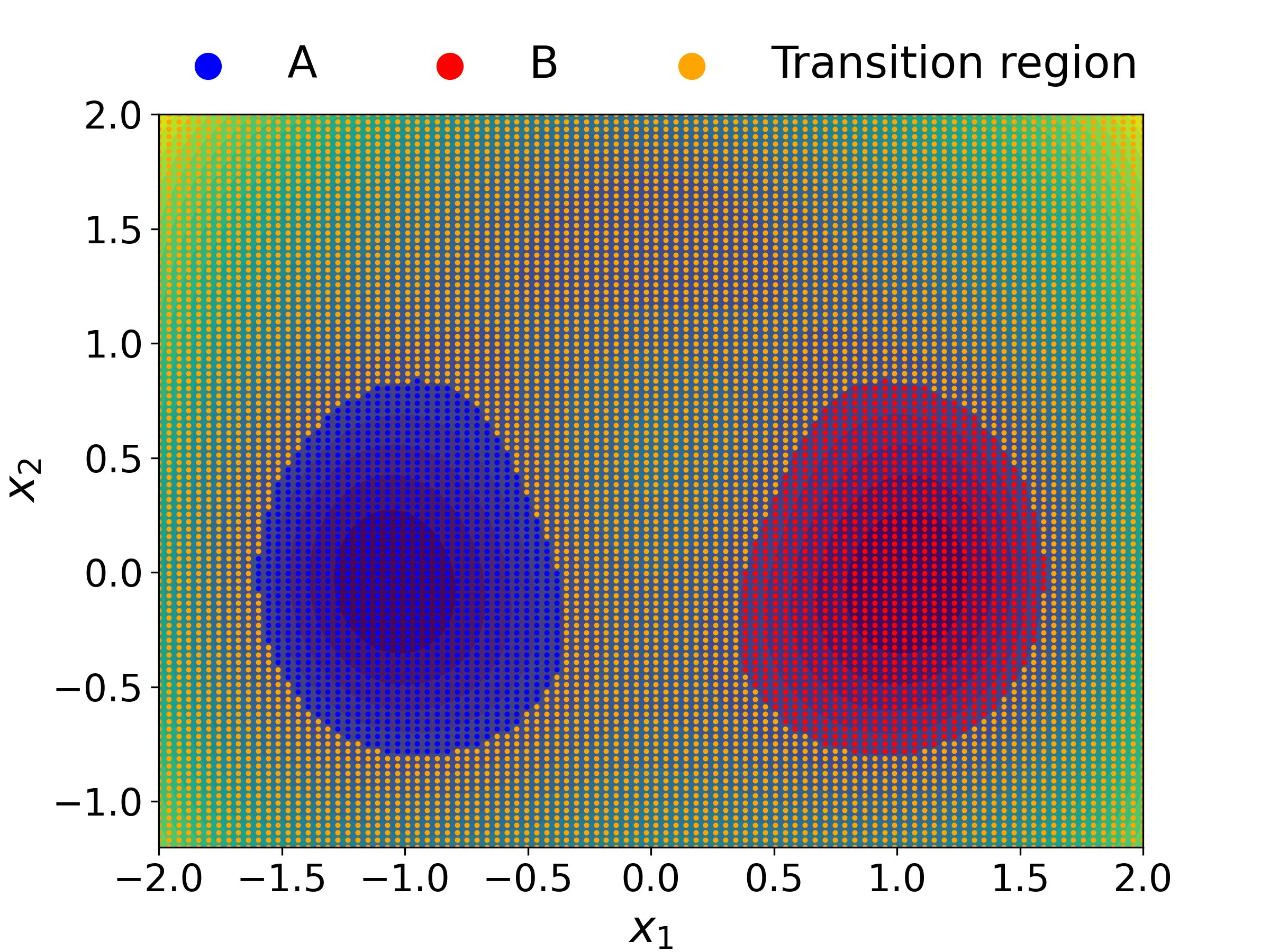}}
\subfigure[Overdamped Langevin ($\beta=1.67$)]{\label{fig:Tri-langvin-beta1p67}\includegraphics[width=0.32\linewidth ]{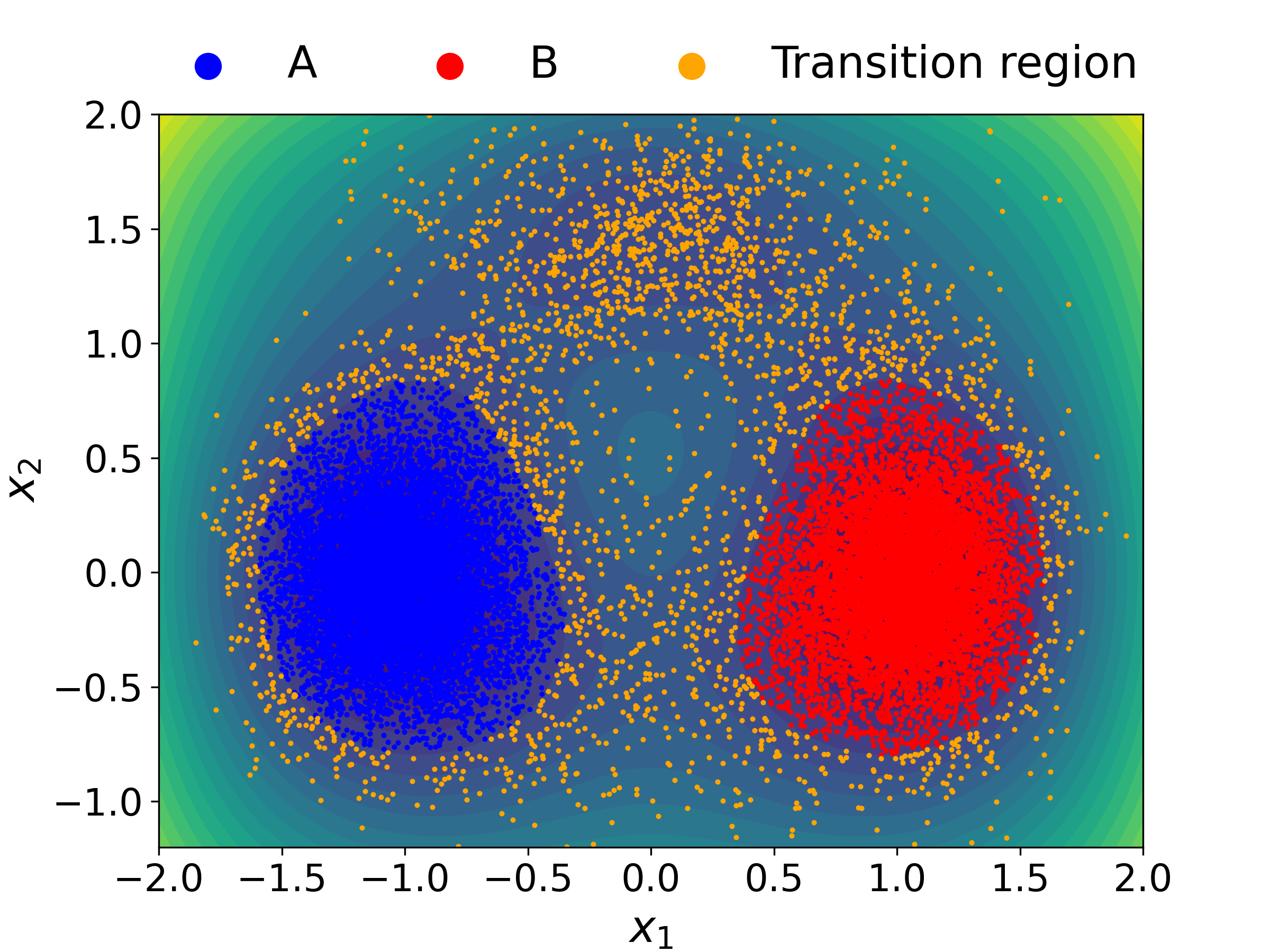}}
\subfigure[Shooting from CC]{\includegraphics[width=0.32\linewidth ]{fig_triple_well/tri-shooting-beta-6.67.png}}
\caption{Different training datasets used for NN solutions of the BKE for the triple well potential with $\beta=6.67$. 
(a) Uniform samples from $\Omega$. (b) Samples along a overdamped Langevin dynamics trajectory with $\beta=1.67$. (c) Configurations generated by shooting trajectories from the identified connective configurations. } \label{fig:Tridataset}
\end{figure*}

\begin{figure*}[ht]
\centering
\subfigure[MSE of the NN loss function,]{\label{fig:Tri-eqnloss}\includegraphics[width=0.24\linewidth ]{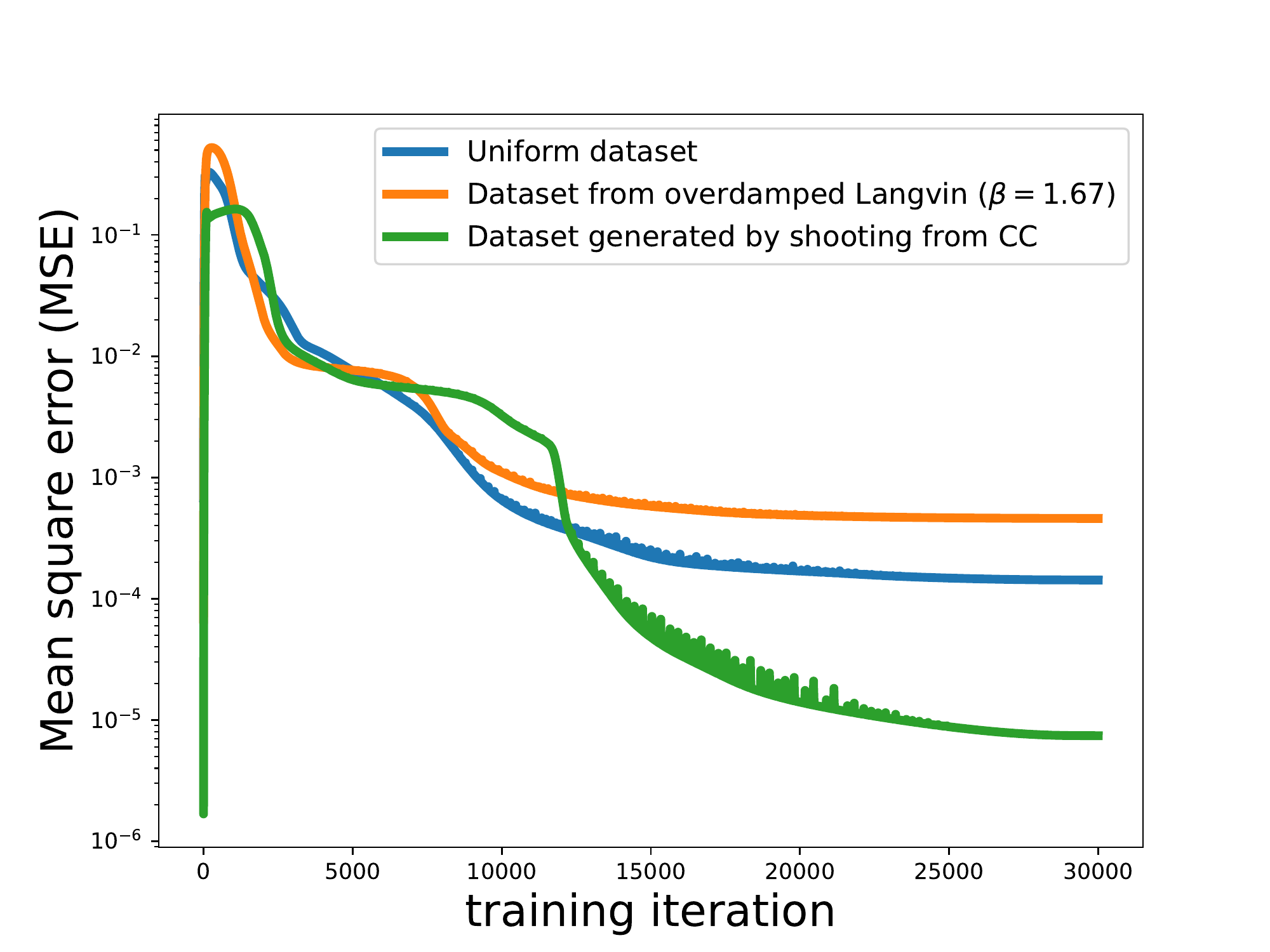}}
\subfigure[Uniformly sampled in $\Omega$.]{\includegraphics[width=0.24\linewidth ]{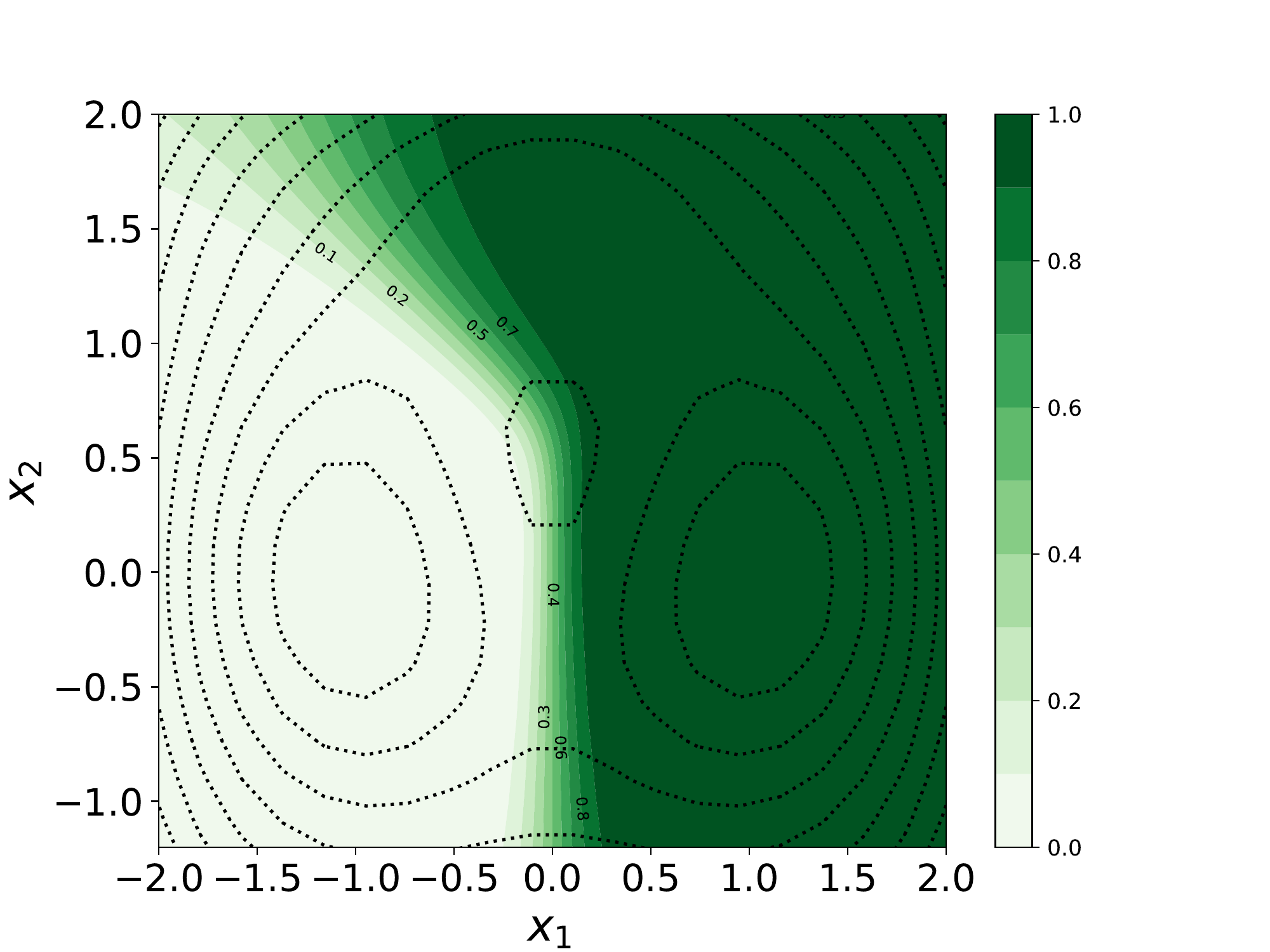}}
\subfigure[Overdamped Langevin ($\beta=1.67$)]{\includegraphics[width=0.24\linewidth ]{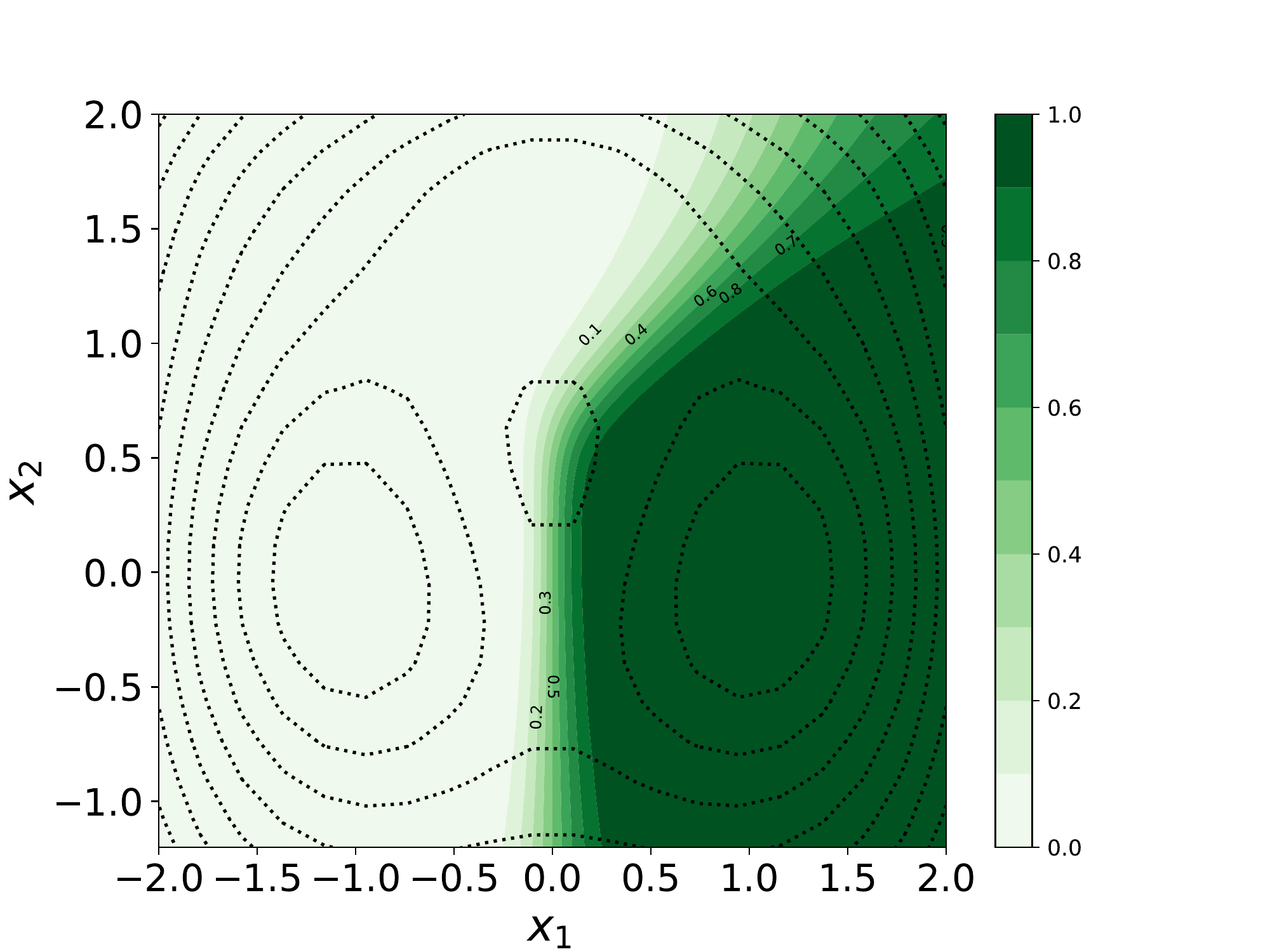}}
\subfigure[Shooting from CC]{\includegraphics[width=0.24\linewidth ]{fig_triple_well/NN_beta6p67.pdf}}
\caption{(a) Mean square error of the training loss at configurations sampled from the transition region of the triple well potential. (b-d) NN solutions  obtained from different training datasets.  }
\label{fig:Tri-BKEloss}
\end{figure*}

\begin{figure*}[ht]
\centering
\subfigure[Uniformly sampled in $\Omega$.]{\includegraphics[width=0.32\linewidth ]{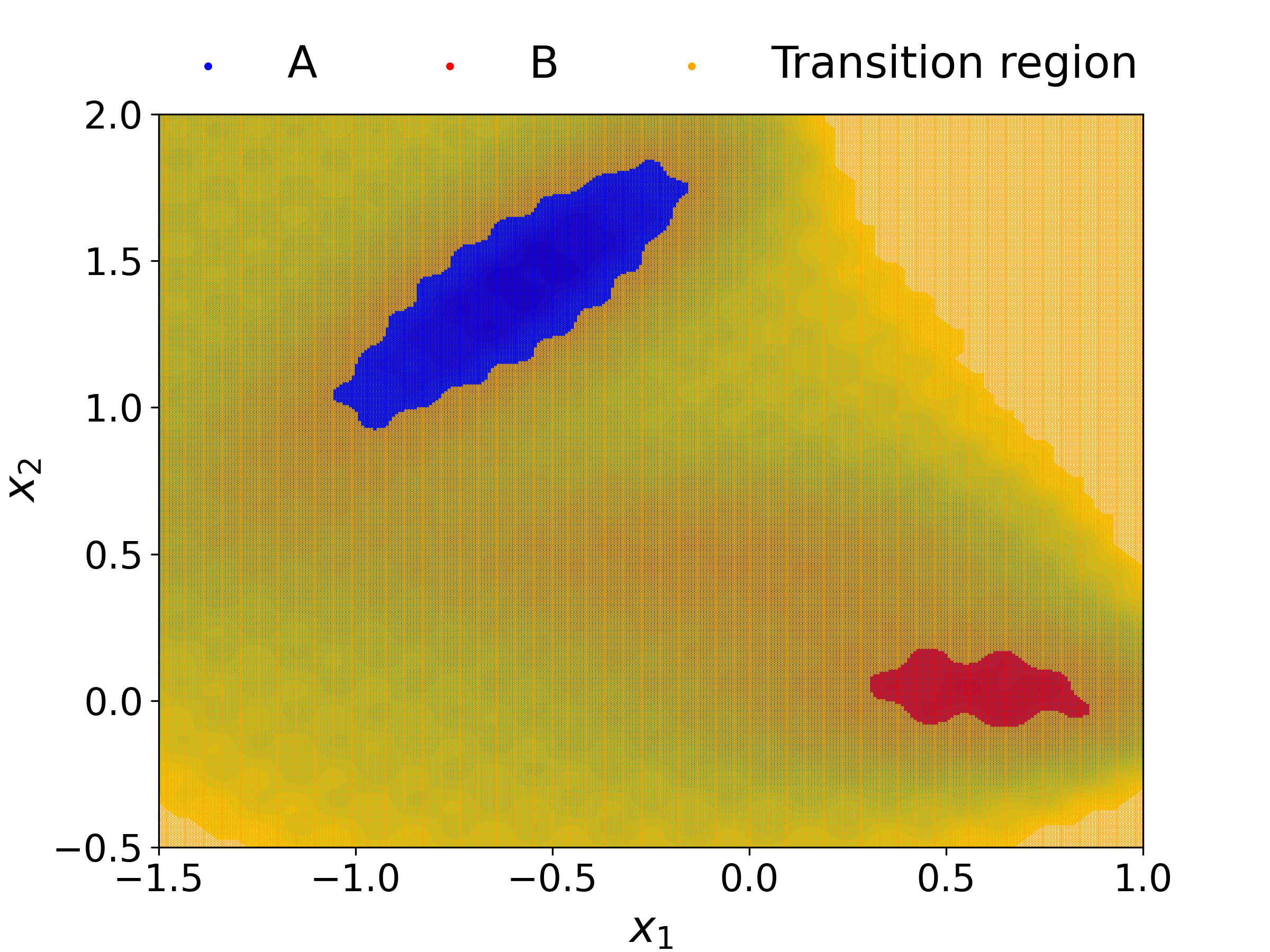}}
\subfigure[Overdamped Langevin ($\beta=0.05$)]{\includegraphics[width=0.32\linewidth ]{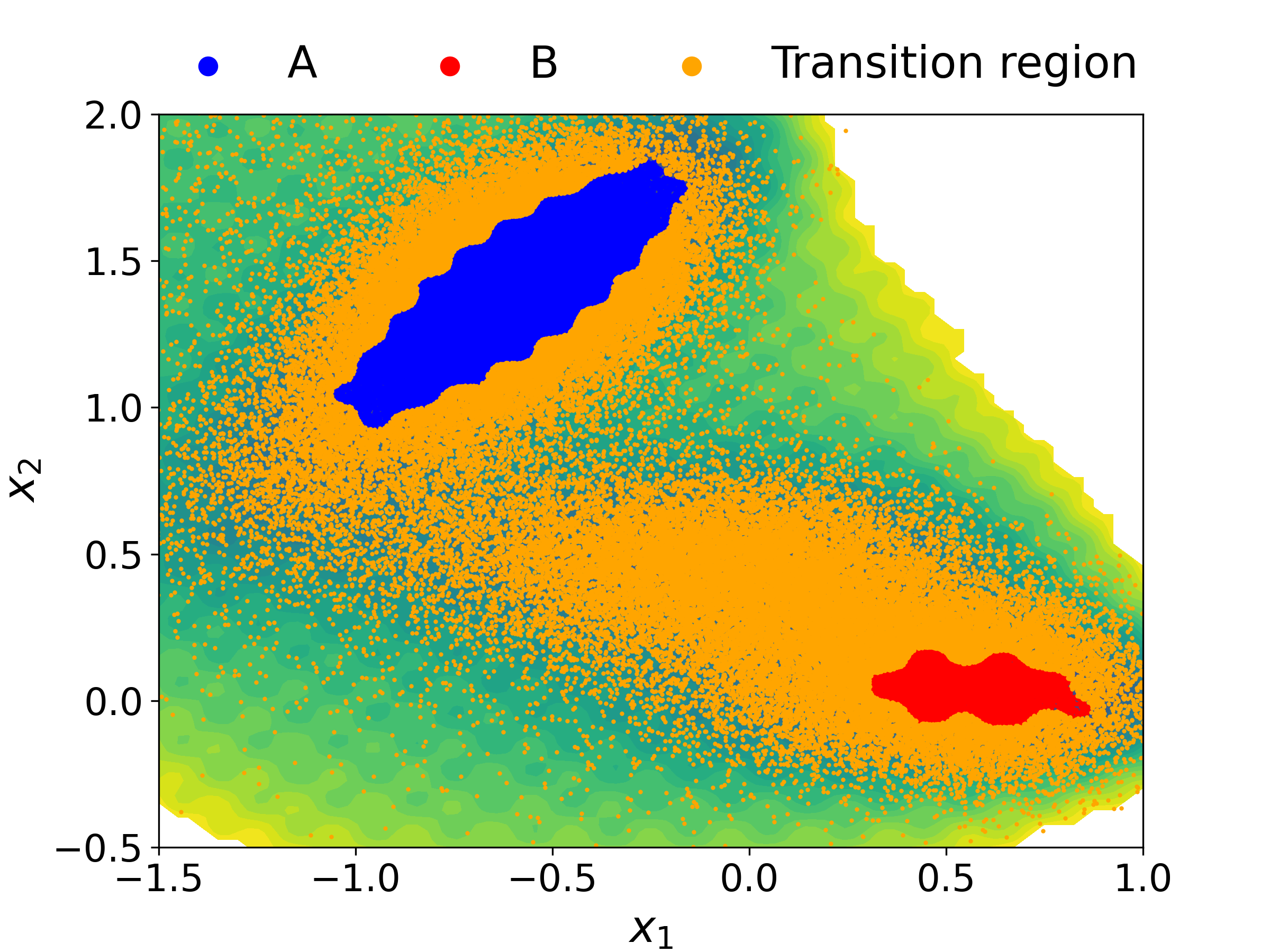}}
\subfigure[Shooting from CC]{\includegraphics[width=0.32\linewidth ]{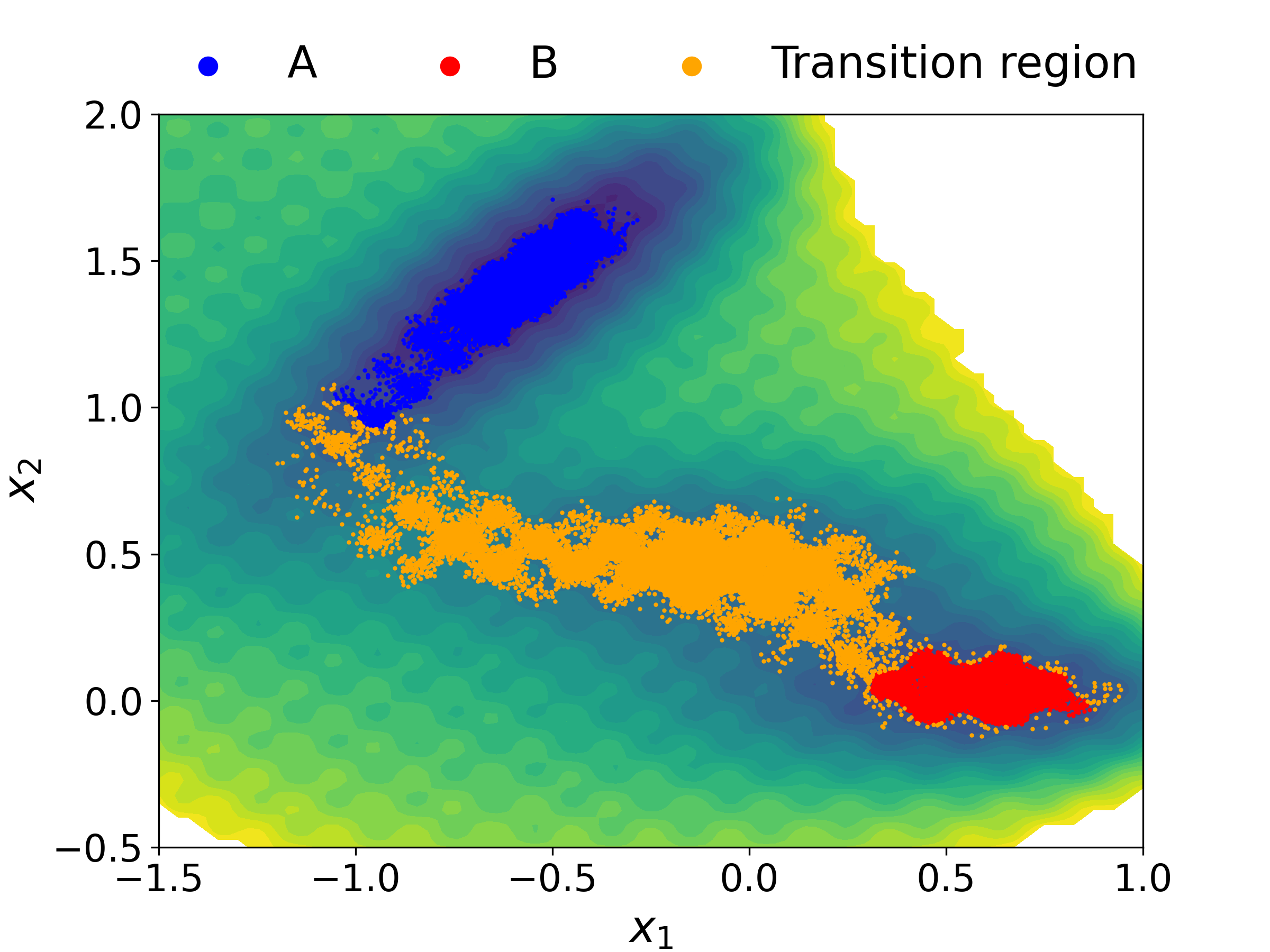}}
\caption{Different training datasets used for NN solutions of the BKE for the Rugged Muller potential with $\beta=0.25$. (a) Uniformly sampled in $\Omega$. (b) Sample along a overdamped Langevin dynamics trajectory with $\beta=0.05$. (c) Configurations generated by shooting from the identified connective configurations.}
\label{fig:RMdataset}
\end{figure*}

\begin{figure*}[ht]
\centering
\subfigure[MSE for BKE]{\includegraphics[width=0.24\linewidth ]{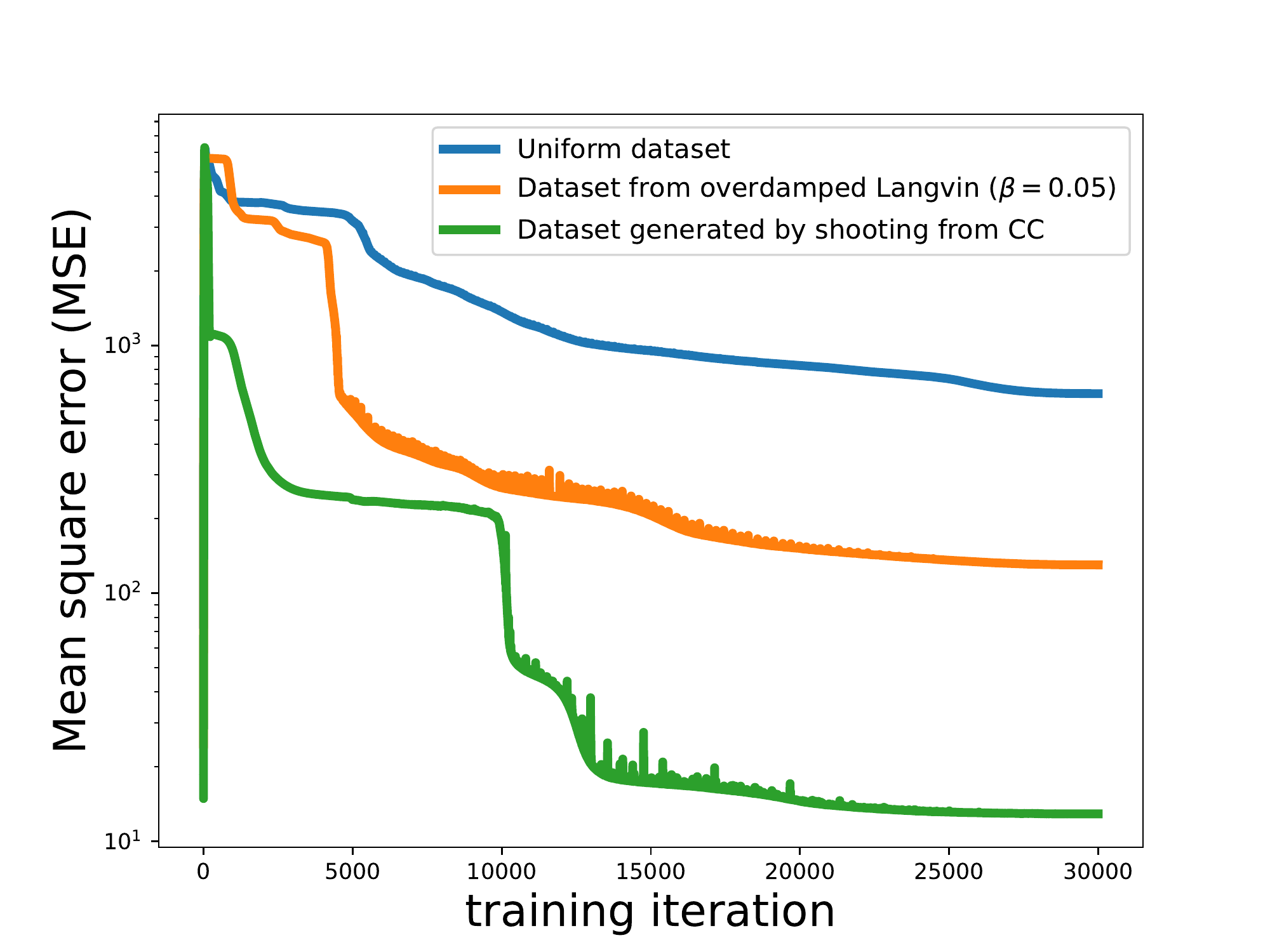}}
\subfigure[Uniform dataset]{\label{fig:Tri-uniform}\includegraphics[width=0.24\linewidth ]{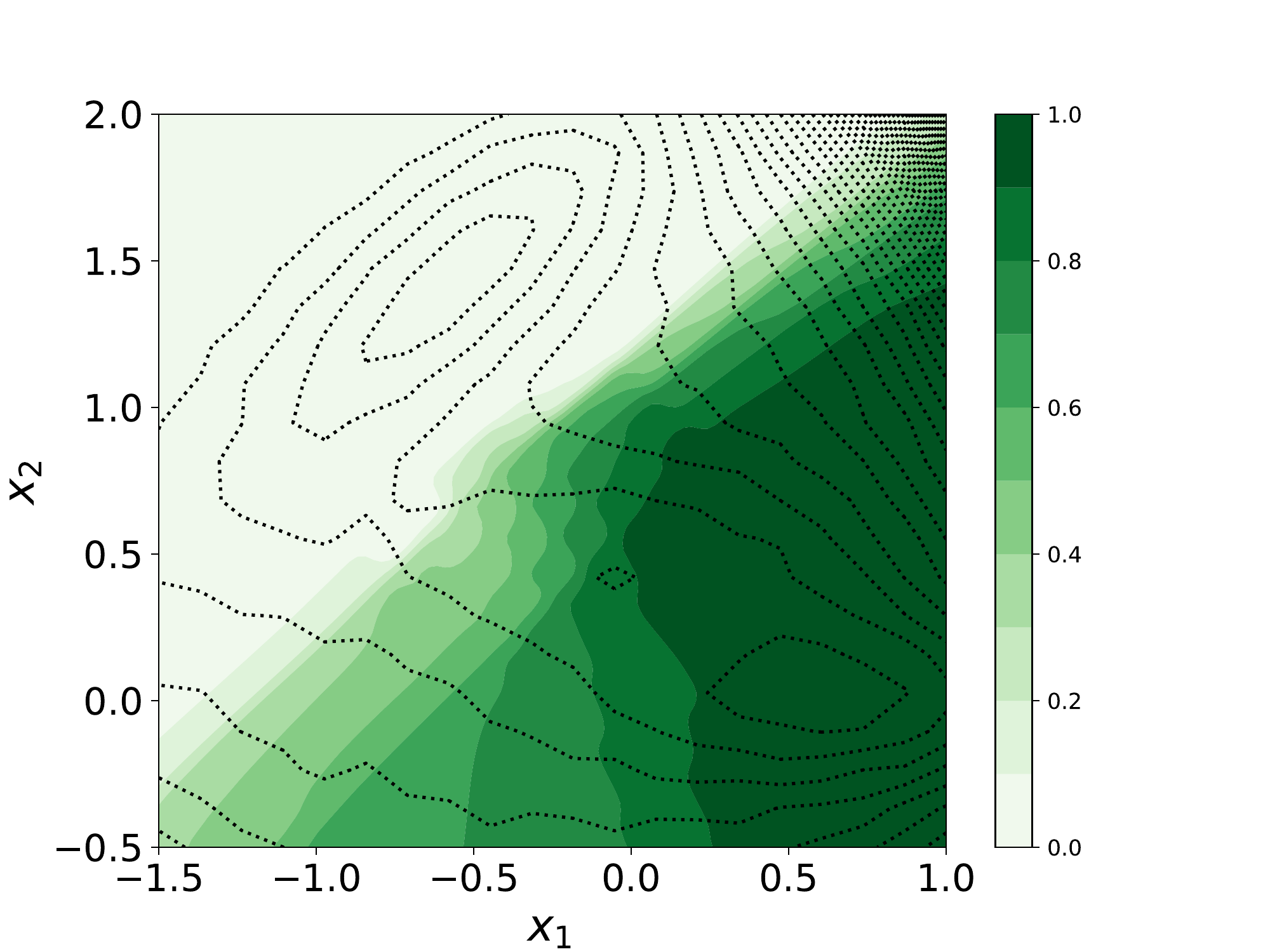}}
\subfigure[Overdamped Langevin ($\beta=0.05$)]{\includegraphics[width=0.24\linewidth ]{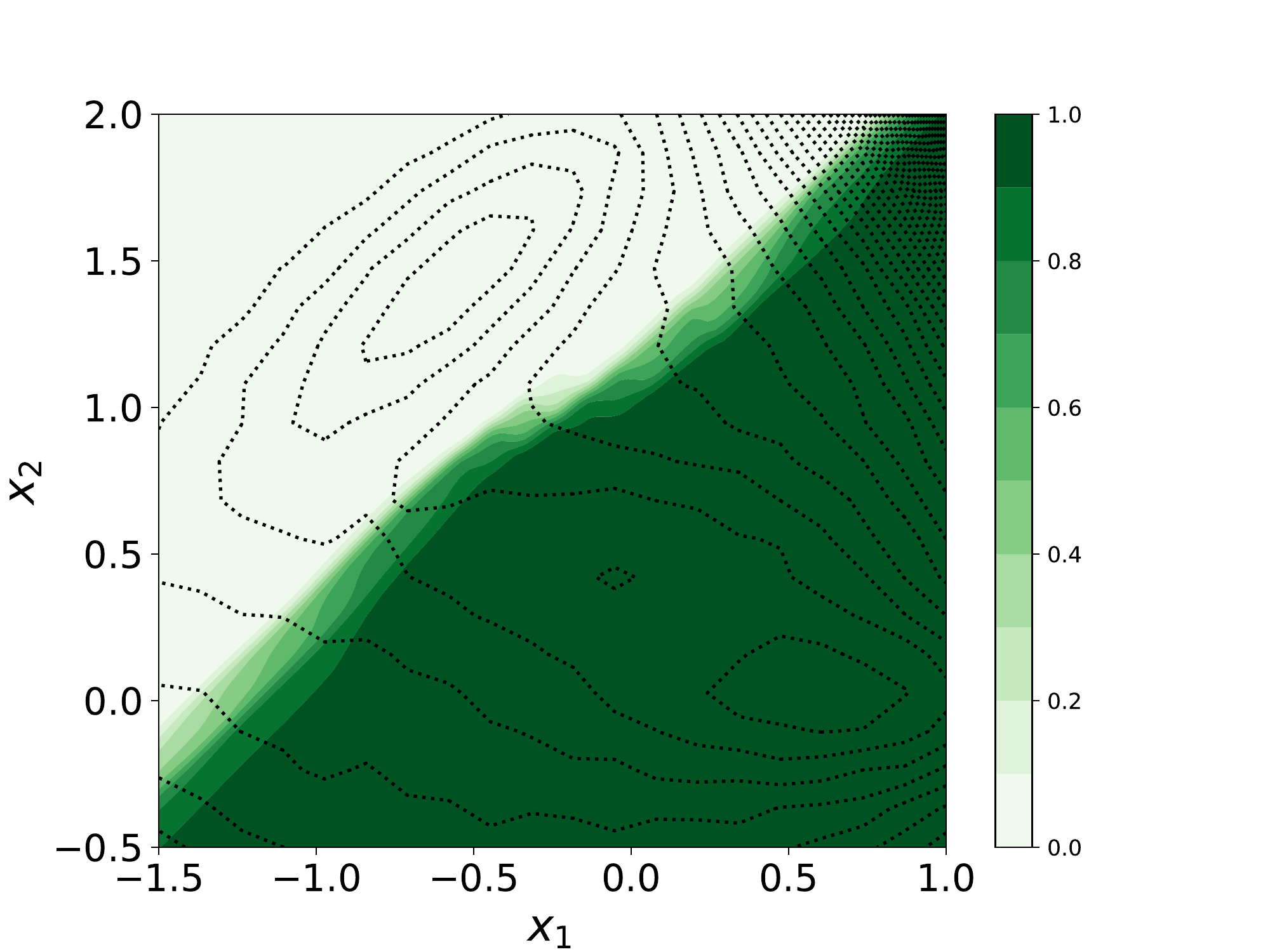}}
\subfigure[Shooting from CC]{\includegraphics[width=0.24\linewidth ]{fig_muller/RM_NN_sampled_pts_beta0p25.pdf}}
\caption{(a) Mean square error of the training loss at configurations sampled from the transition region of the rugged Muller-Brown potential. (b-d) NN solutions obtained from different training datasets. 
}
\label{fig:RM-BKEloss}
\end{figure*}

The utilization of NN-based optimization methods is often associated with an implicit bias toward fitting smooth functions that exhibit fast decay in the frequency domain.~\cite{FP} Consequently, training NN models that can be used to approximate the committor function can be challenging when we attempt to capture drastic changes in the committor function. Such implicit bias can be mitigated by using an appropriate training dataset.~\cite{liang2022stiffness} By carefully selecting the training dataset to provide the necessary samples that encompass the desired variations in the committor function, we can overcome the limitations posed by the implicit bias of NN-based optimization. Here we compare NN solutions of the BKE trained on various datasets. These datasets consist of configurations sampled uniformly on $\Omega$, configurations obtained from overdamped Langevin dynamics ran at a higher temperature, and  configurations generated by shooting from CC's, as depicted in Figures~\ref{fig:Tridataset} and~\ref{fig:RMdataset} respectively. We evaluate the performance of the NN on two examples (a triple-well potential and a Rugged Muller potential) at a low temperature. Figures~\ref{fig:Tri-BKEloss} and~\ref{fig:RM-BKEloss} show the mean square error (MSE) of the training loss   and the NN solutions obtained from different training datasets.

From these results, it is evident that the NN solution trained on the dataset generated by shooting trajectories from CC's achieves a lower MSE and yields a more accurate approximation. 

\end{document}